\documentclass[a4paper,11pt]{article}

\usepackage{amsmath}
\usepackage{amsfonts}
\usepackage{amssymb}
\usepackage{amsthm}
\usepackage{jheppubA}
\usepackage{graphicx}
\usepackage{hyperref}
\usepackage[numbers,sort&compress]{natbib}

\newcommand{\I}{\mathrm{i}}
\newcommand{\D}{\mathrm{d}}
\newcommand{\C}{\mathbb{C}}

\renewcommand{\O}{\mathcal{O}}
\newcommand{\<}{\langle}
\renewcommand{\>}{\rangle}
\newcommand{\bs}[1]{\boldsymbol{#1}}
\newcommand{\nn}{\nonumber}
\newcommand{\lla}{\langle \! \langle}
\newcommand{\rra}{\rangle \! \rangle}

\newcommand{\z}{\zeta}
\newcommand{\dreg}{\hat{d}}
\newcommand{\Dreg}{\hat{\Delta}}
\newcommand{\g}{\gamma}

\DeclareMathOperator{\re}{Re}

\DeclareMathOperator{\Li}{Li}

\newcommand{\Ast}{\scalebox{2.0}{\raisebox{-0.2ex}{$\ast$}}}

\title{Dimensional renormalization in AdS/CFT}
\author{Adam Bzowski}
\affiliation{Institute for Theoretical Physics, KU Leuven, Celestijnenlaan 200D, 3001 Leuven, Belgium}
\emailAdd{adam.bzowski@kuleuven.be}

\numberwithin{equation}{section}
\begin{document}

\abstract{
In this paper we present a dimensional renormalization scheme suitable for holographic theories. We use the bulk physics in the supergravity limit as a definition of the dual CFT. Similar to the perturbative quantization of a QFT, one is free to choose a convenient renormalization scheme, and the holographic renormalization method is only one such choice. We show how the bulk theory can be rendered finite with a renormalization scheme that corresponds to dimensional renormalization in the dual CFT. The method does not require any cut-offs and does not introduce any dimensionful parameters. It delivers a one-to-one map between bulk and boundary counterterms and leads to an exact and unambiguous identification of field theoretical objects in terms of bulk data. In particular, we resolve long standing issues regarding the identification of the renormalization scale and beta functions on both sides of the AdS/CFT correspondence. Furthermore, the method is considerably simpler than standard holographic renormalization on a practical level when evaluating correlation functions.
}

\maketitle

\section{Introduction}

In the supergravity limit, the AdS/CFT correspondence provides an exact equivalence between a classical gravitational field theory living in the bulk of an asymptotically locally AdS space (AlAdS) and a quantum field theory (QFT) living in one less dimension. The AdS/CFT dictionary in this limit consists of two postulates. First, for every bulk field $\Phi$ there exists a primary single trace operator $\O$ in the boundary theory and its source $\phi_0$ is equal to the value of $\Phi$ at the boundary, up to a conformal factor. Second, if $S[\phi_0]$ denotes the on-shell action for the bulk field $\Phi$ with the boundary condition prescribed by $\phi_0$ and $W[\phi_0]$ denotes the generating functional of connected correlation functions of the dual operator $\O$, then the AdS/CFT correspondence postulates
\begin{equation} \label{e:adscft}
S[\phi_{0}] = - W[\phi_{0}],
\end{equation}
for finite, renormalized expressions. However, when a specific problem is considered, usually both the generating functional and the on-shell action diverge and require renormalization. The renormalization of quantum field theories is a broad, well-studied, and well-understood topic. Among various methods, two renormalization schemes are among most common and therefore most important: cut-off and dimensional renormalization. As for the renormalization of the AdS action, the standard procedure has been holographic renormalization.

Holographic renormalization \cite{Henningson:1998gx,Henningson:1998ey,Emparan:1999pm,deHaro:2000vlm,Bianchi:2001de,Bianchi:2001kw} is a powerful procedure suitable for renormalization of holographic theories. Since divergences in the on-shell action in \eqref{e:adscft} emerge from the asymptotic region of the bulk spacetime, the procedure introduces an appropriate cut-off. Next, suitable counterterms located on a cut-off surface are added, so that eventually the cut-off can be removed, yielding finite renormalized correlation functions. Despite its appeal, a successful execution of holographic renormalization can be technically involved as several counterterms are usually required. Furthermore, the form of counterterms changes from case to case and hence each system requires an execution of the entire procedure. Most treatments concentrate on the holographic renormalization of a bulk system consisting of gravity coupled to a specific scalar field \cite{Gubser:1998bc,deHaro:2000vlm,Bianchi:2001de,Bianchi:2001kw,Kalkkinen:2001vg,Martelli:2002sp,Papadimitriou:2004ap,Papadimitriou:2004rz,Papadimitriou:2007sj,Papadimitriou:2010as,Papadimitriou:2011qb,Bourdier:2013axa,Elvang:2016tzz}, where the dual operator is relevant or the analysis is applicable to 2-point functions only.

In this paper we propose a dimensional renormalization scheme, where the regulator shifts the spacetime dimension $d$ and conformal dimensions $\Delta$ of operators without the introduction of a cut-off. We treat objects such as correlation functions as functions of dimensions and derive them in a certain region of the parameter space $(d, \Delta)$. Then we use the principle of analytic continuation to extend the result beyond this region. An advantage of the proposed method relative to holographic renormalization is that it does not introduce any additional scales and required fewer counterterms, if any. In this paper we analyze a simple model of a real, interacting bulk scalar field on a fixed AdS background dual to an operator of arbitrary dimension $\Delta$. The method, however, is applicable in more general settings involving other bulk fields such as gravity or gauge fields, non-trivial backgrounds, or more complicated Lagrangians. Here we elucidate the details of the proposed dimensional renormalization method in a simple context rather than struggle with technicalities not directly related to the procedure. For comparison, a comprehensive analysis of generic irrelevant scalar operators in the context of holographic renormalization is rather cumbersome and can be found in \cite{vanRees:2011fr,vanRees:2011ir}.

This idea of using analytic continuation for the derivation of position space 3- and 4-point functions of various operators was used extensively in the literature, \cite{Witten:1998qj,Muck:1998rr,Freedman:1998tz,Liu:1998bu,Mueck:1998ug,Liu:1998ty,D'Hoker:1999pj,vanRees:2011fr,Raju:2011mp}. While generally expressions presented in these papers may be defined by the analytic continuation in dimensions, for special values of $d$ and $\Delta$ one encounters singularities that must be dealt with. In the framework of holographic renormalization these singularities signal an appearance of special, logarithmic counterterms and this observation was used in \cite{Schwimmer:2000cu,Schwimmer:2003eq} to rederive holographic anomalies. Here, we show how dimensional renormalization simplifies calculations of 3- and 4-point functions. Since properties such as locality of counterterms are manifest in momentum space, we employ the momentum space formalism in our calculations.

On a conceptual level, the additional, radial direction in the bulk should correspond to the RG scale of the dual field theory. Despite significant progress, the details of such a correspondence remain elusive. A successful approach to the holographic renormalization group \cite{deBoer:1999xf,Fukuma:2000bz,Becchi:2002kj} is based on the relation between the Hamilton-Jacobi equation in the bulk and the Callan-Symanzik equation in the dual QFT. It is shown that rescaling the bulk metric corresponds to changing the renormalization scale of the dual QFT. More recently, there have been investigations into a connection between holographic RG flow and Wilsonian flow, \cite{Heemskerk:2010hk,Faulkner:2010jy,Sin:2011yh,Balasubramanian:2012hb,Radicevic:2011py,Grozdanov:2011aa,Behr:2015yna}. This approach addresses a relation between an effective scale of the dual QFT and the radial cut-off in the bulk.

In this paper we return to a more fundamental concept in textbook QFT where the renormalization scale is introduced via QFT counterterms, as considered in \cite{Skenderis:2002wp}. We show that the proposed procedure is equivalent to the dimensional renormalization of the boundary QFT in the sense that counterterms on both sides of the correspondence match. Since it is the counterterms that introduce the renormalization scale, one can unambiguously address long standing questions regarding the structure of beta functions and RG flows. In particular, we identify unambiguously the renormalization scale, beta functions and anomalies of the dimensionally renormalized theory in terms of the bulk data.

\section{Notation and outlook}

\subsection{Notation and definitions}

In this paper we will consider a single bulk scalar field $\Phi$ in a Euclidean $(d+1)$-dimensional AdS background in the Poincar\'{e} patch,
\begin{equation}
\D s^2 = \frac{1}{z^2} \left[ \D z^2 + \D \bs{x}^2 \right].
\end{equation}
The field $\Phi$ of mass $m$ is dual to a conformal primary operator $\O$ of dimension $\Delta$ satisfying $m^2 = \Delta(\Delta - d)$. Throughout this paper we assume $\Delta > d/2$. 

Furthermore, we consider a specific model described by a real bulk scalar field $\Phi$ with a single interaction of the form $\Phi^M$, with integer $M \geq 3$,
\begin{equation} \label{e:introS}
S = \int_0^\infty \D z \int \D^{d} \bs{x} \sqrt{g} \left[ \frac{1}{2} \partial_\mu \Phi \partial^\mu \Phi + \frac{1}{2} m^2 \Phi^2 - \frac{\lambda}{M} \Phi^M \right],
\end{equation}
where $\lambda$ is an arbitrary coupling constant. By adding multiple interaction terms our method is applicable to any potential $V(\Phi)$ with a Taylor expansion around $\Phi = 0$.

Since we are interested in correlation functions of the dual field theory, our analysis is inherently perturbative in $\lambda$. We do not require the on-shell action to converge in $\lambda$, and equivalently we do not assume that the generating functional converges. We treat these objects as formal power series that keep track of all correlation functions simultaneously. For this reason we solve equations of motion perturbatively in $\lambda$ by expanding the field,
\begin{equation}
\Phi(z, \bs{x}) = \Phi_{\{0\}}(z, \bs{x}) + \lambda \Phi_{\{1\}}(z, \bs{x}) + \lambda^2 \Phi_{\{2\}}(z, \bs{x}) + O(\lambda^3).
\end{equation}
Following the notation of \cite{vanRees:2011fr} by the subscript $\{-\}$ we denote the order of an object with respect to the coupling $\lambda$.

The second part of the AdS/CFT dictionary is the imposition of boundary conditions. Throughout this paper we impose Dirichlet asymptotic boundary conditions on the bulk field. The bulk scalar can be expanded in the radial variable, and by $\phi_{(\alpha)}$ we denote a coefficient of $z^{\alpha}$ in this expansion. In every AlAdS spacetime the expansion contains two universal terms: a source coefficient $\phi_{(d - \Delta)}$ and a vev coefficient $\phi_{(\Delta)}$. Dirichlet boundary conditions identify the CFT source $\phi_0$ with the leading source term in the $\lambda$ expansion,
\begin{equation} \label{e:sources}
\phi_0 =\phi_{\{0\}(d - \Delta)}, \qquad\qquad \Phi = z^{d - \Delta} \phi_0 + \text{subleading in } z.
\end{equation}
The vev coefficient is then related to the 1-point function $\< \O \>_s$ of the dual operator, where the subscript indicates sources turned on.

The dual field theory lives on flat Euclidean spacetime. Given a generating functional $W[\phi_0]$ of the dual field theory, correlation functions read
\begin{equation}
\< \O(\bs{x}_1) \ldots \O(\bs{x}_n) \> = \left. (-1)^{n} \frac{\delta^n W[\phi_0]}{\delta \phi_0(\bs{x}_1) \ldots \delta \phi_0(\bs{x}_n)} \right|_{\phi_0 = 0}
\end{equation}
In this paper we always consider generating functionals of connected correlation functions and by $\< - \>$ we denote connected correlators.

The majority of the calculation is carried out in momentum space. This is similar to the standard dimensional renormalization method applied to momentum space Feynman diagrams. For the holographic theories under consideration, we analyze the momentum space version of Witten diagrams. We follow the notation of \cite{Bzowski:2013sza}: $\bs{x}$ denotes a general vector in position space while $\bs{k}$ denotes a vector in momentum space. By $x = |\bs{x}|$ and $k = | \bs{k} |$ we denote lengths of the vectors. Fourier transform of the $n$-point function is denoted by $\< \O(\bs{k}_1) \ldots \O(\bs{k}_n) \>$. Due to momentum conservation, this object contains a delta function, so we can define
\begin{equation}
\< \O(\bs{k}_1) \ldots \O(\bs{k}_n) \> = (2 \pi)^d \delta \left( \sum_{j=1}^n \bs{k}_j \right) \lla \O(\bs{k}_1) \ldots \O(\bs{k}_n) \rra.
\end{equation}

\subsection{Dimensional regularization}

A typical renormalization procedure consists of two steps: regularization and renormalization. In the first step one introduces a regulator so that undefined quantities become finite. One can choose various regularization methods such as a cut-off or dimensional regularization schemes. Cut-off regularization is very similar to the holographic scheme: one identifies the spacetime region producing the divergence and excises it so that QFT quantities become finite. In contrast, the dimensional regularization method introduces a dimensionless regulator, henceforth denoted by $\epsilon$, that shifts dimensions.

The procedure of dimensional regularization starts with finding a non-empty open region of the parameter space $(d, \Delta)$ such that the on-shell action in \eqref{e:introS} produces finite correlation functions of the dual CFT. We shall prove that all correlators can be regarded as functions of $d$ and $\Delta$ and they are finite and analytic analytic at least in the region
\begin{equation} \label{e:introCond}
\frac{d}{2} < \Delta < \min \left( \frac{d}{2} + 1, \frac{M-1}{M} d \right),
\end{equation}
where $M$ is the order of interaction in the action \eqref{e:introS}. Therefore, in this range all holographic correlation functions are finite and no renormalization is required. Outside this range we invoke the power of analytic continuation. Hence, if one obtains an expression for a correlation function for $d$ and $\Delta$ satisfying \eqref{e:introCond}, it represents the unique valid correlation function for any $d$ and $\Delta$ for which it is well-defined. In particular the standard relation between the 1-point function and the vev coefficient of the bulk field holds,
\begin{equation}
\< \O \>_{s, \text{reg}} = - (2 \Delta - d) \phi_{(\Delta)}.
\end{equation}
The subscript `reg' stands for `regulated', meaning that the equality holds for any $d$ and $\Delta$ as long as the right hand side is well-defined by means of analytic continuation. 

The analytically continued correlation functions may become singular at some points of the parameter space $(d, \Delta)$. A singularity in a correlation function may occur only for some specific descreet set of values of dimensions $d$ and $\Delta$. This is strikingly different to the case of the holographic renormalization, where counterterms are required in almost any case. 

Assume we are interested in the system with dimensions $d$ and $\Delta$ such that some correlation function becomes singular. The regularization procedure shifts these dimensions by small quantities proportional to the regulator,
\begin{equation} \label{e:introReg}
d \longmapsto \dreg = d + u \epsilon, \qquad\qquad \Delta \longmapsto \Dreg = \Delta + v \epsilon,
\end{equation}
where $u$ and $v$ are fixed numbers that indicate a direction of the shift in the parameter plane $(d, \Delta)$. Correlators are now regulated, \textit{i.e.}, finite, but divergent in the $\epsilon \rightarrow 0$ limit. In a local QFT all singularities should be local and may be canceled by the addition of appropriate counterterms. Given a regularization procedure, there exists a set of all possible counterterms. Two different regularization schemes will have two different sets of counterterms. In particular, one usually needs more counterterms in the holographic renormalization scheme than in the dimensional scheme. This stems from the fact that dimensional regularization does not introduce any scale, so there are fewer ways to produce a counterterm of an appropriate dimension.

Typically, counterterms are not uniquely fixed, leading to some scheme-dependence in the renormalized theory. A particular choice of scheme-dependent terms is called a renormalization scheme within the given regularization scheme. For example, in the context of dimensionally regulated perturbative QED the most standard renormalization schemes are the on-shell scheme, MS and $\overline{\text{MS}}$. Furthermore, counterterms depend both on the regulator and an additional scale: a renormalization scale $\mu$.

A fully renormalized QFT does not depend on the choice of regularization or renormalization scheme. For example, perturbative QED is uniquely defined by its classical Lagrangian, irrespectively of the regularization and renormalization methods. Any discrepancies originating from a choice of the renormalization scheme must be local and adjustable by local counterterms. However, a choice of a convenient scheme may significantly simplify the renormalization procedure.

We will say that two renormalization schemes are equivalent if there exists a one-to-one correspondence between counterterms and their respective contributions to correlation functions are equal.\footnote{This is a stronger definition than the standard one, \cite{Collins:1984xc}. Usually one would say that two renormalization schemes are equivalent if they lead to the same quantized theory. Cut-off and dimensional renormalization schemes would then be equivalent.} For example, MS and $\overline{\text{MS}}$ renormalization schemes are trivially equivalent as they share the same set of counterterms. In this paper we show that our proposed dimensional renormalization procedure is equivalent to the standard dimensional renormalization of the dual CFT. Bulk counterterms are covariant in the bulk, built up out of objects such as the bulk fields. The boundary counterterms involve boundary data such as the dual operator $\O$ and its source.

Finally, let us point out that if two renormalization schemes are equivalent, then they have identical beta functions. Indeed, beta functions are regularization-dependent objects, so there is no reason for two beta functions in two different regularization schemes to agree. Nevertheless, within equivalent renormalization schemes, beta functions depend on divergences only and hence are immune to finite counterterms.

\subsection{Renormalization}

For a generating functional of the dual CFT we introduce the following notation: $W_{\text{reg}}[\phi_0; \epsilon]$ denotes a regulated generating functional, where the dimensions $d$ and $\Delta$ are shifted according to \eqref{e:introReg}. The regulated functional is expected to have a singularity at $\epsilon = 0$ and appropriate counterterms need to be added. These can be gathered into a counterterm functional $W_{\text{ct}}[\phi_0; \epsilon, \mu]$, which now -- in addition to the regulator -- depends on a renormalization scale $\mu$ arising purely on dimensional grounds. Finally the renormalized generating functional is
\begin{equation} \label{e:introWren}
W[\phi_0; \mu] = \lim_{\epsilon \rightarrow 0} \left( W_{\text{reg}}[\phi_0; \epsilon] + W_{\text{ct}}[\phi_0; \epsilon, \mu] \right).
\end{equation}
By a slight abuse of language we also refer to the sum $W_{\text{reg}} + W_{\text{ct}} = W + O(\epsilon)$ as a renormalized generating functional. The same notation is applicable to the on-shell action and all correlation functions.

The proposed renormalization procedure is recursive in nature with an induction over the order of $\lambda$. At each step two sources of divergences may be identified. The first occurs when the asymptotic value problem \eqref{e:sources} in the bulk becomes ill-defined. Indeed, only when $\Delta < d$ the source term of order $d - \Delta$ is the most leading term in the radial expansion of the bulk field $\Phi$. Dimensional renormalization leads to an unambiguous extension of the definition \eqref{e:sources} to the case $\Delta \geq d$. In the process, however, divergences may arise, which can be removed by turning on subleading source terms $\phi_{\{n\}(\dreg - \Dreg)}$. These subleading terms depend on the QFT source $\phi_0$ as well as a renormalization scale $\mu$. Their coefficients are necessarily divergent in the $\epsilon \rightarrow 0$, but the divergence is easy to retrieve from the radial expansion of the bulk field.

The second source of divergences follows from an explicit integration appearing in a definition of correlation functions. Their removal requires a local counterterm action, which we call the Wess-Zumino counterterm for its role in the emergence of anomalies. All in all, we show that the renormalized generating functional of the dual field theory may be expressed in terms of the bulk data as
\begin{align} \label{e:introW}
W[\phi_0; \mu] & = W_{\text{reg}}[\phi_{(\dreg - \Dreg)}[\phi_0; \epsilon, \mu]; \epsilon] + W_{WZ}[\phi_{(\dreg - \Dreg)}[\phi_0; \epsilon, \mu]; \epsilon, \mu] \nn\\
& = \< \exp \left( - \int \D^{\dreg} \bs{x} \phi_{(\dreg - \Dreg)}[\phi_0; \epsilon, \mu] \O \right) \>_{\text{reg}} + W_{WZ}[\phi_{(\dreg - \Dreg)}[\phi_0; \epsilon, \mu]; \epsilon, \mu].
\end{align}
From the point of view of the dimensionally regulated dual QFT, the redefined source $\phi_{(\dreg - \Dreg)}$ is identified with the bare source, while $\phi_0$ is the renormalized source. This form of the renormalized generating functional is the most general form arising from a multiplicative renormalization of the dual QFT.

In the context of a perturbative QFT the first term in \eqref{e:introW} leads to a beta function. Hence, the beta function of a genuine dimensional renormalization procedure can be immediately read off of this expression. For example, in the case of a marginal operator satisfying $\Delta = d$ one finds
\begin{equation}
\beta_{\phi_0} = - \left( \frac{\partial \phi_{(\dreg - \Dreg)}}{\partial \phi_0} \right)^{-1} \mu \frac{\partial}{\partial \mu} \phi_{(\dreg - \Dreg)}.
\end{equation}

The Wess-Zumino counterterm is responsible for the emergence of anomalies. It can be written both in terms of bulk fields as well as QFT data. Its most general form is
\begin{align} \label{e:introSren}
W_{WZ} & = \int \D^{\dreg} \bs{x} \mu^{u \epsilon} \mathcal{L}_{WZ}[\phi_{(\dreg - \Dreg)} \mu^{-(u-v)\epsilon}, \phi_{(\Dreg)} \mu^{-v \epsilon}, \partial_j] \nn\\
& = \lim_{z \rightarrow 0} \int_z \D^{\dreg} \bs{x} \sqrt{\gamma_z} (z \mu)^{u \epsilon} \mathcal{L}_{WZ}[(z \mu)^{-(u-v)\epsilon} \Phi, - (2 \Dreg - \dreg)^{-1} (z \mu)^{-v \epsilon} \Pi, \nabla_j],
\end{align}
for some local functional $\mathcal{L}_{WZ}$, where $\Pi$ denotes a canonical momentum, $\Pi = -z \partial_z \Phi + (\dreg - \Dreg) \Phi$. The second line expresses the counterterm covariantly in terms of the bulk data. Mathematically, to make the second line well-defined, one needs to evaluate it at some cut-off surface of constant $z$ and then take the limit $z \rightarrow 0$. Since in the dimensional renormalization procedure no cut-off is available, this turns out to be a defining property of all possible bulk counterterms: they must be covariant in the bulk fields and supported on the boundary. In other words the leading term in the radial expansion of the second line of \eqref{e:introSren} is proportional to $z^0$, so that the finite limit exists.

The counterterm depends on the renormalization scale both explicitly and via the redefined source $\phi_{(\dreg - \Dreg)}$. The total scaling anomaly $\mathcal{A}$ appears due to the explicit dependence and one finds,
\begin{equation} \label{e:introA}
\mathcal{A} = \mu \frac{\partial}{\partial \mu} \left[ \mu^{u \epsilon} \mathcal{L}_{WZ} \right].
\end{equation}
Since the dependence of \eqref{e:introSren} on the renromalization scale $\mu$ is identical to its explicit dependence on the radial variable $z$, we find,
\begin{equation}
\mu \frac{\partial}{\partial \mu} = z \frac{\partial}{\partial z}.
\end{equation}
This is a long sought relation between the radial direction of the bulk and the renormalization scale of the boundary theory. From the point of view of the bulk theory one is forced to use explicit radial coordinate $z$ in the counterterm \eqref{e:introSren}, which breaks diffeomorphism invariance and hence it introduces the anomaly. Furthermore, since the bulk theory is classical and no scale is introduced via renormalization, the renormalization scale $\mu$ must be related to the only available scale: the AdS radius $L$. By dimensional analysis this simply requires $L \sim \mu^{-1}$ up to a proportionality factor that can be fixed to one.

\section{Scalar 2-point functions}

This section is devoted to the analysis of 2-point functions both from the point of view of the bulk theory as well as the dual CFT. The topic is well-understood since the first days of the AdS/CFT correspondence and hence it allows us to emphasize differences between the holographic renormalization procedure and the proposed dimensional renormalization.

We start by presenting a comparison between the cut-off and the dimensional renormalization procedure of a 2-point function directly in a CFT. Then we discuss the holographic 2-point function obtained by means of the standard holographic renormalization and then we move to the proposed dimensional renormalization. We conclude by showing exact correspondence between the dimensional renormalization directly in the CFT and the dimensional renormalization in AdS. In particular we show the exact correspondence between the counterterms and equality of their local, scheme-dependent contributions. Finally, we show how the scaling anomaly emerges from the holographic context.

Given a dimension $\Delta$ of a scalar operator and $d$ denoting spacetime dimension, we consider two cases depending on the value of $2 \Delta - d$. If $2 \Delta - d \neq 2 n$ for a non-negative integer $n$, then we talk about \emph{a generic case}. On the other hand we will refer to the case of $2 \Delta - d = 2 n$ for a non-negative integer $n$ as \emph{a spacial case}. We always assume than $n$ is a non-negative integer.

\subsection{Renormalization of a CFT}

It is a well-known fact that the conformal symmetry of the dual CFT fixes the form of the 2-point function to be
\begin{equation}
\< \O(\bs{x}) \O(0) \> = \frac{C_{\O}}{x^{2 \Delta}},
\end{equation}
where $C_{\O}$ is a theory-specific constant. Its renormalization properties are best exposed by looking at the Fourier transform. Since the integral 
\begin{equation} \label{e:fourier_trans}
\int \D^d \bs{x} \: e^{-\I \bs{k} \cdot \bs{x}} \frac{1}{x^{2 \Delta}} = \frac{\pi^{d/2} 2^{d - 2 \Delta}}{\Gamma(\Delta)} \Gamma \left( \frac{d}{2} - \Delta \right) k^{2 \Delta - d}
\end{equation}
converges only for $0 < 2 \Delta < d$, a regularization is necessary outside this region.

\subsubsection{Cut-off renormalization} \label{sec:cutoff}

In the cut-off regularization one can regulate the Fourier transform by cutting out a small ball around $x = 0$. If $\delta$ denotes the cut-off, then one finds
\begin{align} \label{e:fourier_transcut}
\int_{x > \delta} \D^d \bs{x} \: e^{-\I \bs{k} \cdot \bs{x}} \frac{1}{x^{2 \Delta}} & = \frac{\pi^{d/2} 2^{d - 2 \Delta}}{\Gamma(\Delta)} \Gamma \left( \frac{d}{2} - \Delta \right) k^{2 \Delta - d} \nn\\
& \qquad - \: \frac{2 \pi^{d/2} \delta^{d - 2 \Delta}}{(d - 2 \Delta) \Gamma \left( \frac{d}{2} \right)} {}_1 F_2 \left( \frac{d}{2} - \Delta; \frac{d}{2}, \frac{d}{2} - \Delta + 1; - \frac{k^2 \delta^2}{4} \right)
\end{align}
assuming $2 \Delta - d \neq 2 n$. As expected, the second term has a finite $\delta \rightarrow 0$ limit only if $2 \Delta < d$. Otherwise, the hypergeometric function can be expanded around $\delta = 0$ leading to a finite number of divergent terms. In particular first two terms are
\begin{equation}
\delta^{d - 2 \Delta} \frac{\pi^{d/2}}{\Gamma \left( \frac{d}{2} \right)} \left[ - \frac{2}{d - 2 \Delta} + \frac{k^2 \delta^2}{d (d - 2 \Delta + 2)} + O(k^4 \delta^4) \right].
\end{equation}
In any case all terms are local, \textit{i.e.}, they have the momentum dependence of the form $(k \delta)^{2 m}$ for an integer $m$, as follows from the definition of hypergeometric functions. Terms divergent in the $\delta \rightarrow 0$ limit are these with $2 m < 2 \Delta - d$ and hence there is only finitely many of them. From their locality it then follows that they can be removed by appropriate counterterms. For example the two listed terms are removable by terms proportional to $\int \D^d \bs{x} \: \phi_0^2$ and $\int \D^d \bs{x} \: \phi_0 \partial^2 \phi_0$, where $\phi_0$ is the source for the operator $\O$. Eventually, in a generic case the 2-point function is given by
\begin{equation} \label{e:2ptcutoff}
\lla \O(\bs{k}) \O(-\bs{k}) \rra = \frac{C_{\O} \pi^{d/2} 2^{d - 2 \Delta}}{\Gamma(\Delta)} \Gamma \left( \frac{d}{2} - \Delta \right) k^{2 \Delta - d}
\end{equation}
for a theory-dependent constant $C_{\O}$. Notice that this results appears on the right hand side of the expression \eqref{e:fourier_trans}, which is well-defined for all values of $d$ and $\Delta$ for which $2 \Delta - d \neq 2 n$. One can therefore ask whether it was really necessary to consider any counterterms.

While the cut-off regularization is a perfectly good renormalization scheme, it is definitely not the simplest one. The regularized expression \eqref{e:fourier_transcut} becomes complicated and requires utilization of special functions. Moreover, as we can see, the counterterms are required in almost any physically interesting case. To be precise, given $d$ and $\Delta$ one needs exactly
\begin{equation} \label{e:number}
\text{number of counterterms} = \max \left( \Big\lfloor \Delta - \frac{d}{2} \Big\rfloor + 1, 0 \right),
\end{equation}
where $\lfloor x \rfloor$ is a floor function rounding $x$ down to the largest integer not exceeding $x$. We only assume $\Delta > 0$ here.

Further inconvenience is created by the fact that the form of counterterms depends strongly on the values of parameters $d$ and $\Delta$. In the special case of $2 \Delta - d = 2 n$, the expression \eqref{e:fourier_transcut} changes and logarithmic terms appear. For example, in case of $n = 1$, \textit{i.e.}, $\Delta = d/2 + 1$ one finds
\begin{equation} \label{e:2ptcutoffreg}
\lla \O(\bs{k}) \O(-\bs{k}) \rra_{\text{reg}} = \frac{C_{\O} \pi^{d/2}}{\Gamma \left( \frac{d}{2} \right)} \left[ \frac{1}{\delta^2} + \frac{k^2}{2 d} \left( - 1 + \gamma_E - \psi \left( \frac{d}{2} + 1 \right) + \log \left( \frac{k^2 \delta^2}{4} \right) \right) + O(\delta^2) \right],
\end{equation}
where $\psi$ denotes the digamma function. The first term is removed by the counterterm as before, but the removal of the logarithmic terms requires a counterterm
\begin{equation}
W_{\text{ct}}[\phi_0] = - \frac{C_{\O} \pi^{d/2}}{4 d \Gamma \left( \frac{d}{2} \right)} \left[ \log ( \delta^2 \mu^2 ) + a_0 \right] \int \D^d \bs{x} \: \phi_0 \partial^2 \phi_0
\end{equation}
to be added to the generating functional $W_{\text{reg}}$. The renormalization scale $\mu$ appears due to dimensional reasons and $a_0$ is an undetermined constant. The 2-point function then reads
\begin{equation} \label{e:2ptrencut}
\lla \O(\bs{k}) \O(-\bs{k}) \rra = c_{\O} k^2 \left[ \log \left( \frac{k^2}{\mu^2} \right) + a_0' \right],
\end{equation}
where
\begin{equation}
c_{\O} = \frac{C_{\O} \pi^{d/2}}{2 d \Gamma \left( \frac{d}{2} \right)}, \qquad\qquad a_0' = -1 + \gamma_E - \psi \left( \frac{d}{2} + 1 \right) - 2 \log 2 - a_0.
\end{equation}
Constant $a_0'$ is scheme-dependent and can be adjusted at will by changing either the renormalization scale $\mu$ or the free counterterm constant $a_0$.

\subsubsection{Dimensional renormalization} \label{sec:dimreg}

In many applications the dimensional renormalization turns out to be a simpler procedure than the cut-off renormalization. One reason is a smaller number of counterterms required, another reason is generally simpler expressions. We have seen that for the cut-off renormalization of a 2-point function one needs a sequence of counterterms whenever $\Delta \geq d/2$. What is more the regularized expressions such as \eqref{e:fourier_transcut} and \eqref{e:2ptcutoffreg} change their form depending on a relation between dimensions $d$ and $\Delta$. On the other hand, as we will see, in the dimensional regularization procedure the regulated expression has a fixed form and the counterterms are required in discreet special cases $2 \Delta - d = 2 n$ only.

Since the expression \eqref{e:fourier_trans} is analytic in $d$ and $\Delta$, one can extend it to any values of parameters if the function is well-defined. This leads to the regulated 2-point function,
\begin{equation} \label{e:2pt}
\lla \O(\bs{k}) \O(-\bs{k}) \rra_{\text{reg}} = c_{\O} k^{2 \Delta - d},
\end{equation}
where we gathered the constant term into a new theory-dependent constant $c_{\O}$. Furthermore, if $2 \Delta - d \neq 2 n$, no counterterms are required and the fully renormalized 2-point function is immediately given by \eqref{e:2pt}.

If, however, $2 \Delta - d = 2 n$, then $k^{2 \Delta - d} = k^{2 n}$ is a local function, being the Fourier transform of $\partial^{2 n} \delta(\bs{x})$. In such a case one can introduce a regulator $\epsilon$ and shift dimensions according to \eqref{e:introReg} for fixed numbers $u$ and $v$. As we will see not all choices of $u$ and $v$ regulate the theory, but a set of such bad pairs is sufficiently small.

In order to regulate the 2-point function \eqref{e:2pt}, one needs to choose a scheme such that the regulated power of momentum, $2 \Dreg - \dreg$, is no longer integral. This requires use of any scheme such that $u \neq 2v$. Then the constant $c_{\O}$ in \eqref{e:2pt} can be expanded in the regulator,
\begin{equation}
c_{\O}(u, v; \epsilon) = \frac{2 c_{\O}^{(-1)}(u, v)}{(2v - u) \epsilon} + c_{\O}^{(0)}(u, v) + O(\epsilon)
\end{equation}
and the regulated 2-point function now reads
\begin{equation} \label{e:2ptreg}
\lla \O(\bs{k}) \O(-\bs{k}) \rra_{\text{reg}} = k^{2 n} \left[ \frac{2 c_{\O}^{(-1)}}{(2v - u) \epsilon} + c_{\O}^{(-1)} \log k^2 + c_{\O}^{(0)} + O(\epsilon) \right].
\end{equation}
The specific factor of $2/(2v - u)$ is chosen in such a way that the leading $c^{(-1)}_{\O}$ constant turns out to be a coefficient of a physically relevant term $k^{2n} \log k^2$.

The leading divergence can be removed by the addition of the counterterm
\begin{equation} \label{e:Sct}
W_{\text{ct}}[\phi_0] = a_{\text{ct}}(u, v; \epsilon) \int \D^{\dreg} \bs{x} \: \phi_0 \partial^{2 n} \phi_0 \mu^{- (u - 2 v) \epsilon}
\end{equation}
to the regulated generating functional $W_{\text{reg}}[\phi_0]$. Here $\phi_0$ denotes the source of the operator $\O$ and the renormalization scale $\mu$ appears due to dimensional reasons. The value of the counterterm constant is only partially determined by the divergence of the regulated 2-point function. By calculating the contribution of the counterterm to the 2-point function one finds that the divergence is cancelled if
\begin{equation} \label{e:act}
a_{\text{ct}}(u, v; \epsilon) = \frac{(-1)^{n} c_{\O}^{(-1)}(u, v)}{(2v - u) \epsilon} + a_0(u, v) + O(\epsilon),
\end{equation}
where $a_0$ is an arbitrary constant. The finite $\epsilon \rightarrow 0$ limit exists and leads to the renormalized correlation function
\begin{align} \label{e:2ptren}
\lla \O(\bs{k}) \O(-\bs{k}) \rra & = k^{2 n} \left[ c_{\O}^{(-1)} \log \left( \frac{k^2}{\mu^2} \right) + c_{\O}^{(0)} + 2 (-1)^{n+1} a_0 \right] \nn\\
& = c'_{\O} k^{2 n} \left[ \log \left( \frac{k^2}{\mu^2} \right) + a_0' \right].
\end{align}
In the last line we redefined theory-specific and scheme-dependent constants to match the form found in the cut-off renormalization \eqref{e:2ptrencut}. This result, however, is valid for any $n$. Furthermore note that in the dimensional renormalization a single counterterm was sufficient for the successful renormalization, regardless of the value of $n$.

\subsection{Holography}

In this section we first review the usual holographic renormalization procedure for the scalar 2-point function in the AdS background. Next, we will show how the dimensional renormalization method can be applied to the evaluation of the 2-point function from the bulk theory. We will discuss how the bulk and boundary counterterms match, leading to the equivalence of the proposed bulk renormalization and the usual dimensional renormalization procedure in the CFT. Finally we show how anomalies emerge in both formalisms and that they both match.

\subsubsection{Set-up}

For the analysis of the scalar 2-point function we need to consider a single bulk scalar field $\Phi_c$ governed by the action,
\begin{equation} \label{e:freeScan}
S = \frac{1}{2} \int_0^\infty \D z \int \D^{d} \bs{x} \sqrt{g_c} \left[ g_c^{\mu\nu} \partial_\mu \Phi_c \partial_\nu \Phi_c + m_c^2 \Phi_c^2 \right]
\end{equation}
on the fixed (Euclidean) AdS background of radius $L$,
\begin{equation}
g^c_{\mu\nu} \D x^\mu \D x^\nu = \frac{L^2}{z^2} \left[ \D z^2 + \D \bs{x}^2 \right].
\end{equation}
Coordinates have dimensions of $L$, which we denote as $[ L ] = [ z ] = [ \bs{x} ] = 1$. The total action remains dimensionless which assigns standard dimensions,
\begin{equation}
[ g^c_{\mu\nu} ] = 0, \qquad [ \Phi_c ] = - \frac{d - 1}{2}, \qquad [ m_c^2 ] = - 2.
\end{equation}
In particular $\Phi_c$ has the standard dimension of the free field (note that $d$ is the spacetime dimension of the boundary theory). Furthermore, from the CFT analysis we know that dimensions of an operator $\O$ and its source $\phi_0$ are
\begin{equation}
[ \phi_{0} ] = - (d - \Delta), \qquad [ \O ] = -\Delta.
\end{equation}

Usually $L$ is set to one, but we anticipate that this scale will be essential for the renormalization of the theory. Nevertheless, one can normalize fields differently, so that the scale will almost entirely disappear. Define field $\Phi$, the metric $g_{\mu\nu}$ and dimensionless mass $m^2$ according to
\begin{equation} \label{e:BulkFieldRedef}
\Phi_c = L^{-\frac{1}{2}(d - 1)} \Phi, \qquad g^c_{\mu\nu} = L^2 g_{\mu\nu}, \qquad m_c^2 = m^2 L^{-2}.
\end{equation}
New fields have the following dimensions
\begin{equation}
[ \Phi ] = 0, \qquad [ g_{\mu\nu} ] = -2, \qquad [ \D^d \bs{x} \sqrt{g} ] = 0, \qquad [m^2] = 0.
\end{equation}
Note that an unusual dimension of the metric remains consistent with the fact that in Feffermann-Graham gauge the dynamical part of the metric $\gamma_{ij}$, defined by
\begin{equation}
\D s^2 = g_{\mu\nu} \D x^{\mu} \D x^{\nu} = \frac{1}{z^2} \left[ \D z^2 + \gamma_{ij} \D x^i \D x^j \right]
\end{equation}
remains dimensionless, $[ \gamma_{ij} ] = 0$. Hence it sources an operator of dimension $d$ in the dual field theory, as it should.

With the above redefinitions the action takes form
\begin{equation} \label{e:freeS}
S = \frac{1}{2} \int_0^\infty \D z \int \D^{d} \bs{x} \sqrt{g} \left[ g^{\mu\nu} \partial_\mu \Phi \partial_\nu \Phi + m^2 \Phi^2 \right].
\end{equation}
The redefined mass $m^2$ is dimensionless and we use the parametrization,
\begin{equation}
m^2 = \Delta (\Delta - d), \qquad\qquad \Delta > \frac{d}{2}.
\end{equation}
From now on we will always assume $\Delta > d/2$.

Notice that all parameters and dynamical fields in the action \eqref{e:freeS} are dimensionless. This is in agreement with the fact that the dual field theory is conformal and hence does not contain any scale. The equation of motion following from this action reads
\begin{equation} \label{e:freeeqofmo}
\left( - \Box_g + m^2 \right) \Phi = 0,
\end{equation}
where the Laplacian is taken with respect to the metric $g_{\mu\nu}$.

In this paper we mostly work in the momentum space by Fourier transforming along the $\bs{x}$-directions parallel to the boundary. By $\Phi(z, \bs{k})$ we denote the Fourier transform of the scalar field along $\bs{x}$. The equation of motion then expands into
\begin{equation} \label{eqofmo}
\left[ - z^2 \partial_z \partial_z + (d - 1) z \partial_z + ( m^2 + k^2 z^2 ) \right] \Phi(z, \bs{k}) = 0.
\end{equation}
The most general solution satisfying the regularity condition $\Phi \sim e^{-k z}$ at $z = \infty$ is given by the Bessel function $K$,
\begin{equation} \label{e:freesolA}
\Phi(z, \bs{k}) = A(\bs{k}) z^{d/2} K_{\Delta - \frac{d}{2}}(k z),
\end{equation}
where $A(\bs{k})$ is an undetermined functions. 

The near-boundary expansion of the field depends on whether $\Delta - d/2$ is an integer or not. Since throughout this paper we assume $\Delta > d/2$ we can limit ourselves to the following cases,
\begin{equation} \label{e:freeExp}
\Phi(z, \bs{k}) = z^{d - \Delta} \sum_{j=0}^\infty \phi_{(d - \Delta + 2j)} z^{2j} + \left\{ \begin{array}{ll} z^{\Delta} \sum_{j=0}^\infty \phi_{(\Delta + 2j)} z^{2j} & \text{ if } 2 \Delta - d \neq 2 n, \\
z^{\Delta} \log z^2 \sum_{j=n}^\infty \tilde{\phi}_{(\Delta - n + 2j)} z^{2j} & \text{ if } 2 \Delta - d = 2 n, \end{array} \right.
\end{equation}
where $n$ is a non-negative integer. We  assume standard Dirichlet boundary conditions, which singles out the coefficient $\phi_{(d - \Delta)}$ to be the source $\phi_0$ of the dual operator $\O$, and hence we write $\phi_{(d - \Delta)} = \phi_0$. For simplicity, we refer to the coefficient $\phi_{(\Delta)}$ as the vev term, despite the fact that its relation to the actual expectation value $\< \O \>_s$ of the dual operator requires the execution of the renormalization procedure. All remaining coefficients are determined locally in terms of $\phi_{(d - \Delta)}$ and $\phi_{(\Delta)}$. For example, in the generic case,
\begin{equation} \label{e:phidDelta2}
\phi_{(d - \Delta + 2)} = \frac{\partial^2 \phi_{(d - \Delta)}}{2(2 \Delta - d - 2)}.
\end{equation}
For other coefficients relations can be read off of the solution \eqref{e:freesolA} or found in \cite{Skenderis:2002wp,Papadimitriou:2010as}.

The precise form of the solution \eqref{e:freesolA} normalized with the condition 
\begin{equation}
\Phi = \phi_{(d - \Delta)} z^{d-\Delta} + \text{ subleading in } z
\end{equation}
as $z$ approaches the boundary at $z = 0$ reads
\begin{equation} \label{e:freesolPhi}
\Phi(z, \bs{k}) = \phi_{(d - \Delta)}(\bs{k}) \frac{k^{\Delta - \frac{d}{2}} z^{d/2} K_{\Delta - \frac{d}{2}}(k z)}{2^{\Delta - \frac{d}{2} - 1} \Gamma \left( \Delta - \frac{d}{2} \right)}
\end{equation}
for any $\Delta > d/2$. The value of the coefficient of $z^{\Delta}$, however, depends on whether $\Delta - d/2$ is a non-negative integral. One finds
\begin{equation} \label{e:vev_cf}
\phi_{(\Delta)} = \phi_{(d - \Delta)} \times \left\{ \begin{array}{ll} \frac{\Gamma \left( \frac{d}{2} - \Delta \right)}{\Gamma \left( \Delta - \frac{d}{2} \right)} \left( \frac{k}{2} \right)^{2 \Delta - d} & \text{ if } 2 \Delta - d \neq 2 n, \\
\frac{(-1)^n}{n! (n-1)!} \left( \frac{k}{2} \right)^{2n} \left( - \log k^2 + 2 \log 2 - 2 \gamma_E + H_n \right) & \text{ if } 2 \Delta - d = 2 n \end{array} \right.
\end{equation}
where $H_n = \sum_{j=1}^n j^{-1}$ denotes the $n$-th harmonic number.

\subsubsection{Holographic renormalization}

Having discussed basic properties of the cut-off and dimensional renormalization methods in a CFT, we will now discuss elements of the holographic renormalization procedure. Essentially we review classic results of \cite{Bianchi:2001kw,Skenderis:2002wp,Papadimitriou:2004ap,Papadimitriou:2004rz} and compare them to other renormalization methods, including the proposed dimensional renormalization, to be discussed in the next section.

In the holographic renormalization, similarly to the cut-off renormalization in a QFT, dimensions $d$ and $\Delta$ are fixed and the cut-off $\delta$ on the radial variable becomes the regulator. The aim of the procedure is then to add local counterterms localized on the cut-off surface in such a way that the on-shell action or equivalently its derivative, \textit{i.e.}, the 1-point function with sources becomes finite. The form of the counterterms strongly depends on values of $d$ and $\Delta$ and in case of a free massive field in AdS the number of required counterterms grows as the conformal dimension $\Delta$ grows. In particular, the first two counterterms are
\begin{equation} \label{e:Sctholo}
S_{\text{ct}} = \int_{\delta} \D^d \bs{x} \sqrt{\g_\delta} \left[ \frac{1}{2}(d - \Delta) \Phi^2 + \frac{\Phi \Box_{\delta} \Phi}{2(2 \Delta - d - 2)} \right].
\end{equation}
By $\g_{\delta}$ we denote the induced metric on a slice of constant $z = \delta$ and the index on the integral indicates that the integral is taken on such a slice. Similarly, $\Box_{\delta}$ is a Laplacian with respect to the induced metric, $(\g_\delta)_{ij} = z^{-2} \delta_{ij}$. 

The first term in \eqref{e:Sctholo} is always present (we assume $\Delta > d/2$), the second term appears if $\Delta > d/2 + 1$ and more terms are required for $\Delta > d/2 + 2$. The counterterm is written in a covariant way and hence its explicit dependence on the cut-off $\delta$ is implicit. Assuming $\Delta > d/2 + 1$ its expansion in terms of the boundary data gives
\begin{equation}
S_{\text{ct}} = \delta^{d - 2\Delta} \int_{\delta} \D^d \bs{x} \left[ \frac{1}{2} (d - \Delta) \phi_{(d - \Delta)}^2 + \frac{d - \Delta + 1}{2(2 \Delta - d - 2)} \phi_{(d - \Delta)} \delta^2 \partial^2 \phi_{(d - \Delta)} \right].
\end{equation}
Clearly, for $\Delta > d/2 + 1$, both terms diverge in the $\delta \rightarrow 0$ limit. Only when the counterterm action is added to the regulated on-shell action a finite $\delta \rightarrow 0$ limit exists.

In general the number of terms in the counterterm action is given exactly by \eqref{e:number}. Each counterterm has a form $c_j \Phi \Box_\delta^{2 j} \Phi$ for specific values of constants $c_j$ determined by the near-boundary expansion. We will not need the exact form of such counterterms, see \cite{Skenderis:2002wp,Papadimitriou:2007sj} for an example.

All in all, in a generic case the renormalized 1-point function with sources reads,
\begin{equation}
\< \O(\bs{x}) \>_s = - (2 \Delta - d) \phi_{(\Delta)}.
\end{equation}
By taking minus a derivative with respect to the source $\phi_0$ the 2-point function can be calculated. From \eqref{e:vev_cf} we immediately find
\begin{equation} \label{e:2pt_holoren}
\lla \O(\bs{k}) \O(-\bs{k}) \rra = (2 \Delta - d) \frac{\Gamma \left( \frac{d}{2} - \Delta \right)}{\Gamma \left( \Delta - \frac{d}{2} \right)} \left( \frac{k}{2} \right)^{2 \Delta - d}.
\end{equation}
By comparing with \eqref{e:2ptcutoff} or \eqref{e:2pt} we find specific values of the theory-dependent constants $c_{\O}$ and $C_{\O}$.

If, on the other hand $2 \Delta - d = 2 n$, the form of the radial expansion and the form of counterterms changes. According to \eqref{e:freeExp} logarithmic terms appear, which alters the analysis significantly. If one considers $n = 1$, \textit{i.e.}, $\Delta = d/2 + 1$ exactly, then the radial expansion reads,
\begin{equation} \label{e:explog}
\Phi(z, \bs{k}) = z^{\frac{d}{2} - 1} \left[ \phi_{(\frac{d}{2} - 1)} + z^2 \left( \phi_{(\frac{d}{2}+1)} + \tilde{\phi}_{(\frac{d}{2}+1)} \log z^2 \right) + O(z^4) \right].
\end{equation}
The expansion now contains logarithmic terms in addition to powers. $\phi_{(\frac{d}{2} - 1)}$ is the source term and $\phi_{(\frac{d}{2}+1)}$ is the vev term. All remaining terms are local functionals of these coefficients. Furthermore, as discussed in \cite{Skenderis:2002wp}, the 1-point function with sources in this case reads
\begin{equation}
\< \O(\bs{x}) \>_s = - 2 \left[ \phi_{(\frac{d}{2}+1)} + \tilde{\phi}_{(\frac{d}{2}+1)} \right],
\end{equation}
so an additional local contribution appears. Finally, the numerical coefficient in the second term of \eqref{e:Sctholo} diverges and hence the counterterm is replaced by
\begin{equation}
S_{\text{ct}} = \int_{\delta} \D^d \bs{x} \sqrt{\g_\delta} \left[ \frac{d-2}{4} \Phi^2 - \frac{1}{4} \left( \log \frac{\delta}{L} + a_0 \right) \Phi \Box_\delta \Phi \right],
\end{equation}
where $a_0$ is an undetermined scheme-dependent constant. On dimensional grounds we have included the only available scale, the AdS radius $L$, in the construction of the counterterm. Notice that this is the first place where the AdS radius appears explicitly since the redefinition \eqref{e:BulkFieldRedef} took place.

All in all the renormalized 2-point function reads
\begin{equation} \label{e:2ptren1}
\lla \O(\bs{k}) \O(-\bs{k}) \rra = k^2 \left[ \frac{1}{2} \log ( k^2 L^2 ) + \left( \gamma_E - \log 2 - a_0 \right) \right].
\end{equation}
This result agrees with \eqref{e:2ptrencut} and \eqref{e:2ptren} if one identifies the renormalization scale $\mu$ with the AdS radius, $L = \mu^{-1}$.

In the holographic renormalization method, similarly to the cut-off renormalization in a QFT, one usually needs a large number of counterterms in order to renormalize the theory. Furthermore, the form of the counterterms strongly depends on fixed dimensions $d$ and $\Delta$. As we shall see, these issues can be avoided in the dimensional renormalization in AdS.

\subsubsection{Hybrid renormalization}

As pointed out in the previous section, special cases satisfying $2 \Delta - d = 2 n$ require a special treatment in holographic renormalization. For that reason it is popular to use a blend of holographic renormalization in the bulk and dimensional renormalization in the dual field theory. In this approach one considers slightly shifted, regularized dimensions $\dreg$ and $\Dreg$ defined by \eqref{e:introReg} with small $\epsilon$. Since special cases such as $2 \Delta - d = 2 n$ for the 2-point functions constitute a discreet set, one can always shift actual dimensions slightly in order to avoid radial expansions with logarithms such as \eqref{e:explog}. Then one proceeds with the regular holographic renormalization method and obtains expressions for generic, regulated correlation functions in the dual CFT. Finally, the dimensional renormalization in the QFT can be applied to yield finite expressions.

As an example, consider the case of $2 \Delta - d = 2$. If one shifts the dimensions $d$ and $\Delta$ in such a way that $2 \Dreg - \dreg < 2$, then only the first term in \eqref{e:Sctholo} is required to yield the on-shell action finite. Hence the regulated 2-point function \eqref{e:2pt} follows, with regulated dimensions
\begin{equation} \label{e:2ptk0}
\lla \O(\bs{k}) \O(-\bs{k}) \rra_{\text{reg}} = (2 \Dreg - \dreg) \frac{\Gamma \left( \frac{\dreg}{2} - \Dreg \right)}{\Gamma \left( \Dreg - \frac{\dreg}{2} \right)} \left( \frac{k}{2} \right)^{2 \Dreg - \dreg}.
\end{equation}
This expression is a valid 2-point function if $\epsilon$ is kept positive and small. However, it still diverges in $\epsilon \rightarrow 0$ limit and hence it requires further renormalization. Now the QFT counterterm \eqref{e:Sct} is added to the generating functional of the dual QFT such that the finite $\epsilon \rightarrow 0$ limit exists.

This method of renormalization was successfully used for evaluation of correlation functions in generic cases in many papers, including \cite{Witten:1998qj,Muck:1998rr,Freedman:1998tz,Liu:1998bu,Mueck:1998ug,Liu:1998ty,D'Hoker:1999pj,vanRees:2011fr,Raju:2011mp}. In \cite{Schwimmer:2000cu,Schwimmer:2003eq} the authors use the hybrid renormaliztion method to analyze anomalies in the dual CFT from the point of view of bulk diffeomorphisms, recovering in particular results of \cite{Henningson:1998gx,Henningson:1998ey}.

The problem with the hybrid renormalization method is that this is an \textit{ad hoc} procedure, which combines both genuine holographic renormalization in the bulk and dimensional renormalization of the dual field theory. Its advantage is that one does not need to worry about special cases on the bulk side, which would lead to exceptional radial expansions and counterterms. On the other hand the hybrid renormalization requires exactly the same amount of counterterms as the holographic renormalization. What is more some of these counterterms come from the genuine holographic renormalization method while other follow from dimensional regularization. This make the whole procedure much more messy and hinders any relation between the renormalization properties of the bulk and boundary theories.

\subsection{Dimensional renormalization} \label{sec:DIMREG}

The idea behind the dimensional renormalization of a QFT is to treat dimensions $d$ and $\Delta$ as parameters, derive correlation functions in some `region of convergence' and then analytically continue the result. Only when singularities in the continued expression arise, the counterterms are necessary.

Here we want to apply the same reasoning to the holographic theory, \textit{i.e.}, to the correlation functions derived from the bulk action \eqref{e:freeS}. For this to work one would need first to find a region of the parameter space $(d, \Delta)$ such that the on-shell action converges. As it stands, this unfortunately never happens. Therefore we have no much choice but to include the first term of the counterterm action \eqref{e:Sctholo} in our bulk action. Therefore we start our analysis from the action
\begin{equation} \label{e:Sfree}
S_{\text{free}} = \lim_{\delta \rightarrow 0} \left[ \frac{1}{2} \int_{\delta}^\infty \D z \int_{\delta} \D^{d} \bs{x} \sqrt{g} \left( g^{\mu\nu} \partial_\mu \Phi \partial_\nu \Phi + m^2 \Phi^2 \right) + \frac{1}{2}(d - \Delta) \int_{\delta} \D^d \bs{x} \sqrt{\gamma_{\delta}} \Phi^2 \right].
\end{equation}
Let us make two comments here. Firstly, since we always assume $\Delta > d/2$ the included term would always appear in holographic renormalization. Hence it is a universal term and we can simply include it in our defining action if we feel like it. Secondly, notice that the cut-off $\delta$ is introduced for the sake of definition only. Formally, every object in the above expression has a non-trivial radial dependence and every formal definition must deal with this fact. For example, a proper definition of the integral in \eqref{e:freeS} requires an introduction of the cut-off that is eventually sent to zero. To summarize, the entire action $S_{\text{free}}$ does not depend on $\delta$ and the cut-off was introduced merely as a formal tool.

Now, by the analysis of the previous section we know that, when evaluated on a solution to the equation of motion, this action is finite if 
\begin{equation} \label{e:FreeRange}
\frac{d}{2} < \Delta < \frac{d}{2} + 1.
\end{equation}
In particular, the 2-point function \eqref{e:2pt_holoren} follows,
\begin{equation} \label{e:2pt_toa}
\lla \O(\bs{k}) \O(-\bs{k}) \rra = (2 \Delta - d) \frac{\Gamma \left( \frac{d}{2} - \Delta \right)}{\Gamma \left( \Delta - \frac{d}{2} \right)} \left( \frac{k}{2} \right)^{2 \Delta - d}.
\end{equation}

The idea of the dimensional renormalization method is to treat the action \eqref{e:Sfree} and the correlators it produces as functions of dimensions $d$ and $\Delta$. The above 2-point function is analytic in $d$ and $\Delta$ as long as $2 \Delta - d \neq 2 n$. Therefore it represents the unique, analytically continued 2-point function for any such case. In particular no counterterms are required in order to obtain a finite correlator.

This may not seem like a substantial improvement over the holographic renormalization method, but this is only because for now we considered such a simple problem. A possibility of evaluating higher-point correlation functions without a need for any counterterms turns out to be a significant simplification as we will see in sections \ref{sec:ExampleEasy} and \ref{sec:ExampleHard}. Furthermore, as we will ses in the following subsection, in cases when renormalization is required, only a single counterterm is needed and a direct identification between bulk and boundary counterterms is possible.

\subsubsection{Regularization}

As we can see from \eqref{e:2pt_toa}, the 2-point function exhibits singularities only in the special case when $2 \Delta - d = 2 n$ for a non-negative integer $n$ (we assume $\Delta > d/2$). Let us assume now that this is the case. As in the dimensional renormalization in QFT or hybrid renormalization we consider shifted dimensions $\dreg$ and $\Dreg$ defined by \eqref{e:introReg}. Notice also that this parameterization differs from the one used in \cite{Bzowski:2013sza,Bzowski:2015pba,Bzowski:2015yxv} as here we regulate actual dimensions. As before, not all choices of $u$ and $v$ regularize the 2-point function and we need to work in a regularization scheme with $u \neq 2 v$. The regulated 2-point function then reads
\begin{align} \label{e:2ptk}
& \lla \O(\bs{k}) \O(-\bs{k}) \rra_{\text{reg}} = (2 \Dreg - \dreg) \frac{\Gamma \left( \frac{\dreg}{2} - \Dreg \right)}{\Gamma \left( \Dreg - \frac{\dreg}{2} \right)} \left( \frac{k}{2} \right)^{2 \Dreg - \dreg} \nn\\
& \qquad\qquad = \frac{(-1)^n k^{2n}}{4^{n-1} [(n-1)!]^2} \left[ \frac{1}{(u - 2v) \: \epsilon} + \left( - \tfrac{1}{2} \log k^2 + \log 2 - \gamma_E + H_{n-1} \right) \right] + O(\epsilon),
\end{align}
where $H_{n} = \sum_{j=1}^n j^{-1}$ denotes the $n$-th harmonic number. We assumed $2 \Delta - d = 2n$ and $u \neq 2v$ and hence the expression possesses a single pole in the regulator $\epsilon$.

\subsubsection{Renormalization}

Let us start our discussion from the point of view of the dual QFT. The divergence in \eqref{e:2ptk} can be removed by a counterterm of the form \eqref{e:Sct}, where $\phi_0 = \phi_{(\Delta - d)}$ is the source for the dual operator. To be precise one needs
\begin{equation} \label{e:SCT}
W_{\text{ct}}[\phi_0] = a_{\text{ct}} \int \D^{\dreg} \bs{x} \: \phi_0 \partial^{2 n} \phi_0 \mu^{- (u - 2 v) \epsilon}
\end{equation}
with the value of the counterterm constant $a_{\text{ct}}$ fixed by the divergence of the regulated 2-point function. Hence,
\begin{equation} \label{e:aFreeCT}
a_{\text{ct}} = - \frac{1}{2 \cdot 4^{n-1} [(n-1)!]^2} \left[ \frac{1}{(2v - u) \epsilon} + a_0 + O(\epsilon) \right],
\end{equation}
where $a_0$ is an arbitrary constant, so that the renormalized 2-point function becomes finite and reads
\begin{equation}\label{e:2ptRENmu}
\lla \O(\bs{k}) \O(-\bs{k}) \rra = \frac{(-1)^n k^{2n}}{4^{n-1} [(n-1)!]^2} \left[ - \frac{1}{2} \log \left( \frac{k^2}{\mu^2} \right) + \log 2 - \gamma_E + H_{n-1} + a_0 \right].
\end{equation}
The entire constant can be incorporated into the renormalization scale $\mu$. We see that this expression agrees with the form of the renormalized 2-point function in a CFT given by \eqref{e:2ptren}. For this particular model we find specific values of parameters,
\begin{equation} \label{e:deploc}
c'_{\O} = - \frac{(-1)^n}{2 \cdot 4^{n-1} [(n-1)!]^2}, \qquad\qquad a_0' = 2 (\gamma_E - \log 2 - H_{n-1} - a_0).
\end{equation}
Furthermore, for $n=1$ we recover the result \eqref{e:2ptren1} obtained by the holographic renormalization method. The two expressions agree up to scheme-dependent terms, which can be adjusted by a redefinition of $\mu$.

Now we want to understand the renormalization procedure from the bulk point of view. Consider the following term localized on the boundary of AdS,
\begin{equation} \label{e:toSCT}
\bar{a}_{\text{ct}} \lim_{z \rightarrow 0} \int_z \D^d \bs{x} \sqrt{\g_z} \Phi \Box_{z}^n \Phi
\end{equation}
and add it to the action \eqref{e:Sfree}. The value of constant $\bar{a}_{\text{ct}}$ is yet to be determined. By using expansion \eqref{e:freeExp} one finds
\begin{align}
& \bar{a}_{\text{ct}} \lim_{z \rightarrow 0} \int_z \D^d \bs{x} \sqrt{g_z} \Phi \Box_{z}^n \Phi = \nn\\
& \qquad = \bar{a}_{\text{ct}} \lim_{z \rightarrow 0} \int_z \D^d \bs{x} z^{-d + 2n} \Phi \partial^{2n} \Phi \nn\\
& \qquad = \bar{a}_{\text{ct}} \lim_{z \rightarrow 0} \int_z \D^d \bs{x} \left[ \phi_{(d - \Delta)} \partial^{2n} \phi_{(d - \Delta)} + \text{subleading in } z \right] \nn\\
& \qquad = \bar{a}_{\text{ct}} \int \D^d \bs{x} \phi_{(d - \Delta)} \partial^{2n} \phi_{(d - \Delta)}.
\end{align}
The key observation here is that, when the covariant term in the first line is expanded in the radial variable, it results in a finite term supported on the boundary of AdS. For this to hold, however, it is essential that dimensions match correctly, \textit{i.e.}, it is necessary that $2 \Delta - d = 2 n$ for a non-negative integer $n$. Only then the finite $z \rightarrow 0$ limit correctly reproduces the CFT counterterm \eqref{e:Sct} with $\epsilon = 0$. Therefore by adding a single counterterm \eqref{e:toSCT} one should be able to render the 2-point function finite. 

However, the above expression is valid in the bare, unregularized theory. In the theory regularized one finds instead
\begin{equation} \label{e:aproblem}
\bar{a}_{\text{ct}} \int_z \D^{\dreg} \bs{x} \sqrt{\g_z} \: \Phi \Box_z^n \Phi = \bar{a}_{\text{ct}} z^{(u - 2 v) \epsilon} \int_z \D^{\dreg} \bs{x} \left[ \phi_{(\dreg - \Dreg)} \partial^{2n} \phi_{(\dreg - \Dreg)} + \text{subleading in } z \right],
\end{equation}
since dimensions $d$ and $\Delta$ are shifted according to \eqref{e:introReg}, while the value of $n$ remained fixed by their unperturbed values. Hence the double limit $z, \epsilon \rightarrow 0$ is undefined. Note that in the regulated theory field $\Phi$ remains dimensionless as it follows from the requirement that the action remains dimensionless.

We can resolve this issue by inserting a compensating factor of $z^{-(u - 2 v)\epsilon}$ in the regularized theory. This, however, would violate the requirement that the action remains dimensionless for any dimensions $d$ and $\Delta$. Indeed, all coordinates have the dimension of the AdS radius $L$ and the dimension of the source $\phi_{(d - \Delta)}$ equals $\Delta - d$. Therefore the counterterm \eqref{e:SCT} remains dimensionless even after the regularization procedure. Hence the compensating factor must use the only available scale: the radius of AdS, $L$. We conclude, that the correct form of the regularized counterterm having a finite $z, \epsilon \rightarrow 0$ limit is
\begin{align} \label{e:bulkCT}
S_{\text{ct}} & = \bar{a}_{\text{ct}} \lim_{z \rightarrow 0} \int_z \D^{\dreg} \bs{x} \sqrt{g_z} \: \Phi \Box_z^n \Phi \left( \frac{z}{L} \right)^{-(u - 2v) \epsilon} \nn\\
& = \bar{a}_{\text{ct}} \int \D^{\dreg} \bs{x} \: \phi_{(\dreg - \Dreg)} \partial^{2n} \phi_{(\dreg - \Dreg)}  L^{(u - 2 v) \epsilon}.
\end{align}
Clearly, we have traded the problematic $z$ factor in \eqref{e:aproblem} for a scale-dependence of the boundary counterterm!

Due to the additional minus sign in \eqref{e:adscft} we have $W_{\text{ct}} = - S_{\text{ct}}$. Hence, for the right hand side of the bulk counterterm \eqref{e:bulkCT} to match minus the boundary counterterm \eqref{e:SCT} we can set $\bar{a}_{\text{ct}} = - a_{\text{ct}}$ as well as
\begin{equation} \label{e:Lmu}
L = \frac{1}{\mu}.
\end{equation}
This is the fundamental relation between an emergent renormalization scale $\mu$ of the quantum theory leaving on the boundary of the AdS spacetime and the AdS radius $L$ present in the classical bulk theory. Notice that this is not a postulate of any kind, but rather a derived result. To be precise one can maintain the equality of \eqref{e:SCT} and \eqref{e:bulkCT} by introducing an arbitrary positive factor $\alpha$ in \eqref{e:Lmu}, $L = \alpha/\mu$, while simultaneously redefining the subleading counterterm constant $a_0$ in \eqref{e:aFreeCT} by an $\alpha$-dependent factor. Therefore an introduction of a proportionality factor $\alpha \neq 1$ results in finite, local, scheme-dependent terms removable by adjustments of counterterms. For that reason we may fix $\alpha = 1$.

\subsubsection{Equivalence between counterterms} \label{sec:equiv}

As we have seen above, in the special case $2 \Delta - d = 2 n$ a single counterterm \eqref{e:bulkCT} is necessary in order to render the 2-point function finite. This is also a feature of the dimensional renormalization of a QFT as discussed in section \ref{sec:dimreg}. Furthermore, the two counterterms \eqref{e:bulkCT} and \eqref{e:SCT} are equal when the identification of scales \eqref{e:Lmu} is made.

The equivalence of the dimensional renormalization schemes in the bulk and the dual QFT is the statement that bulk and boundary counterterms match. We have seen that the two counterterms 
\begin{align}
S_{\text{ct}} & = \lim_{z \rightarrow 0} \int_z \D^{\dreg} \bs{x} \sqrt{\g_z} \: \Phi \Box_z^n \Phi \left( \frac{z}{L} \right)^{-(u - 2v) \epsilon}, \label{e:genSct2} \\
W_{\text{ct}} & = \int \D^{\dreg} \bs{x} \phi_{(\dreg - \Dreg)} \partial^{2n} \phi_{(\dreg - \Dreg)} \mu^{- (u - 2 v) \epsilon} \label{e:genWct2}
\end{align}
are equal provided \eqref{e:Lmu} holds. The existence of counterterms requires the condition $2 \Delta - d = 2 n$ to be satisfied for a non-negative integer $n$. Therefore the map implementing equivalence of the two schemes assigns 
\begin{equation}
S_{\text{ct}}[\Phi; \epsilon, L] \longmapsto - W_{\text{ct}}[\phi_{(\dreg - \Dreg)}; \epsilon, \mu^{-1}].
\end{equation}
The awkward minus sign arises due to the minus sign in the AdS/CFT defining equation \eqref{e:adscft}. Without the minus sign their contributions to correlation functions would have opposite signs.

First let us argue that \eqref{e:genSct2} and \eqref{e:genWct2} are indeed the most general counterterms contributing to the 2-point functions. From the point of view of the dual CFT all possible counterterms are classified by dimensions. For a counterterm to contribute to the 2-point function of an operator $\O$, one needs exactly two sources $\phi_0$. Locality implies that an even number of derivatives may appear. Any term containing two sources of conformal dimension $\Delta - d$ and $2 n$ derivatives has a total dimension $2(\Delta - d) + 2 n$. For the entire counterterm action to be scale-invariant one needs this to be equal to $-d$. Hence we arrive at the conclusion that the most general counterterm in the regulated theory takes form \eqref{e:genWct2} and requires $2 \Delta - d = 2 n$ for a non-negative integer $n$. The renormalization scale $\mu$ appears on dimensional grounds so that the entire expression remains dimensionless in the regularized theory.

From the bulk point of view every counterterm should be built up from covariant quantities such as the bulk field $\Phi$, so that Poincar\'{e} symmetries of the dual theory are maintained. The counterterm contributing to the 2-point function requires two boundary sources and from linearity of the equation of motion \eqref{e:freeeqofmo} it requires two bulk fields. Locality demands a dependence on an even number of derivatives. While in the holographic renormalization scheme there is plenty of terms satisfying these conditions, in the dimensional scheme one has an important additional constraint: no scale. The AdS radius $L$ may appear only with powers proportional to the regulator $\epsilon$ since otherwise the dual QFT would contain an explicit dependence on the scale. And the lack of any cut-off requires that a valid bulk counterterm in the dimensional regularization scheme must live on the boundary! This can happen only if fields have an appropriate radial expansion. The leading term of the bulk field $\Phi$ is the source term proportional to $z^{d - \Delta}$. Each covariant derivative induced on a slice of constant $z$ comes with the factor of $z$ and $\sqrt{\gamma_z} = z^{-d}$. Altogether the integrand of \eqref{e:genWct2} is supported on the boundary only if $2 \Delta - d = 2 n$ is satisfied for a non-negative integer $n$.

This shows that \eqref{e:genSct2} and \eqref{e:genWct2} represent all possible counterterms in dimensionally regulated theories. From \eqref{e:bulkCT} we also know that $S_{\text{ct}}$ reduces to $W_{\text{ct}}$ when fields are expanded in the radial variable. 

A last missing piece of information is how to reconstruct the bulk counterterm from the boundary one. To do it, one needs to invert expansions of bulk fields in terms of boundary data. However, due to the limit in \eqref{e:genSct2}, it is always the most leading term in the expansion of the bulk field that matters. Hence, given the boundary counterterm \eqref{e:genWct2} its bulk covariant version is obtained by substitutions,
\begin{equation} \label{e:covariantize}
\phi_{(\dreg - \Dreg)} \mapsto \Phi z^{-(\dreg - \Dreg)}, \qquad \partial_i \mapsto z^{-1} \nabla_i, \qquad \D^{\dreg} \bs{x} \mapsto \D^{\dreg} \bs{x} \sqrt{\g_\zeta}
\end{equation}
together with \eqref{e:Lmu}. Explicit powers of the radial variable cancel leading to \eqref{e:genSct2}.

This finalizes the proof of the equivalence of the bulk and boundary dimensional renormalization procedures as far as 2-point functions go. In the following sections we generalize this procedure to higher-point correlation functions and we show that the equivalence holds true.

The emergence of the renormalization scale in the holographic renormalization is usually attributed to the cut-off, \cite{Henningson:1998ey,Henningson:1998gx,Skenderis:2002wp}. In the dimensional renormalization scheme the renormalization scale is related to the AdS scale $L$ by \eqref{e:Lmu}. This is a very natural phenomenon. The AdS theory is a classical theory and does not require any additional scales. Furthermore the dimensional regularization method does not introduce any additional scales, so the only possibility is a relation between the AdS radius $L$ and the renormalization scale $\mu$. This is what we found in equation \eqref{e:Lmu}.

\subsubsection{Scaling anomaly}

In the dimensional renormalization the scaling anomaly arises as a failure of the counterterms to be scale-invariant. Consider an infinitesimal rescaling of the boundary according to $\delta_{\sigma} \bs{x} = \sigma \bs{x}$. Since the theory is conformal, the infinitesimal transformation of the source $\phi_0$ is,
\begin{equation}
\delta_{\sigma} \phi_0(\bs{x}) = \sigma \left[ x^j \partial_j \phi_0 - ( \Delta - d ) \phi_0 \right],
\end{equation}
Therefore the scale variation of the CFT counterterm \eqref{e:SCT} in the regulated theory reads
\begin{equation}
\delta_{\sigma} W_{\text{ct}} = (2 v - u) \epsilon \sigma W_{\text{ct}} = \mu \frac{\D}{\D \mu} W_{\text{ct}}.
\end{equation}
Since the counterterm constant \eqref{e:aFreeCT} multiplying such a counterterm is of order $1/\epsilon$, this leads to the violation of scale invariance. Indeed, the 2-point function \eqref{e:2ptRENmu} depends on the renormalization scale $\mu$,
\begin{equation}
\mu \frac{\D}{\D \mu} \lla \O(\bs{k}) \O(-\bs{k}) \rra = - L \frac{\D}{\D L} \lla \O(\bs{k}) \O(-\bs{k}) \rra = \frac{(-1)^{n-1} k^{2n}}{4^{n-1} [(n-1)!]^2}.
\end{equation}

From the point of view of the bulk theory the failure of the dual CFT to be non-anomalous stems from the fact that the corresponding counterterm \eqref{e:genSct2} fails to be invariant under the corresponding isometry of AdS. Indeed, an explicit dependence on the radial coordinate breaks the invariance under the scalings, $\delta_{\sigma} \bs{x} = \sigma \bs{x}$ and $\delta_\sigma z = \sigma z$. The variation of the counterterm \eqref{e:bulkCT} then reads
\begin{equation}
\delta_{\sigma} S_{\text{ct}} = (2 v - u) \epsilon \sigma S_{\text{ct}} = - L \frac{\D}{\D L} S_{\text{ct}}.
\end{equation}
As expected, anomalies derived in AdS and in the dual CFT agree provided \eqref{e:Lmu} is used.

Let us also comment to what extent the customarily assumed relation between the renormalization scale $\mu$ and the radial variable $z$ holds. Notice that by introducing the compensating factor of $(z/L)^{-(u - 2v) \epsilon}$ in \eqref{e:genSct2} we effectively exchanged the $z^{(u - 2 v) \epsilon}$ factor in the expansion \eqref{e:aproblem} for $L^{(u - 2 v) \epsilon}$. In other words an explicit dependence of counterterms on the radial variable $z$ in \eqref{e:genSct2} is the same as an explicit dependence of \eqref{e:genWct2} on the renormalization scale $\mu$. Recall that the regulated theory is non-anomalous and hence the only source of the scale dependence in the renormalized theory is via counterterms, as indicated by equation \eqref{e:introWren}. We arrive at the conclusion that
\begin{equation}
\mu \frac{\partial}{\partial \mu} = -L \frac{\partial}{\partial L} = z \frac{\partial}{\partial z}.
\end{equation}
These can be applied to a renormalized on-shell action or a renormalized generating functional. One needs to remember about a relative sign in \eqref{e:adscft}.

\section{Higher-point functions}

Now we start investigations into dimensional renormalization of higher-point correlation functions of a holographic theory. In this section we lay ground for the subsequent renormalization procedure and deal with all technicalities. This involves the analysis of a radial expansion of bulk fields, an extraction of vev coeffiecients, regularization procedure and analytic continuation of correlation functions. In particular we show that regulated correlation functions of the dual field theory are analytic functions of dimensions and hence methods of complex analysis can be applied. Finally we identify and classify all divergences that may appear in the regulated correlators.

\subsection{Bulk problem}

We consider a real scalar bulk field $\Phi$ governed by the action with a potential $V(\Phi)$ having a Taylor series expansion around $\Phi = 0$. However, for simplicity, we will limit our considerations to a single monomial in the potential, \textit{i.e.}, we consider the action
\begin{equation} \label{e:preS}
S = \int_0^\infty \D z \int \D^{d} \bs{x} \sqrt{g} \left[ \frac{1}{2} \partial_\mu \Phi \partial^\mu \Phi + \frac{1}{2} m^2 \Phi^2 - \frac{\lambda}{M} \Phi^M \right],
\end{equation}
for a fixed integer $M \geq 3$. As it stands this action is useful for the derivation of the equations of motion,
\begin{equation} \label{e:fulleqofmo}
\left( - \Box_g + m^2 - \lambda \Phi^{M-2} \right) \Phi = 0.
\end{equation}
The coupling constant $\lambda$ is dimensionless, $[ \lambda ] = 0$. We assume Dirichlet boundary condition, \textit{i.e.}, the asymptotic boundary condition
\begin{equation}
\Phi = z^{d - \Delta} \phi_{(d - \Delta)} + \text{ subleading in } z.
\end{equation}
This prescription is valid if the source term $\phi_{(d - \Delta)}$ is the most leading term in the radial expansion, which happens for $\Delta < d$. Note that this is a genuine asymptotic problem and no cut-off of any sort is required.

In order to proceed with the dimensional renormalization method we must be a little more careful about the definition of the action \eqref{e:preS}. In order to add interactions to the bulk system, we should add them to the free theory action \eqref{e:Sfree}. In this way out starting point is
\begin{equation} \label{e:S}
S = \lim_{\delta \rightarrow 0} \left[ \int_{\delta}^\infty \D z \int_{\delta} \D^{d} \bs{x} \sqrt{g} \left( \frac{1}{2} \partial_\mu \Phi \partial^\mu \Phi + \frac{1}{2} m^2 \Phi^2 - \frac{\lambda}{M} \Phi^M \right) + \frac{1}{2}(d - \Delta) \int_{\delta} \D^d \bs{x} \sqrt{\gamma_{\delta}} \Phi^2 \right].
\end{equation}
As before, the action does not depend on the cut-off, which is used merely as a tool for the sake of the definition. The boundary term does not alter equations of motion.

In section \ref{sec:DIMREG} we have argued that the free theory action \eqref{e:Sfree} converges in an non-empty, open range of parameter space defined by \eqref{e:FreeRange}. In the following sections we will show that the interacting action \eqref{e:S} converges whenever the following inequalities are satisfied,
\begin{equation} \label{e:cond}
\frac{d}{2} < \Delta < \min \left( \frac{d}{2} + 1, \frac{M-1}{M} d \right).
\end{equation}
We state this condition here, since through the remainder of this paper it will reappear repeatedly on a regular basis. Notice that the condition produces a non-empty subset of the parameter space $(d, \Delta)$ for any $M > 2$ and hence our further conclusions are applicable for any analytic potential $V(\Phi)$.

\subsubsection{Perturbative solution}

The equation of motion \eqref{e:fulleqofmo} can be solved perturbatively in $\lambda$. Throughout our analysis we treat $\lambda$ as a formal, `infinitesimal' parameter and we are oblivious to the issues of convergence of the perturbative series. This is a perfectly justified approach if we are interested in correlation functions of the dual CFT. 

From now on, we follow the notation of \cite{vanRees:2011fr} and by $\{ - \}$ we denote the order of any function with respect to the coupling constant $\lambda$. With this convention we may write
\begin{equation}
\Phi(z, \bs{x}) = \Phi_{\{0\}}(z, \bs{x}) + \lambda \Phi_{\{1\}}(z, \bs{x}) + \lambda^2 \Phi_{\{2\}}(z, \bs{x}) + O(\lambda^3)
\end{equation}
and substitute it to the equation of motion \eqref{e:fulleqofmo}. This leads to the infinite tower of equations with the first three being,
\begin{align}
( - \Box_g + m^2) \Phi_{\{0\}} & = 0, \nn\\
( - \Box_g + m^2) \Phi_{\{1\}} & = \Phi_{\{0\}}^{M-1}, \nn\\
( - \Box_g + m^2) \Phi_{\{2\}} & = (M-1) \Phi_{\{0\}}^{M-2} \Phi_{\{1\}},
\end{align}
and so on. For $n \geq 1$ the equation satisfied by $\Phi_{\{n\}}$ reads
\begin{equation} \label{e:PertEq}
( - \Box_g + m^2) \Phi_{\{n\}} = \sum_{(n_j)} \prod_{j=1}^{M-1} \Phi_{\{n_j\}},
\end{equation}
where the sum on the right hand side is taken over all sequences $(n_1, n_2, \ldots, n_{M-1})$ of non-negative integers such that $\sum_{j=1}^{M-1} n_j = n - 1$.
When Fourier transformed to momentum space, each equation becomes an inhomogeneous ordinary differential equation. Therefore each $\Phi_{\{n\}}$ is given by a sum of a homogeneous solution and an inhomogeneous one. The homogeneous part takes form of the free field equation and is determined by the boundary condition of order $\lambda^n$, \textit{i.e.}, $\phi_{\{n\}(d - \Delta)}$. Therefore, perturbative solution of order $\lambda^n$ can be written as
\begin{equation} \label{e:field_decomp}
\Phi_{\{n\}}[\phi_{(d - \Delta)}] = \Phi_{\{n\}}^{\text{inhom}}[\phi_{(d - \Delta)}] + \Phi_{\{0\}}[\phi_{\{n\}(d - \Delta)}].
\end{equation}
The inhomogeneous part of the solution $\Phi_{\{n\}}^{\text{inhom}}$ depends on lower order sources $\phi_{\{k\}(d - \Delta)}$ for $k < n$. This part of the solution can be expressed in terms of Witten diagrams. At each step a homogeneous part $\Phi_{\{0\}}[\phi_{\{n\}(d - \Delta)}]$ can be added, which is sourced by $\phi_{\{n\}(d - \Delta)}$.

From the point of view of the dual CFT the source $\phi_0$ is identified with the leading source term $\phi_{\{0\}(d - \Delta)} = \phi_0$. In principle, one could also turn on subleading sources in $\lambda$, but these will contribute locally to correlation functions and from the point of view of the dual CFT can be removed by source redefinition. On the other hand -- as we will see -- divergent subleading sources will essentially appear when non-trivial renormalization is required. Therefore, we will allow for subleading terms in the source coefficient to be turned on.

Let us continue the discussion of the perturbative solution to the equation of motion. The inhomogeneous part of the solution can be represented in terms of propagators. Bulk-to-bulk propagator in position space is denoted by $G_{d,\Delta}(z, \bs{x}; z', \bs{x}')$ and bulk-to-boundary propagator by $K_{d,\Delta}(z, \bs{x}; \bs{x}')$. The bulk-to-boundary propagator $K_{d, \Delta}$ is defined by
\begin{equation}
\left\{ \begin{array}{l}
(-\Box_g + m^2) K_{d, \Delta}(z, \bs{x}; \bs{x}') = \delta(z) \delta(\bs{x} - \bs{x}'), \\[1ex]
\lim_{z \rightarrow 0} [z^{-(d - \Delta)} K_{d, \Delta}(z, \bs{x}; \bs{x}')] = 1, \\[1ex]
K_{d, \Delta}(\infty, \bs{x}; \bs{x}') = 0. \end{array} \right.
\end{equation}
while the bulk-to-bulk propagator $G_{d, \Delta}$ solves
\begin{equation}
\left\{ \begin{array}{l}
(-\Box_g + m^2) G_{d, \Delta}(z, \bs{x}; z', \bs{x}') = \frac{1}{\sqrt{g}} \delta(z - z') \delta(\bs{x} - \bs{x}'), \\[1ex]
\lim_{z \rightarrow 0} [z^{-(d - \Delta)} G_{d, \Delta}(z, \bs{x}; z', \bs{x}')] = 0, \\[1ex] 
G_{d, \Delta}(\infty, \bs{x}; z', \bs{x}') = 0.
\end{array} \right.
\end{equation}
Position space solutions to these equations can be found in standard lecture notes on AdS/CFT, see for example \cite{Skenderis:2002wp}. When Fourier transformed in boundary directions one obtains their momentum space versions. The bulk-to-boundary propagator is
\begin{equation} \label{e:bulktobnd}
K_{d, \Delta}(z, k) = \frac{2^{\frac{d}{2} - \Delta+1}}{\Gamma \left( \Delta - \frac{d}{2} \right)} k^{\Delta - \frac{d}{2}} z^{\frac{d}{2}} K_{\Delta - \frac{d}{2}}(k z)
\end{equation}
while for the bulk-to-boundary propagator one finds
\begin{equation} \label{e:bulktobulk}
G_{d, \Delta}(z, k; \zeta) = \left\{ \begin{array}{ll}
(z \zeta)^{d/2} I_{\Delta - \frac{d}{2}}(k z) K_{\Delta - \frac{d}{2}}(k \zeta) & \text{for } z \leq \zeta, \\
(z \zeta)^{d/2} I_{\Delta - \frac{d}{2}}(k \zeta) K_{\Delta - \frac{d}{2}}(k z) & \text{for } z > \zeta, \end{array} \right.
\end{equation}

\begin{figure}[ht]
	\includegraphics[width=0.70\textwidth]{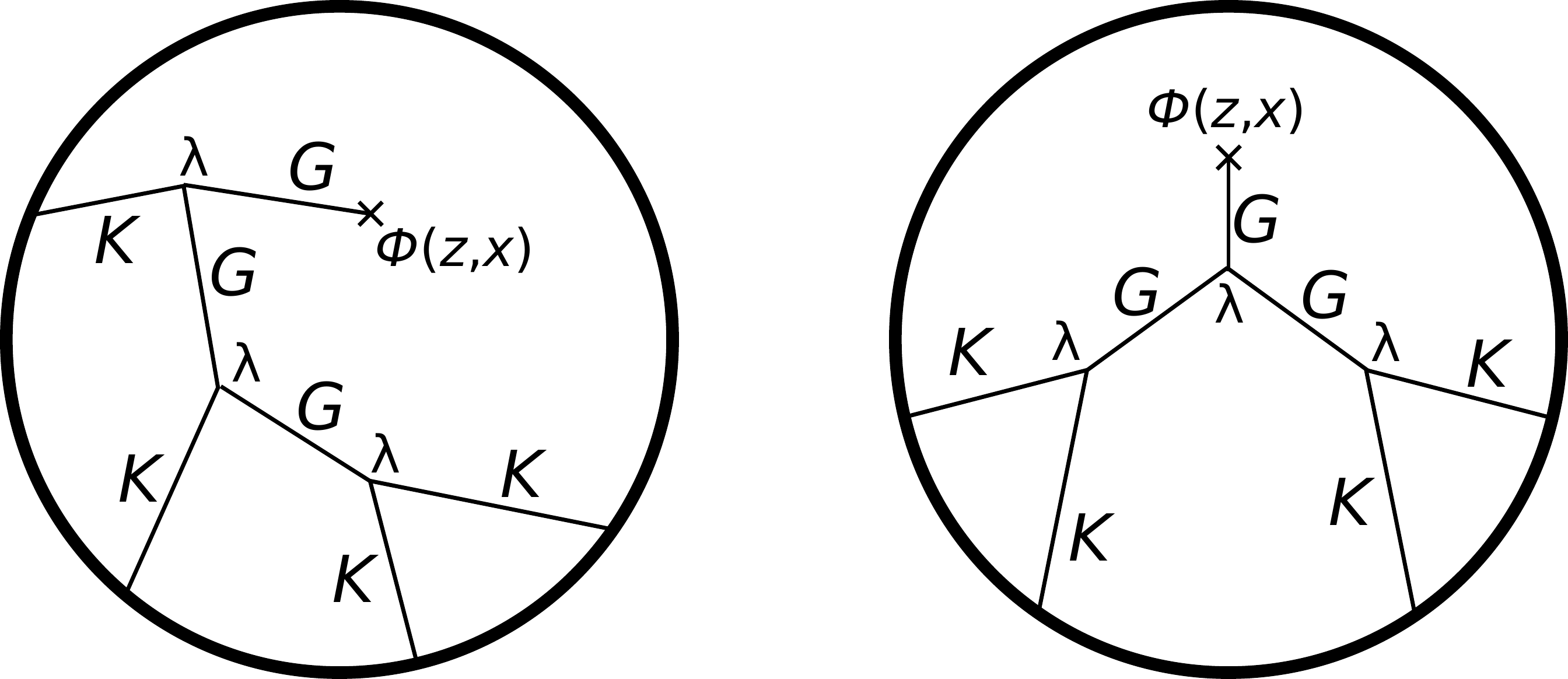}
	\centering
	\caption{Two topologically distinct Witten diagrams representing $\Phi_{\{3\}}$. Each $K$ denotes a bulk-to-boundary propagator and each $G$ stands for the bulk-to-bulk propagator.}
	\label{fig:W}
\end{figure}

In terms of the propagators, perturbative solutions to the equation of motion with the asymptotic boundary condition are
\begin{align}
\Phi_{\{0\}}(z, \bs{x}) & = \int \D^d \bs{x}' \sqrt{\g(\bs{x}')} K_{d,\Delta}(z, \bs{x}; \bs{x}') \phi_{\{0\}(d-\Delta)}(\bs{x}'), \label{e:sol0x} \\
\Phi^{\text{inhom}}_{\{n\}}(z, \bs{x}) & = \int \D z' \D^d \bs{x}' \sqrt{g(z', \bs{x}')} G_{d,\Delta}(z, \bs{x}; z', \bs{x}') \sum_{(n_j)} \prod_{j=1}^{M-1} \Phi_{(n_j)}(z', \bs{x}') \label{e:solnx}
\end{align}
These are position space expressions, which Fourier transform into
\begin{align}
\Phi_{\{0\}}(z, \bs{k}) & = K_{d,\Delta}(z, k) \phi_{\{0\}(d-\Delta)}(\bs{k}), \label{e:solk0} \\[1ex]
\Phi^{\text{inhom}}_{\{n\}}(z, \bs{k}) & = \int_0^{\infty} \D \zeta \: \sqrt{g(\zeta)} \: G_{d,\Delta}(z, k; \zeta) \sum_{(n_j)} \left( \Ast_{j=1}^{M-1} \Phi_{\{n_j\}} \right) (\zeta, \bs{k}),  \label{e:solkn}
\end{align}
where $\Ast_{j=1}^{M-1}$ denotes an $(M-2)$-fold convolution with respect to momentum,
\begin{align} \label{e:convolution}
\left( \Ast_{j=1}^{M-1} \Phi_{\{n_j\}} \right)(\zeta, \bs{k}) & = \int \frac{\D^d \bs{q}_1}{(2 \pi)^d} \ldots \int \frac{\D^d \bs{q}_{M-2}}{(2 \pi)^d} \Phi_{\{n_1\}}(\zeta, \bs{q}_1) \Phi_{\{n_2\}}(\zeta, \bs{q}_2 - \bs{q}_1) \ldots \times\nn\\
& \qquad\qquad \times \Phi_{\{n_{M-2}\}}(\zeta, \bs{q}_{M-2} - \bs{q}_{M-3}) \Phi_{\{n_{M-1}\}}(\zeta, \bs{k} - \bs{q}_{M-2}).
\end{align}
As in \eqref{e:PertEq} the sums are taken over all sequences $(n_1, n_2, \ldots, n_{M-1})$ of non-negative integers such that $\sum_{j=1}^{M-1} n_j = n - 1$. The right hand sides of these equations depend on lower order perturbative solutions $\Phi_{\{k\}}$ with $k < n$. Hence, they can be solved recursively order by order in $\lambda$. The recursive nature of \eqref{e:solkn} is best visible in its graphical representation by means of the Witten diagrams in Figure \ref{fig:W}.

By expanding the bulk-to-bulk propagator in \eqref{e:solkn} according to \eqref{e:bulktobulk}, the perturbative solution reads
\begin{align} \label{e:Phinexp}
\Phi^{\text{inhom}}_{\{n\}}(z, \bs{k}) & = z^{d/2} K_{\Delta - \frac{d}{2}}(k z) \int_0^z \D \zeta \: \zeta^{-d/2-1} I_{\Delta - \frac{d}{2}}(k \zeta) \sum_{\{n_j\}} \left( \Ast_{j=1}^{M-1} \Phi_{\{n_j\}} \right) (\zeta, \bs{k}) \nn\\
& \qquad + z^{d/2} I_{\Delta - \frac{d}{2}}(k z) \int_z^\infty \D \zeta \: \zeta^{-d/2-1} K_{\Delta - \frac{d}{2}}(k \zeta) \sum_{\{n_j\}} \left( \Ast_{j=1}^{M-1} \Phi_{\{n_j\}} \right) (\zeta, \bs{k}).
\end{align}
Since both propagators vanish exponentially fast at $z = \infty$ the second integral is always convergent. The first integral, however, may diverge at $\zeta = 0$, but only if $\Delta \geq d$. Indeed, if $\Delta < d$, then the source term $\phi_{(d - \Delta)}$ is the most leading term in each perturbative solution $\Phi_{\{n_j\}}$. Hence the most leading term in the radial expansion of the first integrand is $\zeta^{(M - 2)(d - \Delta) - 1}$ and so the integral converges. If, on the other hand, $\Delta \geq d$, terms more leading than the source may appear and the integral diverges.

\subsubsection{Radial expansion}

First recall that the structure of the asymptotic expansion of $\Phi$ can be easily inferred from the equation of motion, for specific expressions see \cite{Skenderis:2002wp,vanRees:2011fr}. The zeroth order expansion is given by \eqref{e:freeExp}. The exact form of the expansion of subleading terms in $\lambda$ depends on specific values of parameters $d$ and $\Delta$. If certain discreet conditions are met, the expansion would contain logarithmic terms, such as these encountered in \eqref{e:freeExp} for $2 \Delta - d = 2 n$. However, since such cases form a discreet subset of the parameter plane $(d, \Delta)$ and we will regularize our theory by shifting the dimensions, we may concentrate on a generic case where no logarithms appear.

Assuming $\Delta < d$, the most leading term in the radial expansion of $\Phi$ is the source term $\phi_{(d - \Delta)}$. There are, however, three candidates for the first subleading term. These are either the first subleading term in $\Phi_{\{0\}}$ of order $d - \Delta + 2$, the vev term $\phi_{(\Delta)}$ of order $\Delta$, or a leading term of the perturbative solution $\Phi_{\{1\}}$. By looking at the equation of motion \eqref{e:fulleqofmo} we find that the aforementioned terms in the radial expansion of $\Phi$ are
\begin{align} \label{e:introexp}
\Phi & = z^{d - \Delta} \phi_{(d - \Delta)} + z^{d - \Delta + 2} \phi_{(d - \Delta + 2)} + \ldots \nn\\
& \qquad\qquad + \: z^{(M-1)(d - \Delta)} \phi_{((M-1)(d-\Delta))} + \ldots \nn\\
& \qquad\qquad + \: z^{\Delta} \phi_{(\Delta)} + \ldots,
\end{align}
Notice that if $\Delta < d$, then $(M-1)(d-\Delta) > d-\Delta$ and hence all omitted terms are subleading with respect to the ones written explicitly. If, however, $\Delta \geq d$, then $d - \Delta < 0$ and every next perturbative solution $\Phi_{\{n\}}$ contains terms of even more negative powers than $\Phi_{\{n-1\}}$. In such a case the source term is no longer the most leading term in the radial expansion.

Given $\phi_{(d - \Delta)}$ and $\phi_{(\Delta)}$ all remaining coefficients are locally determined by the equations of motion. For example,
\begin{equation} \label{e:ExampleTerm}
\phi_{((M-1)(d-\Delta))} = \frac{\lambda \phi_{(d-\Delta)}^{M-1}}{M (M-2) (d-\Delta) \left[ \frac{M-1}{M}d - \Delta \right]}.
\end{equation}
Not surprisingly this expression becomes ill-defined whenever $\Delta = d$ or $\Delta = (M-1) d/M$, which is part of the upper bound in \eqref{e:cond}.

For further analysis it will be important to classify all terms that can appear in $\Phi$ and in $\Phi_{\{n\}}$ for a given $n$. Since every term $\phi_{(\alpha)}$ in the radial expansion of $\Phi$ is a local functional of the source $\phi_{(d-\Delta)}$ and the vev $\phi_{(\Delta)}$ the only allowed values of $\alpha$ are those satisfying
\begin{equation} \label{e:alphaform}
\alpha = p (d - \Delta) + q \Delta + 2 r,
\end{equation}
where $p,q,r$ are non-negative integers such that
\begin{equation} \label{e:pq}
p + q = k(M - 2) + 1
\end{equation} 
for some non-negative integer $k$ and no restriction on $r$. Clearly, $p$ counts the number of sources, $q$ -- the number of vevs and $2r$ the number of derivatives in the coefficient $\phi_{(\alpha)}$. For example, the term \eqref{e:ExampleTerm} arises as $p=M-1$ and $q=r=0$, which is the first allowed term with $k=1$.

If one limits themselves to the analysis of possible terms appearing in the $n$-th perturbative solution $\Phi_{\{n\}}$, then the condition \eqref{e:pq} is accompanied with the requirement $k \leq n$. All these facts follow from equation \eqref{e:Phinexp} and simple combinatorics.

\subsubsection{Canonical momentum and 1-point function}

The AdS/CFT correspondence \eqref{e:adscft} dictates that the 1-point function in the dual field theory is given by a derivative of the on-shell action with respect to the source $\phi_{\{0\}(d - \Delta)} = \phi_0$,
\begin{equation} \label{e:to2pt}
\< \O(\bs{x}) \>_s = \frac{\delta S}{\delta \phi_{(d - \Delta)}(\bs{x})} = \int \D^d \bs{u} \D z \frac{\delta S}{\delta \Phi(z, \bs{u})} \frac{\delta \Phi(z, \bs{u})}{\delta \phi_{(d - \Delta)}(\bs{x})}.
\end{equation}
While for $\Delta < d$ the second functional derivative produces a delta function with the specific power of the radial variable, $z^{d - \Delta} \delta(\bs{u} - \bs{x}) \delta(z)$, the first term is identified with the canonical momentum following from action \eqref{e:S},
\begin{equation}
\Pi(z, \bs{u}) = \frac{1}{\sqrt{\gamma_z}} \frac{\delta S}{\delta \Phi(z, \bs{u})} = - z \partial_z \Phi(z, \bs{u}) + (d - \Delta) \Phi(z, \bs{u}).
\end{equation}

Assume $\Delta < d$. When expanded in powers of the radial variable the canonical momentum has the same form of the expansion as the bulk field in \eqref{e:introexp},
\begin{align}
\Pi & = 0 \times z^{d - \Delta} \phi_{(d - \Delta)} - 2 z^{d - \Delta + 2} \phi_{(d - \Delta + 2)}  + \ldots \nn\\
& \qquad\qquad - (M - 2)(d - \Delta) z^{(M-1)(d - \Delta)} \phi_{((M-1)(d-\Delta))}  + \ldots \nn\\
& \qquad\qquad - (2 \Delta - d) z^{\Delta} \phi_{(\Delta)} + \ldots,
\end{align}
where in each line the omitted terms are subleading with respect to the ones written explicitly. 

The expression \eqref{e:to2pt} for the 1-point function is well-defined if the leading term in the radial expansion of the canonical momentum is the vev term of order $\Delta$. As we can see this occurs exactly when the condition \eqref{e:cond} is satisfied. This means that the identity
\begin{equation} \label{e:1pt}
\< \O(\bs{x}) \>_s = \lim_{z \rightarrow 0} z^{-\Delta} \Pi(z, \bs{x}) = - (2 \Delta - d) \phi_{(\Delta)}(\bs{x}).
\end{equation}
holds when the condition \eqref{e:cond} is satisfied. Consequently, it must hold in the regulated theory as long as singularities are absent. This statement the vev coefficient $\phi_{(\Delta)}$ to be an analytic function of dimensions, the fact that we will prove in section \ref{sec:anal}.

\subsection{Correlation functions}

Having defined the usual asymptotic problem for the bulk scalar field $\Phi$ and extracting the 1-point function by means of the formula \eqref{e:1pt}, we may extract correlation functions and show their analyticity in dimensions $d$ and $\Delta$.

\subsubsection{Extraction of the vev} \label{sec:convergence}

In lieu of the relation \eqref{e:1pt} one needs to evaluate the vev coefficient $\phi_{(\Delta)}$ of the $n$-th perturbative solution \eqref{e:Phinexp}. The radial expansion of the second integral in \eqref{e:Phinexp} can be obtained by noticing that
\begin{equation}
(z \zeta)^{d/2} I_{\Delta - \frac{d}{2}}(k z) K_{\Delta - \frac{d}{2}}(k \zeta) = \frac{z^\Delta}{2 \Delta - d} K_{d, \Delta}(\zeta, k) + O(z^{\Delta + 2}),
\end{equation}
where $K_{d, \Delta}$ is the bulk-to-boundary propagator. Hence, due to the prefactor of $z^{\Delta}$, one needs to extract a constant term $z^0$ from the actual integral. This formally amounts to taking the lower integration limit $z = 0$. On the level of Witten diagrams in Figure \ref{fig:W} it corresponds to moving the bulk point $(z, \bs{x})$ to the boundary, leading to the expression,
\begin{equation} \label{e:vevk}
\phi_{\{n\}(\Delta)}(\bs{k}) = \frac{1}{2 \Delta - d} \int_0^{\infty} \D \zeta \: \sqrt{g(\zeta)} K_{d, \Delta}(\zeta, k) \sum_{(n_j)} \left( \Ast_{j=1}^{M-1} \Phi_{\{n_j\}} \right) (\zeta, \bs{k}).
\end{equation}
By using \eqref{e:1pt} we arrive at the 1-point function of order $\lambda^n$,
\begin{equation} \label{e:1ptk}
\< \O(\bs{k}) \>_{\{n\}, s} = - \int_0^{\infty} \D \zeta \: \sqrt{g(\zeta)} K_{d, \Delta}(\zeta, k) \sum_{(n_j)} \left( \Ast_{j=1}^{M-1} \Phi_{\{n_j\}} \right) (\zeta, \bs{k}).
\end{equation}
Let us recall that the sum is taken over all sequences $(n_1, n_2, \ldots, n_{M-1})$ of non-negative integers such that $\sum_{j=1}^{M-1} n_j = n - 1$.

This standard procedure \cite{Witten:1998qj} is valid under two assumptions. The integral in \eqref{e:vevk} must converge and the first integral in \eqref{e:Phinexp} cannot contribute to the vev coefficient. To check the first condition one can count powers of the radial variable in the expansion of the integrand in \eqref{e:vevk}. Assuming $\Delta < d$, the most leading term of any $\Phi_{\{n_j\}}$ does not exceed a source term. Hence, the most leading term in the power expansion of the integrand is $\zeta^{(M - 1)d - M \Delta - 1}$. Therefore the integral converges for $\Delta < (M-1)d/M$. Once again we recover the condition \eqref{e:cond}.

The same method can be applied to the analysis of the first integral in \eqref{e:Phinexp}. By power counting one finds that this expression does not contain terms of order $z^{\Delta}$ as long as the condition \eqref{e:cond} is satisfied. Hence we see that equation \eqref{e:1ptk} holds for any dimensions satisfying condition \eqref{e:cond}.

\subsubsection{Functional dependence} \label{sec:fundep}

In the next section we will analytically continue the theory in $d$ and $\Delta$. For that to work we have to actually define what we mean by such a continuation. For example, in case of the usual Feynman diagram calculations in a QFT it is results of momentum space integrals that are continued in $d$. In a similar fashion we want to regard expressions such as \eqref{e:1ptk} as a function of $d$ and $\Delta$ and analytically continue the result.

To make this precise, consider the on-shell action or the generating functional $W$ as a functional of the CFT source $\phi_{\{0\}(d - \Delta)} = \phi_0$. When expanded in $\phi_0$, each term represents a correlation function. Whether in position or momentum space, each $N$-point function is a function of $N$ vectors: the `arguments' of the dual operators $\O(\bs{x}_j)$. However, due to homogeneity and isotropy of constant radial slices of the AdS spacetime, any correlation function depends on scalar distances only $x_{ij} = | \bs{x}_i - \bs{x}_j|$. In momentum space this translates into the dependence on scalar products $k_{ij} = \bs{k}_i \cdot \bs{k}_j$ with $i \neq j$. Therefore we may consider any $N$-point function to be a scalar function of $N(N-1)/2$ scalar parameters. These parameters and their number do not depend on dimensions $d$ or $\Delta$. At least for $d$ sufficiently large no geometrical dependencies between different $x_{ij}$ exist and so all these variables are independent. Hence, if values of $x_{ij}$ or $k_{ij}$ are kept fixed, the correlation function becomes a function of $d$ and $\Delta$. It will be the task of the next section to show that this dependence is in fact analytic.

By simple counting of legs in a Witten diagram one finds that for a given $n$ the perturbative solution $\Phi_{\{n\}}$, and hence also its coefficient $\phi_{\{n\}(\Delta)}$, depend functionally on exactly 
\begin{equation} \label{e:Nn}
N_n = n(M - 2) + 1
\end{equation}
sources $\phi_{\{0\}(d - \Delta)} = \phi_0$. Therefore, if we define kernels $\mathcal{W}_n$ according to
\begin{equation} \label{e:defW}
\< \O(\bs{x}) \>_{\{n\}, s} = \int \D^d \bs{x}_1 \ldots \int \D^d \bs{x}_{N_n} \mathcal{W}_n(\bs{x}; \bs{x}_1, \ldots, \bs{x}_{N_n}) \phi_0(\bs{x}_1) \ldots \phi_0(\bs{x}_{N_n}),
\end{equation}
then the $(N_n + 1)$-point function reads
\begin{align} \label{e:corrfromphi}
\< \O(\bs{x}) \O(\bs{x}_1) \ldots \O(\bs{x}_{N_n}) \> & = (-1)^{N_n} \left. \frac{\delta^{N_n} \< \O(\bs{x}) \>_s}{\phi_0(\bs{x}_1) \ldots \phi_0(\bs{x}_{N_n})} \right|_{\phi_0 = 0}\nn\\
& = (-1)^{N_n} \lambda^n \sum_{\sigma \in S_{N_n}} \mathcal{W}_n(\bs{x}; \bs{x}_{\sigma(1)}, \ldots, \bs{x}_{\sigma(N_n)}) \nn\\
& = (-1)^{N_n} N_n! \lambda^n \mathcal{W}_n(\bs{x}; \bs{x}_1, \ldots, \bs{x}_{N_n}).
\end{align}
The sum extends over all permutations $\sigma$ of the set $\{1, 2, \ldots, N_n\}$ and in the last line we have used the fact that kernels $\mathcal{W}_n$ are symmetric in all their arguments. This follows from the symmetry of the defining equation \eqref{e:solnx} best visualized in the complete symmetry of the sum of the corresponding Witten diagrams. Finally, the entire generating functional $W$ can be built up out of $\phi_{\{n\}(\Delta)}$ by combining correlation functions. We find
\begin{align}
W[\phi_0] & = \sum_{n=0}^{\infty} \lambda^n \frac{(-1)^{N_n}}{N_n + 1} \int \D^d \bs{x} \< \O(\bs{x}) \>_{\{n\}, s} \phi_{0}(\bs{x}) \nn\\
& = \sum_{n=0}^{\infty} \lambda^n \frac{(-1)^{N_n + 1}}{N_n + 1} \int \D z \D^d \bs{x}  \sqrt{g} \Phi_{\{0\}} \sum_{(n_j)} \prod_{j=1}^{M-1} \Phi_{\{n_j\}},
\end{align}
where in the last line we used a position space version of \eqref{e:vevk} as well as the zeroth order solution \eqref{e:sol0x}. The generating functional is a formal power series in the coupling $\lambda$ with the term of order $\lambda^n$ containing exactly $N_n + 1$ sources and its kernel is proportional to the $(N_n + 1)$-point function. We conclude that the kernels $\mathcal{W}_n$ defined in equation \eqref{e:defW} are directly proportional to the correlation function and hence they are the objects that can be regarded as functions of dimensions $d$ and $\Delta$. These are the objects whose analytic properties we will analyze in the next section.

\subsubsection{Analytic continuation} \label{sec:anal}

In previous sections we have shown that the expression \eqref{e:1ptk} defining the 1-point function with sources turned on converges if condition \eqref{e:cond} is satisfied. We may regard each vev coefficient $\phi_{\{n\}(\Delta)}$ a function of dimensions $d$ and $\Delta$ by looking at its kernel $\mathcal{W}_n$ defined in \eqref{e:defW}. In this section we want to show that $\mathcal{W}_n$ are analytic functions of $d$ and $\Delta$ in the strip given by the complex version of the condition \eqref{e:cond},
\begin{equation} \label{e:condc1}
\frac{\re d}{2} < \re \Delta < \min \left( \frac{\re d}{2} + 1, \frac{M-1}{M} \re d \right).
\end{equation}
Assuming that being true, all correlation functions can be analytically extended beyond this region. Therefore, if one derives an expression for some correlation function as a function of dimensions, this expression is valid for any $d$ and $\Delta$ as long as the expression is non-singular.

The analyticity follows from a rather standard combination of Hartog, Morera, Cauchy and Fubini theorems. The Hartog theorem states that analyticity in two variables is equivalent to the analyticity in each variable separately. The Cauchy theorem states that in a simply-connected domain all integrals of an analytic function over closed curves vanish. The Morera theorem implies the opposite: if an integral of a function vanishes for any closed curve in a domain, then this function is analytic in this domain. The Fubini theorem allows for a change of the order of integration of a function $f$ of several variables: for example the change is allowed if any iterated integral of $|f|$ exists.

Consider the expression \eqref{e:vevk} that defines $\phi_{\{n\}(\Delta)}$ and its kernel $\mathcal{W}_n$ in \eqref{e:defW}. We want to show that for any closed curve $C$ in the parameter space $(d, \Delta) \in \C^2$ completely contained in the region \eqref{e:condc1},
\begin{equation} \label{e:to_anal}
\int_C \mathcal{W}_n(d, \Delta) \D d = 0, \qquad\qquad \int_C \mathcal{W}_n(d, \Delta) \D \Delta = 0.
\end{equation}
Then, by the combination of Hartog and Morera theorem $\mathcal{W}_n(d, \Delta)$ would be analytic in \eqref{e:condc1}. 

Consider expression \eqref{e:vevk} for the vev coefficient. By using \eqref{e:solk0} and \eqref{e:solkn} recursively, one can rewrite it as a long expression containing a number of integrals of propagators and determinants $\sqrt{g}$ and depending on the source $\phi_{\{0\}(d - \Delta)}$. In order to calculate \eqref{e:to_anal} we have to commute the contour integrals through all integrals appearing in the expansion of \eqref{e:vevk}. By Fubini theorem this would be allowed if the series of integrals in \eqref{e:vevk} converges when each term is replaced by its absolute value. Indeed, first notice that for a real $\nu$ the Bessel functions $K_\nu(x)$ and $I_\nu(x)$ are non-negative for all $x > 0$ and hence each term in the full expansion is positive in such case. Then, if $\Delta - d/2$ becomes complex, one can use the estimates
\begin{equation}
| K_{a+b \I}(x) | \leq 2 K_{a}(x), \qquad\qquad |I_{a+b \I}(x) | \leq \sqrt{ \cosh(\pi b) } I_a(x),
\end{equation}
valid for any $x,a > 0$ and real $b$. Hence, the change of the order of integration is allowed.

Finally, we need to prove that the integrand in the full expansion of \eqref{e:vevk} is analytic, so that the contour integrals \eqref{e:to_anal} vanish by Cauchy theorem. Notice that each propagator is an explicit combination of Bessel and power functions and all three functions $\nu \mapsto K_{\nu}(z)$, $\nu \mapsto I_{\nu}(z)$ and $\nu \mapsto z^\nu$ are analytic for $\re z > 0$ and $\re \nu > 0$. Hence the product of propagators and power functions is analytic in $d$ and $\Delta$ and by the Cauchy theorem \eqref{e:to_anal} vanishes. This proves the analyticity of the kernels \eqref{e:defW} and hence the analyticity of all regulated correlation functions of the dual CFT with respect to dimensions $d$ and $\Delta$ in the region \eqref{e:condc1}. 

The conclusion is this: whatever quantity one calculates in a specific system, first calculate it for general $d$ and $\Delta$. If the expression is well-defined inside the range \eqref{e:cond}, then its analytic continuation is unique. For example, if one is interested in correlation functions of some operator of dimension $\Delta \geq d$, then one calculates the correlation function in this range and the solution can be analytically continued outside the region. The obtained expression is a valid correlation function for the case of interest -- if it exists.

\subsection{Divergences} \label{sec:Divs}

Now that we have shown analytic properties of all correlation functions, we may look for singularities of their analytically continued expressions. Correlation functions are given recursively by a composition of equations \eqref{e:solk0}, \eqref{e:solkn} and \eqref{e:vevk}. First, notice that both bulk-to-boundary and bulk-to-bulk propagators are continuous functions of dimensions $d$ and $\Delta$, as one clearly sees from exact expressions \eqref{e:bulktobnd} and \eqref{e:bulktobulk}. This means that the zeroth order solution $\Phi_{\{0\}}$ given by \eqref{e:freesolPhi} is never singular and it is a continuous function of dimensions. Note that this does not imply that all coefficients in the radial expansion of $\Phi_{\{0\}}$ are finite. For example a divergence in the 2-point function \eqref{e:2ptk} stems from the divergence of the vev coefficient $\phi_{\{0\}(\Dreg)}$ despite the fact that the bulk solution is regular for all $d$ and $\Delta$. We will use this observation in sections \ref{sec:extract_source} and \ref{sec:extract_vev} in order to extract divergences of certain coefficients in the radial expansions.

In general the above statement fails for higher perturbative solutions and for correlation functions. However, continuity of $\Phi_{\{0\}}$ and both propagators implies that singularities in the regulator $\epsilon$ emerge from integrals in \eqref{e:solkn} and \eqref{e:vevk} only. All integrals converge at $z = \infty$, due to the exponential fall-off of propagators. Therefore all singularities emerge from the lower, near-boundary region of integration.

Consider a divergence arising in a perturbative solution $\Phi_{\{n\}}$ due to the first integral in \eqref{e:Phinexp}. A divergence is manifested as follows: for a fixed bulk point $(z, \bs{x})$ the perturbative solution $\Phi_{\{n\}}(z, \bs{x})$ is a function of dimensions $d$ and $\Delta$. We say that this solution is divergent at some point $(d, \Delta)$ if $\Phi_{\{n\}}(z, \bs{x})$ has a singularity there.

Consider now a divergent integral either in \eqref{e:Phinexp} or \eqref{e:vevk}. We may expand the integrand around $z = 0$ and try to integrate the series term by term. In principle every integral of $z^{\alpha}$ for $\alpha \leq -1$ will diverge while for $\alpha > -1$ one finds
\begin{equation} \label{e:generic_int}
\int_0^{\Lambda^{-1}} \D z \: z^{\alpha} = \frac{\Lambda^{-\alpha - 1}}{\alpha + 1}.
\end{equation}
The upper limit $\Lambda^{-1}$ is arbitrary and does not contribute to the singularity. The function on the right hand side is nevertheless singular at $\alpha = -1$ only. Since the integral is convergent and the result is analytic in $\alpha$ on an open and non-empty subset of the complex plane, the function
\begin{equation}
f(\alpha) = \frac{\Lambda^{-\alpha - 1}}{\alpha + 1}
\end{equation}
is the unique extension of the integral to any $\alpha \in \C \backslash \{ -1 \}$. Therefore, after the analytic continuation only terms of order equal exactly to $-1$ in the integrals of \eqref{e:solkn} and \eqref{e:vevk} contribute to the divergence of the correlation function.

In the remainder of this section we want to show that if some correlation function or a perturbative bulk solutions becomes singular, then there exists a triple $(p,q,r)$ of non-negative integers such that
\begin{equation} \label{e:pqr}
p(d - \Delta) + q \Delta + 2 r = d, \qquad\qquad p > \max(1, q),
\end{equation}
As usual, we assume $\Delta > d/2$. This will simply follow from the knowledge of the form \eqref{e:alphaform} of all possible powers in the radial expansion of bulk fields.

As pointed out above these singularities come either from the first integral in \eqref{e:Phinexp} or from the explicit integration in \eqref{e:vevk}. We will discuss these two cases separately, but in both we find that a singularity implies condition \eqref{e:pqr}.

\subsubsection{Asymptotic value problem} \label{sec:asym}

As discussed above both bulk-to-bulk and bulk-to-boundary propagators are well-defined and continuous for any values of dimensions $d$ and $\Delta$ (we always assume $\Delta > d/2$). This means that the zeroth order solution $\Phi_{\{0\}}$ is well-defined and obeys asymptotic condition $\Phi_{\{0\}} = z^{d-\Delta} \phi_{(d - \Delta)} + \text{subleading in } z$. For an interacting theory the asymptotic value problem is well-defined when the source term $\phi_{(d - \Delta)}$ is the most leading term in the radial expansion, which requires $\Delta < d$.

For $\Delta \geq d$ we may use analytic continuation to set-up the asymptotic problem. We start with the expression \eqref{e:Phinexp} for all $\Phi_{\{n\}}$ and then we use analytic continuation. The prescription may fail at certain values of $d$ and $\Delta$, but otherwise defines a valid solution. The asymptotic value problem becomes a statement that the term of order $d - \Delta$ in the bulk solution $\Phi$ must be equal to the source of the dual field theory even if $\Phi$ contains more leading terms. Again, this intuitive prescription does not require any cut-offs.

A singularity in a dimensionally continued version of \eqref{e:Phinexp} may only appear in the first integral, in its lower limit. This occurs when the integrand contains a term of order $-1$. Each of the perturbative solutions $\Phi_{\{n_j\}}$ contains terms of order \eqref{e:alphaform} for some triples $(p_j,q_j,r_j)$, while the factor in front expands into $\z^{\Delta - d + 2n - 1}$ for a non-negative integer $n$. In total we see that if the integrand contains a term of order $-1$, then the condition \eqref{e:pqr} is satisfied with
\begin{equation}
p = \sum_{j=1}^{M-1} p_j, \qquad q = 1 + \sum_{j=1}^{M-1} q_j, \qquad r = n + \sum_{j=1}^{M-1} q_j.
\end{equation}
Hence, assuming that the perturbative solution $\Phi_{\{n\}}$ becomes singular after its analytic continuation, we arrive at the following conclusions:
\begin{enumerate}
\item Condition \eqref{e:pqr} is satisfied with $q \geq 1$.
\item Each $\Phi_{\{n_j\}}$ must contribute at least one source or vev coefficient, \text{i.e.}, $p_j > 0$ or $q_j > 0$ for all $j$. Hence $p + q \geq 3$.
\item The expression multiplying the term $\zeta^{-1}$ in the integrand is a local functional of sources and vevs and contains exactly $p$ sources $\phi_{(\dreg - \Dreg)}$ and $q - 1$ vevs $\phi_{(\Delta)}$.
\end{enumerate}
We will use all these properties in following sections.

\subsubsection{Remaining divergences} \label{sec:rest}

Remaining divergences in the correlation function \eqref{e:vevk} follow from the explicit integral there. Again, by expanding the integrand in the radial variable we find that if a singularity occurs, then the following statements are true:
\begin{enumerate}
\item Condition \eqref{e:pqr} is satisfied with $q \geq 0$.
\item Each $\Phi_{\{n_j\}}$ in \eqref{e:vevk} must contribute at least one source or vev coefficient, \text{i.e.}, $p_j > 0$ or $q_j > 0$ for all $j$. Hence again $p + q \geq 3$.
\item The expression multiplying the term $\zeta^{-1}$ in the integrand is a local functional of sources and vevs and contains exactly $p$ sources $\phi_{(\dreg - \Dreg)}$ and $q$ vevs $\phi_{(\Delta)}$.
\end{enumerate}
As we can see there is a difference in $q$ when comparing these statements to the ones listed in previous subsection.

Additional inequalities in \eqref{e:pqr} follow from the assumption that $\Delta > d/2 > 0$ and the observation that in any case $p + q \geq 3$. Assume that it is possible that $q \geq p$. This means that for $\Delta > 0$, $p(d - \Delta) + q \Delta + 2 r \geq p d + 2 r$. Hence for the condition \eqref{e:pqr} to be satisfied with $q \geq p$ one needs $p \in \{0,1\}$. These cases can be analyzed separately.

For $p = 0$ or $p = 1$ we need $q \geq 2$. For $p = 0$ the condition \eqref{e:pqr} requires $\Delta = \frac{d - 2 r}{q} < \frac{d}{2}$ contradicting $\Delta > d/2$. Similarly for $p = 1$ we have $\Delta = \frac{- 2 r}{q - 1} < 0$.

As we will see in following sections numbers $p$, $q$ and $r$ in \eqref{e:pqr} correspond to a number of source terms $\phi_{(d - \Delta)}$, vev terms $\phi_{(\Delta)}$ and derivatives included in an appropriate counterterm canceling the divergence in a correlation function. For that reason we will include in our analysis the special case encountered in the analysis of 2-point functions, $2 \Delta - d = 2 n$, for a non-negative integer $n$. This corresponds to $p = 2$, $q = 0$ and $r = n$.

\subsubsection{Regularization}

Assume now that the condition \eqref{e:pqr} is satisfied and a singularity in some correlation function appears. In such a case we introduce a regulator $\epsilon$ by shifting dimensions according to \eqref{e:introReg},
\begin{equation} \label{e:reg}
d \longmapsto \dreg = d + u \epsilon, \qquad\qquad \Delta \longmapsto \Dreg = \Delta + v \epsilon,
\end{equation}
for fixed complex numbers $u$ and $v$. As we shall see not all choices of $u$ and $v$ regulate the action, but a set of good, regulating choices is uncountable and dense in the plane $(u,v) \in \C^2$. Remember that this parameterization differs from the one used in \cite{Bzowski:2013sza,Bzowski:2015pba,Bzowski:2015yxv} as here we regulate actual dimensions and not their combinations.

Assume \eqref{e:reg} regulates the correlation functions, \textit{i.e.}, no term of power exactly $-1$ occurs underneath any of the integrals in \eqref{e:vevk} nor \eqref{e:Phinexp} for all $n$. This means that every term $z^{-1}$ gets regulated to $z^{-1 + O(\epsilon)}$. This introduces a divergence in $\epsilon$, since now the integral \eqref{e:generic_int} reads
\begin{equation}
\int_0^{\Lambda^{-1}} \D z z^{-1 + p(\dreg - \Dreg) + q \Dreg + 2 r - \dreg} = \int_0^{\Lambda^{-1}} \D z z^{-1 + a_{pq} \epsilon} = \frac{\Lambda^{-a_{pq} \epsilon}}{a_{pq} \epsilon},
\end{equation}
where
\begin{equation} \label{e:a_val}
a_{pq} = p(u - v) + q v - u, \qquad p > \max(1, q).
\end{equation}

For the above expression to be valid we must make sure that the regularization \eqref{e:reg} actually regulates the divergences, \textit{i.e.}, that $a_{pq} \neq 0$ for all $p > \max(1, q)$. As one can see from \eqref{e:alphaform} one can always guarantee this by certain choices of $u$ and $v$. For example, one can choose $u$ rational and $v$ irrational or \textit{vice versa}. Two useful regularization schemes are:
\begin{itemize}
\item $u = v$, which gives $a_{pq} = q - 1$. This choice regularizes every case with $q \neq 1$. Its advantage is that the dimension of the source, $\dreg - \Dreg$, remains unchanged. An important drawback is that this scheme does not renormalize singularities with $q = 1$, which is crucial in the analysis of the marginal case $d = \Delta$.
\item $u = 2 v$, which gives $a_{pq} = v (p - q - 2)$. This choice regularizes every case apart from $(p,q)=(2,0)$. This means that it regulates all correlation functions but not 2-point functions. A huge advantage of this scheme is that the combination $\Dreg - \dreg/2$ appearing as an order of Bessel functions in propagators remains unchanged.
\end{itemize}

\section{Dimensional renormalization}

As discussed in the previous section there exist two sources of divergences in holographic correlation functions: a divergence in a perturbative solution $\Phi_{\{n\}}$ arising in the first integral of \eqref{e:Phinexp} and the divergences in the lower limit of the integral in \eqref{e:vevk}. Now we will show how these singularities can be removed by appropriate counterterms.

\subsection{Perturbative bulk solutions}

In this section we want to show that divergences in perturbative bulk solutions can be removed by source redefinition, \textit{i.e.}, by turning on subleading sources $\phi_{\{n\}(\dreg - \Dreg)}$. The procedure is inductive over the order $n$ of $\lambda$. The zeroth order solution $\Phi_{\{0\}}$ is given by the bulk-to-boundary propagator in \eqref{e:solk0} and hence it is well-defined for all $d$ and $\Delta$. Hence we define a trivially renormalized bulk solution, $\Phi_{\{0\}}^{\text{ren}}[\phi_0] = \Phi_{\{0\}}[\phi_0]$.

Assume now that we have a renormalized bulk solutions $\Phi_{\{k\}}^{\text{ren}}$ for all $k < n$ for some fixed $n$. This was achieved by turning on some sources of order $k < n$, so that
\begin{equation}
\Phi_{\{k\}}^{\text{ren}}[\phi_0] = \Phi_{\{k\}}[\phi_{(\dreg - \Dreg)}]
\end{equation}
is divergence-free. This means that for given $\dreg$ and $\Dreg$ the renormalized bulk solutions $\Phi_{\{k\}}^{\text{ren}}$ are finite when the regulator $\epsilon$ is taken to zero. By $\Phi_{\{k\}}[\phi_{(\dreg - \Dreg)}]$ we denote the $k$-th order solution \eqref{e:solkn} sourced by its argument, the redefined source.

Consider the only source of divergence in $\Phi_{\{n\}}$: the first integral in \eqref{e:Phinexp}. The integrand now contains renormalized bulk solutions $\Phi_{\{k\}}^{\text{ren}}$ for $k < n$. Hence the integrand is finite in $\epsilon \rightarrow 0$ limit and the divergence emerging from the integral is at most linear. From the discussion of previous section we know that the divergence occurs if the condition \eqref{e:pqr} is satisfied with $p > q \geq 1$ and we may rewrite the integral as follows,
\begin{align} \label{e:subleading_source1}
& z^{\dreg/2} K_{\Dreg - \frac{\dreg}{2}}(k z) \int_0^z \D \zeta \: \zeta^{-\dreg/2-1} I_{\Dreg - \frac{\dreg}{2}}(k \zeta) \sum_{(n_j)} \left( \Ast_{j=1}^{M-1} \Phi^{\text{ren}}_{\{n_j\}} \right) (\zeta, \bs{k}) \nn\\
& = K_{\dreg, \Dreg}(k, z) \times 2^{\Dreg - \frac{\dreg}{2} - 1} \Gamma \left( \Dreg - \frac{\dreg}{2} \right) \int_0^z \D \zeta \: \zeta^{-\dreg/2-1} k^{-(\Dreg - \frac{\dreg}{2})} I_{\Dreg - \frac{\dreg}{2}}(k \zeta) \sum_{\{n_j\}} \left( \Ast_{j=1}^{M-1} \Phi^{\text{ren}}_{(n_j)} \right) (\zeta, \bs{k}) \nn\\
& = K_{\dreg, \Dreg}(z, k) \times \left( \sum_{p > q \geq 1} \frac{z^{a_{pq} \epsilon}}{a_{pq} \epsilon} \mathcal{N}_{(a_{pq} \epsilon)} + \text{finite} \right)
\end{align}
In the last line we have used the bulk-to-boundary propagator $K_{d,\Delta}$ and represented an overall divergence in terms of some coefficients $\mathcal{N}_{(a_{pq} \epsilon)}$, where $a_{pq}$ is defined in \eqref{e:a_val}. We also know that these coefficients are locally expressible in terms of the source $\phi_{(\dreg - \Dreg)}$ and vev coefficients $\phi_{\{k\}(\Dreg)}^{\text{ren}}$ for $k < n$. Furthermore, the overall power of momentum is equal to an even non-negative integer as the series expansion of $k^{-\nu} I_{\nu}(k x)$ contains only powers $k^{2n}$ for a non-negative integer $n$. 

\subsubsection{Renormalization} \label{sec:extract_source}

The reasoning above shows that all divergences in a perturbative solution $\Phi_{\{n\}}$ are local. Furthermore, it delivers a simple procedure for their extraction. Since the bulk-to-boundary propagator satisfies $K_{\dreg, \Dreg} = z^{\dreg - \Dreg} + \text{subleading in } z$, the divergence is completely determined by coefficients in $\Phi_{\{n\}}$ close to the source term, namely those of $z^{\dreg - \Dreg + a_{pq} \epsilon}$ for $a_{pq}$ defined in \eqref{e:a_val}.

The divergence may therefore be removed by turning on a source of order $\lambda^n$. The value of the source is given by the expression in brackets in \eqref{e:subleading_source1}. Equivalently, the subleading source is such that
\begin{equation} \label{e:tosrcrenorm}
z^{\dreg - \Dreg} \phi_{\{n\}(\dreg - \Dreg)} + \sum_{p > q \geq 1} z^{\dreg - \Dreg + a_{pq} \epsilon} \phi_{\{n\}(\dreg - \Dreg + a_{pq} \epsilon)}^{\text{ren}} = O(\epsilon^0).
\end{equation}
All terms $\phi_{\{n\}(\dreg - \Dreg + a_{pq} \epsilon)}^{\text{ren}}$ are locally expressible in terms of lower order sources and vevs and follow from equations of motion. We assume here that the renormalization procedure has been carried out up to order $n-1$, so $\phi_{\{n\}(\dreg - \Dreg + a_{pq} \epsilon)}^{\text{ren}}$ are coefficients in $\Phi_{\{n\}}$ that are sourced by the truncated source $\phi_{(\dreg - \Dreg)}^{(n-1)}$ defined as
\begin{equation}
\phi_{(\dreg - \Dreg)}^{(k)} = \phi_0 + \sum_{j=1}^{k} \lambda^j \phi_{\{j\}(\dreg - \Dreg)}.
\end{equation} 
This lets us conclude that we need
\begin{align} \label{e:RedefSrc}
\phi_{\{n\}(\dreg - \Dreg)} & = - \sum_{p > q \geq 1} \mu^{-a_{pq} \epsilon} \phi^{\text{ren}}_{\{n\}(\dreg - \Dreg + a_{pq} \epsilon)} + O(\epsilon^0) \nn\\
& = - \sum_{p > q \geq 1} \mu^{(u-v) \epsilon} \phi^{\text{ren}}_{\{n\}(\dreg - \Dreg + a_{pq} \epsilon)}[\phi_0 \mu^{-(u-v)\epsilon}, \phi_{(\Dreg)} \mu^{-v \epsilon}, \partial_j] + O(\epsilon^0).
\end{align}
The renormalization scale $\mu$ was introduced on dimensional grounds so that the right hand side has a correct dimension of the source. From point 3 in section \ref{sec:asym} we know that each $\mathcal{N}_{(a_{pq} \epsilon)}$, and hence each $\phi^{\text{ren}}_{\{n\}(\dreg - \Dreg + a_{pq} \epsilon)}$, depends on exactly $p$ sources $\phi_0$ and $q-1$ vev coefficients $\phi_{(\Dreg)}$. This allows us to redistribute the renormalization scale accordingly. This redistribution is precisely such that dimensions of $\phi_0 \mu^{-(u-v)\epsilon}$ and $\phi_{(\Dreg)} \mu^{-v \epsilon}$ are equal to unrenormalized $d$ and $\Delta$ respectively.

By turning on the subleading source we render the perturbative solution finite, according to \eqref{e:field_decomp}. By carrying out this procedure order by order in $\lambda$ we renormalize all perturbative bulk solutions by turning on subsequent sources building up full $\phi_{(\dreg - \Dreg)}$. Now each $\Phi_{\{n\}}[\phi_{(\dreg - \Dreg)}]$ has a finite limit $\epsilon \rightarrow 0$. In this way we have traded a singularity in the bulk solutions for a singularity in sources. As we will see this effectively induces a beta function in the dual theory.

In the remaining part of the paper we will abandon a superscript `ren' on perturbative bulk solutions, assuming instead that the redefined source $\phi_{(\dreg - \Dreg)}$ is turned on.

\subsection{Remaining divergences}

Having redefined sources according to \eqref{e:RedefSrc} we may may look at singularities coming from the final integral in \eqref{e:vevk}. In other words, we look for singularities of the vev coefficient $\phi_{\{n\}(\Dreg)}[\phi_{(\dreg - \Dreg)}]$ in each bulk solution $\Phi_{\{n\}}$ now sourced by $\phi_{(\dreg - \Dreg)}$. By the source redefinition procedure these bulk fields are finite at $\epsilon \rightarrow 0$. Hence the integral in \eqref{e:vevk} may produce at most linear singularity from its lower limit.

A divergence can be retrieved by radial expanding the integrand in \eqref{e:vevk} and keeping terms of order $-1 + O(\epsilon)$ as usual. Clearly, each term is a local functional of the source $\phi_{(\dreg - \Dreg)}$ and lower order vev coefficients $\phi_{\{k\}(\Dreg)}$ for $k < n$. In section \ref{sec:fundep} we combined all terms $\phi_{\{n\}(\Delta)}$ into a generating functional $W$. We can therefore consider divergences in all regulated correlation functions by defining divergent part of the generating functional,
\begin{equation} \label{e:Wdiv}
W_{\text{div}}(\epsilon) = \int \D^{\dreg} \bs{x} \sum_{p > \max(1, q)} \frac{\mu^{- a_{pq} \epsilon}}{a_{pq} \epsilon} \mathcal{D}_{(a_{pq} \epsilon)} + O(\epsilon^0).
\end{equation}
The sum is organized in terms of the powers of the renormalization scale $\mu$ and $\mathcal{D}_{(a_{pq} \epsilon)}$ are some coefficients that have a finite, non-zero limit $\epsilon \rightarrow 0$ limit. The constants $a_{pq}$ are defined in \eqref{e:a_val}. Precise values of $\mathcal{D}_{(a_{pq} \epsilon)}$ can be simply retrieved by expanding the bulk field in \eqref{e:vevk} in the radial variable. As no additional singularities are present, the final divergence is at most a single pole in $\epsilon$.

In the remainder of this section we will show how the coefficients $\mathcal{D}_{(a_{pq} \epsilon)}$ can be extracted from the bulk solutions and the on-shell action itself.

\subsubsection{Extraction from bulk solutions} \label{sec:extract_vev}

While the method for the extraction of divergences from expanding \eqref{e:vevk} is perfectly valid, it would be useful to retrieve the divergences from the canonical momentum or -- even better -- the on-shell action. This is achieved in a manner similar to that of section \ref{sec:extract_source}.

Assume that subleading sources have been turned on, so that each perturbative bulk solution $\Phi_{\{n\}}$ is regular at $\epsilon = 0$. It does not imply, however, that each coefficient in their radial expansions are such continuous functions. Nevertheless, it does imply certain cancellations between singularities.

Consider the vev coefficient $\phi_{(\Dreg)}$ in the regulated theory. If some correlation function has a singularity in $\epsilon$, so does the vev coefficient. However, as $\epsilon \rightarrow 0$, the full bulk solution $\Phi$ must remain continuous. This means that divergences between the vev coefficient $\phi_{(\Dreg)}$ and all other terms $\epsilon$-close to the vev term, $\phi_{(\Dreg + O(\epsilon))}$, must combine into a finite expression,
\begin{equation} \label{e:loctermspi}
z^{\Dreg} \phi_{(\Dreg)} + \sum_{p > \max(1, q)} z^{\Dreg + a_{pq} \epsilon} \phi_{(\Dreg + a_{pq} \epsilon)} = O(\epsilon^0),
\end{equation}
where $a_{pq}$ are defined in \eqref{e:a_val}. According to the results of section \ref{sec:rest} the sum contains all possible terms $\epsilon$-close to the vev term. This leads to a simple expression for singularities of the vev coefficient
\begin{equation} \label{e:DivPi}
\phi_{(\Dreg)} = -\sum_{p > \max(1, q)} \phi_{(\Dreg + a_{pq} \epsilon)} + O(\epsilon^0).
\end{equation}
By point 3 of section \ref{sec:rest} each term on the right hand side is a local functional of exactly $p$ sources $\phi_0$ and $q$ vevs of lower order.

Notice that this method of extraction of the divergence is identical to the one described in section \ref{sec:extract_source} and hence equation \eqref{e:loctermspi} has the same form as equation \eqref{e:tosrcrenorm}. There is, however, a difference when it comes to the renormalization. While before we could have simply redefined sources by means of equation \eqref{e:RedefSrc}, a similar procedure is not possible here. The renormalization will be achieved by an addition of a counterterm.

\subsubsection{Extraction from on-shell action} \label{sec:extract_S}

In general the canonical momentum can be functionally integrated to the on-shell action. This procedure, however, may be difficult, especially for the case of irrelevant operators. Furthermore, it would be much more convenient to be able to extract the divergence from the bulk action directly, as is the case for the holographic renormalization method, at least for $\Delta < d$.

The idea here is that the vev coefficient $\phi_{(\Dreg)}$ follows solely from the constant term roughly of the form $\phi_{(\dreg - \Dreg)} \phi_{(\Dreg)}$ in the on-shell action. This is the $p = q = 1$ term and hence the power of the radial variable in the on-shell action equals $-1$ exactly and is immune to any regularization \eqref{e:reg}. After a regularization every other term multiplied by $z^{-1 + a_{pq} \epsilon}$ in the on-shell action corresponds to a local term of order $z^{\Dreg + a_{pq} \epsilon}$ in the canonical momentum.

Let us first consider the free theory action \eqref{e:Sfree}. When integrated by parts this action becomes
\begin{equation}
S_{\text{free}} = \frac{1}{2} \lim_{z \rightarrow 0} \int_{z} \D^d \bs{x} \sqrt{\gamma_{z}} \Phi \Pi
\end{equation}
and posses the finite limit when \eqref{e:cond} is satisfied. The product $\Phi \Pi$ in this expression contains a term $\phi_{(d - \Delta)} \phi_{(\Delta)}$ multiplied by $z^0 = 1$. This is the term which produces the 1-point function when differentiated with respect to the source. Any other term of the order $z^0$ is then related to the local terms in \eqref{e:loctermspi} encoding the divergences of the vev.

When regulated according to \eqref{e:reg}, only the vev producing term $\phi_{(\dreg - \Dreg)} \phi_{(\Dreg)}$ remains unregulated. Every other local term is now multiplied by a power of the radial variable shifted by a factor proportional to the regulator $\epsilon$. In particular terms of order $O(\epsilon)$ must encode singularities of all correlation functions. Indeed, when the on-shell action is differentiated with respect to the field $\Phi$ it produces the canonical momentum where all singularities of the vev coefficient $\phi_{(\Dreg)}$ are encoded in local terms $\phi_{(\Dreg + O(\epsilon))}$.

There is a straighforward method to remove the vev producing term $\phi_{(d - \Delta)} \phi_{(\Delta)}$ from the action, while simultaneously keeping all divergent terms. Write,
\begin{equation}
S_{\text{div}} = - \frac{1}{2} \int_0^{\infty} \D z \frac{\partial}{\partial z} \int \D^{\dreg} \bs{x} \sqrt{\gamma_z} \Phi \Pi + O(\epsilon^0).
\end{equation}
The constant term is now removed by the derivative, while the dependence of the remaining terms on $\epsilon$ is unchanged. Furthermore, in this way we have rewritten the analytically continued on-shell action as an integral over the radial variable.

Consider now the full action \eqref{e:S}. By using the equations of motion \eqref{e:fulleqofmo} we arrive at
\begin{equation} \label{e:DivS}
W_{\text{div}} = - S_{\text{div}} = \frac{1}{2} \int_0^{\mu^{-1}} \D z \left[ \frac{\partial}{\partial z} \int \D^{\dreg} \bs{x} \sqrt{\gamma_z} \Phi \Pi + \lambda \frac{M-2}{M} \int \D^{\dreg} \bs{x} \sqrt{g} \Phi^M \right] + O(\epsilon^0).
\end{equation}
By the principle of analytic continuation, one only needs to extract terms of order $z^{-1 + O(\epsilon)}$ in the integrand and then integrate them in order to obtain singularities in $\epsilon$. The divergence does not depend on the choice of $\mu$ since the integral converges at infinity. Note that it is important to redefine sources according to \eqref{e:RedefSrc} before this divergence is evaluated. To see how this procedure works in practice, let us consider an example in a free field theory, which we analyzed in section \ref{sec:DIMREG}.

\subsubsection{Example}

Let us quickly check equations \eqref{e:DivPi} and \eqref{e:DivS} for the case of the 2-point function of an operator of dimension $\Delta = 3$ in $d = 4$ spacetime dimensions as discussed in section \ref{sec:DIMREG}. The radial expansion is given by \eqref{e:explog}, which in this case reads
\begin{equation}
\Phi = z^{1 + (u -v)\epsilon} \phi_0 + z^{3 + v \epsilon} \phi_{(3 + v \epsilon)} + z^{3 + (u -v)\epsilon} \phi_{(3 + (u - v)\epsilon)} + O(z^5).
\end{equation}
The source term is $\phi_{(1 + (u - v)\epsilon)} = \phi_0$, while $\phi_{(3 + v \epsilon)}$ is the vev term. The third coefficient, locally expressible in term of the source, is given by \eqref{e:phidDelta2}, which here gives
\begin{equation}
\phi_{(3 + (u - v)\epsilon)} = \frac{\partial^2 \phi_0}{2 (2v - u) \epsilon}.
\end{equation}
This is indeed divergent in the $\epsilon \rightarrow 0$ limit and hence equation \eqref{e:DivPi} predicts,
\begin{equation}
\phi_{(3 + v \epsilon)} = - \frac{\partial^2 \phi_{(1 + (u -v)\epsilon)}}{2 (2v - u) \epsilon} + O(\epsilon^0).
\end{equation}
This shows that the regulated 2-point function must be divergent at $\epsilon = 0$ and allows for the extraction of this divergence. By using \eqref{e:1pt} we find that the divergent part of the regulated 2-point function in momentum space is equal to $k^2/((2v - u)\epsilon)$ in agreement with previously found \eqref{e:2ptreg}.

The same result one obtains by the analysis of the on-shell action. The integrand in \eqref{e:DivS} reads
\begin{equation}
-\frac{1}{2} \sqrt{\gamma_z} \Phi \Pi =  \left( 2 + (2v - u) \epsilon \right) \phi_0 \phi_{(3 + v\epsilon)} + \frac{z^{(u-2v)\epsilon}}{2(2v - u) \epsilon} \phi_0 \partial^2 \phi_0 + O(z^2).
\end{equation}
When differentiated with respect to $z$ the first term vanishes and the second one produces a divergence in the generating functional,
\begin{equation}
S_{\text{div}} = \frac{1}{2(2v - u) \epsilon} \phi_0 \partial^2 \phi_0 + O(\epsilon^0).
\end{equation}
By taking minus two functional derivatives with respect to the source $\phi_0$, this produces the correct divergence in the 2-point function equal $k^2/((2v-u)\epsilon)$.

\subsection{Renormalization}

Now we want to understand how the procedure described in the previous subsections impacts the generating functional. From the point of view of the bulk physics, the source redefinition alters the 1-point function. Indeed, the regulated 1-point function in \eqref{e:1pt} was derived under the assumption that the CFT source is given by the entire bulk source term $\phi_{(\dreg - \Dreg)}$. Since now, however, the CFT source $\phi_0 = \phi_{\{0\}(\dreg - \Dreg)}$, we may introduce a \emph{redefined} 1-point function,
\begin{align} \label{e:1ptRen}
\< \O(\bs{x}) \>_{s, \text{redef}} = \frac{\delta S}{\phi_0(\bs{x})} & = \int \D^{\dreg} \bs{u} \frac{\delta S}{\delta \phi_{(\dreg - \Dreg)}(\bs{u})} \frac{\delta \phi_{(\dreg - \Dreg)}(\bs{u})}{\delta \phi_0(\bs{x})} \nn\\
& = \int \D^{\dreg} \bs{u} \< \O(\bs{u}) \>_{s, \text{reg}}[\phi_{(\dreg - \Dreg)}] \frac{\delta \phi_{(\dreg - \Dreg)}(\bs{u})}{\delta \phi_0(\bs{x})}.
\end{align}
The first term underneath the integral is identified with the regulated 2-point function as presented in equation \eqref{e:to2pt}. The second term equals $\delta(\bs{u} - \bs{x}) + O(\lambda)$, where the subleading terms emerge from the source redefinition \eqref{e:RedefSrc}.

The 1-point function with sources turned on is a functional derivative of the generating functional and the expression \eqref{e:1ptRen} is a derivative of
\begin{equation} \label{e:WctZ}
W_{\text{reg}}[\phi_{(\dreg - \Dreg)}[\phi_0]] = \< \exp \left( -\int \D^{\dreg} \bs{x} \phi_{(\dreg - \Dreg)}[\phi_0] \O \right) \>_{\text{reg}}.
\end{equation}
This immediately leads to a CFT interpretation of the result. Since $p \geq 2$ in \eqref{e:pqr}, each subleading source $\phi_{\{n\}(\dreg - \Dreg)}$ depends functionally on at least two sources $\phi_0$. Hence one may use a parametrization,
\begin{equation} \label{e:SrcZ}
\phi_{(\dreg - \Dreg)} = \phi_0 Z[\phi_0 \mu^{-(u-v)\epsilon}, \phi_{(\Dreg)} \mu^{-v \epsilon}, \partial_j], \qquad\qquad Z = 1 + O(\phi_0).
\end{equation}
Now we clearly recognize the multiplicative renormalization factor of a dimensionally regulated QFT. This allows us to identify $\phi_0$ with the renormalized source and $\phi_{(\dreg - \Dreg)}$ with the bare source. Further implications will be analyzed in section \ref{sec:QFT}.

The remaining part of the renormalization procedure is to extract divergences of $W_{\text{reg}}[\phi_{(\dreg - \Dreg)}[\phi_0]]$ and construct a local counterterm functional,
\begin{equation} \label{e:toWZ}
W_{WZ}[\phi_0] = - W_{\text{reg}}[\phi_{(\dreg - \Dreg)}[\phi_0]] + O(\epsilon^0).
\end{equation}
The subscript $WZ$ stands for Wess-Zumino. The functional $W_{WZ}$ is not exactly equal to the Wess-Zumino action, since here it is a quantity divergent in $\epsilon$. It does, however, generate anomalies in the dual field theory and hence its name seems to be appropriate.

The overall renormalized generating functional would be then equal to \eqref{e:introW}. In order to finalize the renormalization procedure we need to show two facts. First, the divergence in $W_{\text{reg}}[\phi_{(\dreg - \Dreg)}[\phi_0]]$ must be local and its contribution to any correlation function should be expressible locally in terms of sources and lower order vev coefficients. Second, there should exist a unique correspondence between bulk and boundary counterterms. We deal with these two issues in the following subsections.

\subsubsection{Derivation}

To derive the Wess-Zumino counterterm, we have to consider the structure of the 1-point function \eqref{e:1ptRen}. We can parametrize the functional derivative there as follows,
\begin{equation}
\frac{\delta \phi_{(\dreg - \Dreg)}(\bs{u})}{\delta \phi_0(\bs{x})} = \delta(\bs{u} - \bs{x}) + \lambda \mathcal{K}[\phi_0](\bs{u} - \bs{x}),
\end{equation}
for some functional $\mathcal{K}$ containing at least one CFT source $\phi_0$. The 1-point function then can be written as
\begin{equation} \label{e:1ptRenA}
\< \O(\bs{x}) \>_{s, \text{redef}} = \< \O(\bs{x}) \>_{s, \text{reg}}[\phi_{(\dreg - \Dreg)}] - \lambda (2 \Dreg - \dreg) \int \D^{\dreg} \bs{u} \phi_{(\Dreg)}[\phi_{(\dreg - \Dreg)}] \mathcal{K}[\phi_0].
\end{equation}
The divergences of the regulated 1-point function with sources are captured by \eqref{e:DivPi}. On the level of the generating functionals, the divergence is captured by $W_{\text{div}}$ in \eqref{e:Wdiv}, which can be retrieved from the on-shell action by means of the formula \eqref{e:DivS}. The second term in \eqref{e:1ptRenA} is subleading in $\lambda$ with respect to the first and is `more local' than the first term. That means that if a contribution to the $n$-point function is considered, the second term depends on lower order correlation functions only. Hence, working order by order in $\lambda$ one can use \eqref{e:DivPi} to derive the divergence in the first term of \eqref{e:1ptRenA} and use lower order result for the second term. This produces a total divergence of the 1-point function $\< \O(\bs{x}) \>_{s, \text{redef}}$ that is local and hence can be gathered into a local generating functional,
\begin{equation} \label{e:WdivRedef}
W_{\text{reg}}[\phi_{(\dreg - \Dreg)}[\phi_0]] = \int \D^{\dreg} \bs{x} \sum_{p > \max(1, q)} \frac{\mu^{- a_{pq} \epsilon}}{a_{pq} \epsilon} \mathcal{D}'_{(a_{pq} \epsilon)} + O(\epsilon^0),
\end{equation}
where $a_{pq}$ are defined in \eqref{e:a_val}. As in the case of \eqref{e:Wdiv} the sum is organized in terms of the powers of the renormalization scale $\mu$ and $\mathcal{D}'_{(a_{pq} \epsilon)}$ are some coefficients that have a finite, non-zero limit $\epsilon \rightarrow 0$ limit. Each term $\mathcal{D}'_{(a_{pq} \epsilon)}$ depends on $p$ source coefficients $\phi_{(\dreg - \Dreg)}$ and $q$ vev coefficients $\phi_{(\Dreg)}$ as it was the case for \eqref{e:Wdiv}. If no source redefinition is required, \textit{i.e.}, $\phi_0 = \phi_{(\dreg - \Dreg)}$, then $\mathcal{D}'_{(a_{pq} \epsilon)} = \mathcal{D}_{(a_{pq} \epsilon)}$ and the on-shell action \eqref{e:DivS} can be used to find the divergences. In such a case $W_{WZ} = - W_{\text{div}} = S_{\text{div}}$.

The Wess-Zumino counterterm is simply \eqref{e:toWZ}. As we have argued, it is a local counterterm, contributing locally to all correlation function. In the next subsection we show how it can be rewritten in terms of the bulk data.

\subsubsection{Equivalence of counterterms}

As discussed above the Wess-Zumino counterterm is local from the point of view of the CFT. Each term $\mathcal{D}'_{(a_{pq} \epsilon)}$ in \eqref{e:WdivRedef} depends on exactly $p$ source coefficients and $q$ vev coefficients and the counterterm can be written similarly to \eqref{e:RedefSrc} as
\begin{align} \label{e:WZ}
W_{WZ}(\epsilon, \mu) & = - \int \D^{\dreg} \bs{x} \sum_{p > \max(1, q)} \frac{\mu^{- a_{pq} \epsilon}}{a_{pq} \epsilon} \mathcal{D}'_{(a_{pq} \epsilon)}[\phi_{(\dreg - \Dreg)}, \phi_{(\Dreg)}, \partial_j] \nn\\
& = - \int \D^{\dreg} \bs{x} \sum_{p > \max(1, q)} \frac{\mu^{u \epsilon}}{a_{pq} \epsilon} \mathcal{D}'_{(a_{pq} \epsilon)}[\phi_{(\dreg - \Dreg)} \mu^{-(u-v)\epsilon}, \phi_{(\Dreg)} \mu^{-v \epsilon}, \partial_j].
\end{align}

In order to obtain the corresponding covariant bulk counterterm, one needs to convariantize this expression. This can be achieved by the same logic as presented in section \ref{sec:equiv}. We have
\begin{align} \label{e:BulkCtrs}
& \sqrt{\gamma_z} (z \mu)^{-a_{pq} \epsilon } \mathcal{D}'_{(a_{pq} \epsilon)}[ \Phi, -(2 \Dreg - \dreg)^{-1} \Pi, \nabla_j ] = \nn\\
& \qquad\qquad = \mu^{-a_{pq} \epsilon } \mathcal{D}'_{(a_{pq} \epsilon)}[ \phi_{(\dreg - \Dreg)} + \text{subleading in } z,\   \phi_{(\Dreg)} + \text{subleading in } z, \ \partial_j ] \nn\\
& \qquad\qquad = \mu^{u \epsilon} \mathcal{D}'_{(a_{pq} \epsilon)}[ \phi_{(\dreg - \Dreg)} \mu^{-(u-v)\epsilon}, \phi_{(\Dreg)}  \mu^{-v \epsilon}, \partial_j ] + \text{subleading in } z
\end{align}
where the omitted terms are subleading in the radial variable. As usual the counterterm contribution should be evaluated for generic $d$ and $\Delta$ in the range \eqref{e:reg} and then analytically continued. In particular all above equalities hold in the region of convergence \eqref{e:reg}.

The bulk-covariant version of the counterterm \eqref{e:WZ} then reads
\begin{align} \label{e:SWZ}
W_{WZ} & = \lim_{z \rightarrow 0} \int \D^{\dreg} \bs{x} \sqrt{\gamma_z} \sum_{p > \max(1, q)} \frac{(\mu z)^{-a_{pq} \epsilon}}{a_{pq} \epsilon} \mathcal{D}'_{(a_{pq} \epsilon)}[ \Phi, -(2 \Dreg - \dreg)^{-1} \Pi, \nabla_j ] \nn\\
& = \lim_{z \rightarrow 0} \int \D^{\dreg} \bs{x} \sqrt{\gamma_z} \sum_{p > \max(1, q)} \frac{1}{a_{pq} \epsilon} \left( \frac{z}{L} \right)^{-a_{pq} \epsilon} \mathcal{D}'_{(a_{pq} \epsilon)}[ \Phi, -(2 \Dreg - \dreg)^{-1} \Pi, \nabla_j ],
\end{align}
where we have used equation \eqref{e:Lmu} to relate the renormalization scale $\mu$ to the AdS radial $L$. Notice that it is precisely the condition \eqref{e:pqr} that ensures that this expression is of order $z^{0}$ in the radial variable and hence it is supported on the boundary. In fact every covariant bulk term that is supported on the boundary requires the condition \eqref{e:pqr} to be satisfied. This follows from the radial expansion of the bulk fields $\Phi$ and $\Pi$ in the range \eqref{e:cond}. It is a straightforward generalization of the argument of section \ref{sec:equiv} for the case of higher-point correlation functions.

All of this shows that expressions: \eqref{e:WZ} and \eqref{e:SWZ} are identical and equal to each other. This established the promised result of exact equality of counterterms both on the CFT and the bulk side of the AdS/CFT correspondence.

\subsection{QFT interpretation} \label{sec:QFT}

By combining results of two previous sections, we have obtained a perturbative procedure for dimensional renormalization of any correlation function. This led us to a conclusion that the renormalized generating functional is the sum of the regulated generating functional with redefined sources \eqref{e:WctZ} and an explicit Wess-Zumino counterterm action \eqref{e:WZ},
\begin{equation} \label{e:WREN}
W[\phi_0; \mu] = W_{\text{reg}}[\phi_{(\dreg - \Dreg)}[\phi_0; \epsilon, \mu]] + W_{WZ}[\phi_{(\dreg - \Dreg)}[\phi_0; \epsilon, \mu]; \epsilon, \mu].
\end{equation}
Here $\phi_{(\dreg - \Dreg)}$ is the redefined source, given recursively by equation \eqref{e:RedefSrc}. Now we want to compare this result with a standard dimensionally regulated QFT.

\subsubsection{Renormalization}

Perturbative dimensional renormalization of a QFT is a standard textbook material. A regulated generating functional (of connected correlation functions) can be written as
\begin{equation}
W_{\text{reg}}[\phi_0] = \< \exp \left( - \int \D^{\dreg} \bs{x} \phi_0 \O \right) \>_{\text{reg}},
\end{equation}
where the expectation value is taken in the regulated theory. The multiplicative renormalization procedure \cite{Collins:1984xc} results in a perturbative renormalization factor $Z$ introduced in the action and a possible anomaly inducing Wess-Zumino counterterm. The renormalized generating functional can then be written as
\begin{equation} \label{e:WrenCFT}
W[\phi_0; \mu] = \< \exp \left( - \int \D^{\dreg} \bs{x} \phi_0 Z[\phi_0, \O; \epsilon, \mu] \O \right) \>_{\text{reg}}  + W_{WZ}[\phi_0, \O; \epsilon, \mu]
\end{equation}
with $Z = 1 + O(\phi_0)$. This expression has precisely the form we have found in previous sections. The first term is produced by a perturbative source redefinition leading to equation \eqref{e:WctZ}. This equation proves an equivalence between the AdS and CFT counterterms. To be precise, equation \eqref{e:SrcZ} shows that the renormalization $Z$ factor obtained there and the standard QFT renormalization factors are equal. The counterterms on two sides of the correspondence match exactly.

The remaining part of the equivalence is to show that the Wess-Zumino counterterms match. A dictionary between the CFT counterterms expressed by means of the CFT data and the bulk counterterms given by the bulk objects is expressed by the equality of \eqref{e:WZ} and \eqref{e:SWZ}. We have also seen from equation \eqref{e:BulkCtrs} that for a valid bulk counterterm to be supported on the boundary, one needs the condition \eqref{e:pqr} to be satisfied. We may now also argue that every CFT counterterm must also satisfy this condition. 

Indeed, in a dimensionally regulated CFT all possible counterterms can be easily classified. As no scale is available, a counterterm must be composed out of sources, operators and derivatives in such a way that it becomes dimensionless. The renormalization scale $\mu$ may appear only with a power that vanishes in the $\epsilon \rightarrow 0$ limit, where $\epsilon$ is a regulator. For example, the counterterm \eqref{e:genWct2} is a valid term only if $2 \Delta - d = 2 n$, for a non-negative integer $n$. 

In general, for a counterterm built up with $p$ sources, $q$ 1-point functions $\< \O \>_s$ and $2r$ derivatives to be dimensionless, one needs the condition \eqref{e:pqr} to be satisfied. This leads to the conclusion that the most general QFT counterterm must indeed be of the form \eqref{e:WZ}.

\subsubsection{Classification of cases}

Depending on dimensions $d$ and $\Delta$ only a certain class of counterterms is allowed. The classification is the standard one:
\begin{itemize}
\item For $\Delta < d$ only counterterms containing the source $\phi_0$ are allowed, \textit{i.e.}, $q = 0$ in \eqref{e:pqr}. Furthermore, this condition can be satisfied for a finitely many triples $(p,q,r)$ only. Indeed, in addition to $q = 0$ one needs $p < d/(d - \Delta)$ and $r < d/2$. Therefore this is a case of a super-renormalizable theory since only a finite number of counterterms may appear.
\item In a marginal case $\Delta = d$ counterterms with a single operator insertion may appear as well. Then the condition \eqref{e:pqr} is satisfied for $q = 1$, $r = 0$ and arbitrary $p \geq 2$. In total a single renormalization factor $Z[\phi_0]$ appears, plus a possible anomaly if the spacetime dimension $d$ is even. This is the case of a renormalizable theory.
\item For $\Delta > d$ the number of different counterterms is unlimited. Since $(d - \Delta) < 0$, with $p$ sufficiently large one can increase $q$ without limits. This is the case of a non-renormalizable theory. Non-renormalizability means here that an infinite number of different counterterms is required to renormalize all correlation functions. The theory remains renormalizable in the sense that every correlator can be renormalized. In particular for any $N$, if one limits the attention to $n$-point functions with $n \leq N$, then a finite number of counterterms is required. This is also the case where terms such as $\Phi^3 \Pi^2$ may appear in the bulk Wess-Zumino counterterm. From the point of view of the holographic renormalization procedure such multitrace counterterms where analyzed in \cite{vanRees:2011fr,vanRees:2011ir}.
\end{itemize}

Obviously, in any case one needs to consider counterterms only if the condition \eqref{e:pqr} is satisfied. Otherwise correlation functions are uniquely determined by analytic continuation from the region of convergence \eqref{e:cond}.

\subsection{Scale-dependence}

Scale-dependence of any local QFT is governed by the Callan-Symanzik equation, 
\begin{equation} \label{e:CS}
\mu \frac{\D}{\D \mu} W[\phi_0; \mu] = \left( \mu \frac{\partial}{\partial \mu} + \int \D^d \bs{x} \beta(\phi_0) \frac{\delta}{\delta \phi_0(\bs{x})} \right) W[\phi_0; \mu] = \int \D^d \bs{x} \mathcal{A}[\phi_0].
\end{equation}
Since dimensions of the operators are fixed by the bulk mass, we do not have anomalous dimensions.

Let us act with the derivative with respect to the scale on the renormalized generating functional \eqref{e:WREN}. The explicit renormalization scale appears either via the redefined source $\phi_{(\dreg - \Dreg)}$ or explicitly in the Wess-Zumino counterterm. The derivative is therefore equal
\begin{align}
\mu \frac{\partial}{\partial \mu} W[\phi_0; \mu] & = \int \D^d \bs{u} \frac{\delta (W_{\text{reg}}+ W_{WZ})}{\delta \phi_{(\dreg - \Dreg)}(\bs{u})}[\phi_{(\dreg - \Dreg)}] \mu \frac{\partial}{\partial \mu} \phi_{(\dreg - \Dreg)}(\bs{u}) + \mu \frac{\partial}{\partial \mu} W_{WZ} \nn\\
& = \int \D^d \bs{x} \frac{\delta W[\phi_{0}; \mu]}{\delta \phi_{0}(\bs{x})} \int \D^d \bs{u} \left(  \frac{\delta \phi_{(\dreg - \Dreg)}(\bs{u})}{\delta \phi_0(\bs{x})} \right)^{-1} \mu \frac{\partial}{\partial \mu} \phi_{(\dreg - \Dreg)}(\bs{u}) + \mu \frac{\partial}{\partial \mu} W_{WZ},
\end{align}
where in the second line we have used a chain rule in order to obtain a functional derivative of the renormalized generating functional with respect to the renormalized source $\phi_0$. By comparing with the RG equation \eqref{e:CS} we immediately identify the beta function and the anomaly,
\begin{align} \label{e:beta}
\beta(\phi_0) & = - \int \D^d \bs{u} \left(  \frac{\delta \phi_{(\dreg - \Dreg)}(\bs{u})}{\delta \phi_0(\bs{x})} \right)^{-1} \mu \frac{\partial}{\partial \mu} \phi_{(\dreg - \Dreg)}(\bs{u}), \\
\int \D^d \bs{x} \mathcal{A} & = \lim_{\epsilon \rightarrow 0} \mu \frac{\partial}{\partial \mu} W_{WZ}(\phi_0). \label{e:an}
\end{align}
The redefined source is given by \eqref{e:RedefSrc}, while the Wess-Zumino counterterm by \eqref{e:WZ}. This is an extremely simple and powerful expression for the field theoretic data in a dimensionally regulated QFT by means of the bulk data of a holographic theory.

\subsubsection{Holographic beta function}

The expression for beta function \eqref{e:beta} agrees with a definition valid in a perturbative QFT,
\begin{equation}
\beta(\phi_0) = \mu \frac{\D}{\D \mu} \phi_0
\end{equation}
since the two expressions are related by the functional version of a derivative of an inverse function, $(f^{-1})' = -1/f'$. Notice that from the point of view of the standard perturbative renormalization it is the renormalized source $\phi_0$ that is scale-dependent, while the bare source $\phi_{(\dreg - \Dreg)}$ is scale-independent. This is despite equation \eqref{e:RedefSrc}, which could suggest that the situation is rather the opposite.

Consider a marginal case where the dimensions satisfy $d = \Delta$. In such a case $\phi_{(\dreg - \Dreg)}$ is a polynomial in the renormalized source $\phi_0$. To be more specific, it has the following expansion,
\begin{equation} \label{e:RedefSrcZ}
\phi_{(\dreg - \Dreg)} = \phi_0 \left[ 1 + \sum_{n=1}^{\infty} \lambda^n a_n \left(\phi_0 \mu^{-(u-v)\epsilon} \right)^n \right] = \phi_0 Z[\lambda \phi_0 \mu^{-(u-v) \epsilon}],
\end{equation}
with the dependence on the renormalization scale $\mu$ that can be argued purely on dimensional grounds. The renormalization factor has an expansion 
\begin{equation}
Z(g) = 1 + \sum_{n=1}^{\infty} a_n g^n,
\end{equation}
where coefficients $a_n$ are the same that appear in the redefined source \eqref{e:RedefSrcZ}. Once again, the standard QFT textbooks identify $\phi_0$ with the renormalized source, while the product $\phi_{(\dreg - \Dreg)} = \phi_0 Z[\lambda \phi_0 \mu^{-(u-v) \epsilon}]$ is a bare source.

If the deformation is marginal, then the functional derivative underneath the integral in \eqref{e:beta} becomes proportional to $\delta(\bs{x} - \bs{u})$ and we find
\begin{equation} \label{e:betaMargin}
\beta_{\phi_0} = - \left( \frac{\partial \phi_{(\dreg - \Dreg)}}{\partial \phi_0} \right)^{-1} \mu \frac{\partial}{\partial \mu} \phi_{(\dreg - \Dreg)}. 
\end{equation}
This expression typically will depend explicitly on the renormalization scale. Indeed, the source $\phi_{(\dreg - \Dreg)}$ is dimensionful and even for a marginal operator its dimension is $[\phi_{(\dreg - \Dreg)}] = (v - u) \epsilon$. In the context of the perturbative QFT one usually defines a dimensionless coupling $g = \phi_0 \mu^{-(u - v)\epsilon}$. The beta function for $g$ then reads
\begin{equation}
\beta_g = - (u - v) \epsilon g + \mu^{-(u-v) \epsilon} \beta_{\phi_0}
\end{equation}
and the right hand side depends on $g = \phi_0 \mu^{-(u - v)\epsilon}$ only.

\subsubsection{Holographic anomalies}

The anomaly given by \eqref{e:an} can be expressed more explicitly in terms of the bulk data via equation \eqref{e:WZ}. It is not clear, however, how the coefficients $\mathcal{D}'_{(a_{pq} \epsilon)}$ can be directly retrieved from the on-shell action. This is due to the effect described by equation \eqref{e:1ptRenA}. While it is straightforward to use \eqref{e:DivS} to obtain a divergence in the regulated vev coefficient $\phi_{(\dreg - \Dreg)}$, the actual Wess-Zumino counterterm requires a careful analysis of the generating functional \eqref{e:WctZ}.

The situation simplifies for relevant operators, $\Delta < d$. In such a case no source redefinitions appear and the only singularity in correlation functions follows from the explicit integral in \eqref{e:vevk}. In such case
\begin{equation}
W_{WZ} = - W_{\text{div}} = S_{\text{div}}
\end{equation}
where the divergent part of the on-shell action is given by \eqref{e:DivS}. A divergence of the Wess-Zumino counterterm must be at most linear for the anomaly to be finite. We find
\begin{equation}
\lim_{\epsilon \rightarrow 0} \mu \frac{\partial}{\partial \mu} W = \int \D^{d} \bs{x} \sum_{p=2}^{\infty} \mathcal{D}_{(a_{p0} \epsilon)}[\phi_{(\dreg - \Dreg)}, \phi_{(\Dreg)}, \partial_j].
\end{equation}
Since this is a case of a relevant operator, only counterterms with $q = 0$ can appear. Constants $a_{pq}$ are defined in \eqref{e:a_val}. Furthermore, there is only a finite number of non-vanishing terms under the sum.

If the divergent part of the action is parametrized as
\begin{equation}
S_{\text{div}} = \int_0^{\mu^{-1}} \D z \int \D^{\dreg} \bs{x} \sum_{p=2}^{\infty} z^{-1 + a_{p0} \epsilon} \mathcal{L}_{(-1 + a_{p0} \epsilon)} + \text{finite},
\end{equation}
then this implies
\begin{equation}
\mathcal{D}_{(a_{p0} \epsilon)} = \mathcal{L}_{(-1 + a_{p0} \epsilon)}.
\end{equation}
The anomaly can be read off of the on-shell action, as it is the case for the holographic renormalization procedure. Recall, however, that a proper derivation of the counterterm requires prior renormalization of sources, the same procedure that one encounters in a perturbative QFT.

Finally, as in the case of 2-point functions, we see that \eqref{e:SWZ} implies
\begin{equation}
\mu \frac{\partial}{\partial \mu} = - L \frac{\partial}{\partial L} = z \frac{\partial}{\partial z}
\end{equation}
relating a change in the radial coordinate with the change of the renormalization scale.

\section{Examples}

Here we present two simple examples in $d=3$ spacetime dimensions. In both cases we consider the action \eqref{e:preS} with cubic interaction,
\begin{equation} \label{e:S3}
S = \int \D^{3} \bs{x} \D z \sqrt{g} \left[ \frac{1}{2} g^{\mu\nu} \partial_\mu \Phi \partial_\nu \Phi + \frac{1}{2} m^2 \Phi^2 - \frac{\lambda}{3} \Phi^3 \right].
\end{equation}
Formally, this action needs to be supplemented with the boundary term as indicated in \eqref{e:S}. 

In the first example we consider the dual operator $\O$ of dimension $\Delta = 2$, while in the second we analyze the case of $\Delta = 3$. The first example is a relevant operator with only a single singularity corresponding to the condition \eqref{e:pqr} satisfied with $p=3$ and $q=r=0$. In the latter case we have a more complicated example of a marginal operator. Its 2- and 3-point function in momentum space was evaluated in \cite{Bzowski:2015pba} by means of holographic renormalization.

In both cases we will carry out a renormalization procedure up to order $\lambda^2$ and compute 2-, 3-, and 4-point functions. In both cases we use the same regularization scheme, $u = 2$ and $v = 1$. The idea behind such a scheme is that the combination $\Dreg - \dreg/2$ appearing as orders of Bessel functions in propagators \eqref{e:bulktobnd} and \eqref{e:bulktobulk} does not regulate, so it is equal to unregulated $\Delta - d/2$. Since Bessel functions of a half-integral order simplify to elementary functions, the propagators simplify considerably.

\subsection{Relevant operator} \label{sec:ExampleEasy}

Here we consider an operator of dimension $\Delta = 2$ in $d=3$ spacetime dimensions. The bulk action is given by \eqref{e:S3}. The propagators in the scheme with $u=2$ and $v=1$ read
\begin{align}
K_{\dreg, \Dreg}(z, k) & = z^{1 + \epsilon} e^{-k z}, \label{e:K1} \\
G_{\dreg, \Dreg}(z, k; \zeta) & = \frac{(z \zeta)^{1 + \epsilon}}{k} \times \left\{ \begin{array}{ll}
e^{-k z} \sinh(k \zeta) & \text{ for } \zeta < z, \\
e^{-k \z} \sinh(k z) & \text{ for } \zeta > z
\end{array} \right. \label{e:G1}
\end{align}

\subsubsection{2-point function}

The coefficient of $z^{2+\epsilon}$ in the bulk-to-boundary propagator equals $-k$ and hence the 2-point function is finite and reads
\begin{equation}
\lla \O(\bs{k}) \O(-\bs{k}) \rra = \lla \O(\bs{k}) \O(-\bs{k}) \rra_{\text{reg}} = - k.
\end{equation}

\subsubsection{3-point function}

The 3-point function is given by the following triple-$K$ integral,
\begin{align} \label{e:ex2reg3}
\lla \O(\bs{k}_1) \O(\bs{k}_2) \O(\bs{k}_3) \rra_{\text{reg}} & = - 2 \lambda \int_0^{\infty} \D z z^{-\dreg-1} K_{\dreg,\Dreg}(z, k_1) K_{\dreg,\Dreg}(z, k_2) K_{\dreg,\Dreg}(z, k_3) \nn\\
& = - 2 \lambda \int_0^{\infty} \D z z^{-1 + \epsilon} e^{-(k_1 + k_2 + k_3) z} \nn\\
& = - 2 \lambda \Gamma(\epsilon) (k_1 + k_2 + k_3)^{\epsilon} \nn\\
& = - \frac{2 \lambda}{\epsilon} + 2 \lambda \left[ \log (k_1 + k_2 + k_3) + \gamma_E \right] + O(\epsilon).
\end{align}
As expected, the 3-point function is singular and from the point of view of the CFT the singularity is canceled by the Wess-Zumino counterterm,
\begin{equation} \label{e:ct3}
W_{WZ}[\phi_0; \epsilon, \mu] = \lambda \left( - \frac{1}{3 \epsilon} + a_0 \right) \int \D^{\dreg} \bs{x} \mu^{-\epsilon} \phi_0^3.
\end{equation}
The total renormalized generating functional is then $W = W_{\text{reg}} + W_{WZ}$. By taking minus three derivatives of this expression we arrive at a finite, renormalized correlation function,
\begin{equation}
\lla \O(\bs{k}_1) \O(\bs{k}_2) \O(\bs{k}_3) \rra = 2 \lambda \left[ \log \left( \frac{k_1 + k_2 + k_3}{\mu} \right) + \gamma_E - 3 a_0 \right].
\end{equation}
The 3-point function is anomalous and the total anomaly in the generating functional is
\begin{equation} \label{e:toAnom}
\lim_{\epsilon \rightarrow 0} \mu \frac{\D}{\D \mu} W[\phi_0; \mu] = \lim_{\epsilon \rightarrow 0} \mu \frac{\partial}{\partial \mu} W_{WZ}[\phi_0; \mu] = \frac{\lambda}{3} \int \D^3 \bs{x} \phi_0^3.
\end{equation}

\subsubsection{Divergences in the 3-point function}

While the computation of the 3-point function is straightforward, let us discuss its divergence. The counterterm \eqref{e:ct3} is written in the language of the dual CFT and we would like to rewrite it in terms of the bulk data. Since $\Phi = z^{1 + \epsilon} \phi_0 + O(z^2)$ the counterterm from the bulk perspective reads
\begin{align}
S_{WZ} = - W_{WZ} & = \lambda \left( \frac{1}{3 \epsilon} - a_0 \right) \lim_{z \rightarrow 0} \int \D^{\dreg} \bs{x} \sqrt{\gamma_z} \Phi^3 (z \mu)^{-\epsilon} \nn\\
& = \lambda \left( \frac{1}{3 \epsilon} - a_0 \right) \lim_{z \rightarrow 0} \int \D^{\dreg} \bs{x} \sqrt{\gamma_z} \Phi^3 \left( \frac{z}{L} \right)^{-\epsilon},
\end{align}
where in the last line we used the relation $L = \mu^{-1}$ to relate the renormalization scale $\mu$ to the only available scale in classical AdS, its radius $L$.

In the above calculations we have simply evaluated the 3-point function together with its divergence. On the other hand one should be able to extract the singularity at $\epsilon = 0$ without the need for the evaluation of the entire correlation function. First, let us see how procedure of section \ref{sec:extract_vev} allows us to extract the divergence in the 3-point function from local terms in the bulk solution. For that one needs to radial expand the bulk field $\Phi$ and find values of coefficients in terms of the source and vev. Up to order $\lambda$, in the scheme $u = 2$ and $v = 1$, the expansion reads
\begin{align}
\Phi & = \phi_0 z^{1 + \epsilon} + \phi_{\{0\}(2 + \epsilon)} z^{2 + \epsilon} + O(z^3) \nn\\
& \qquad\qquad + \lambda \left[ \phi_{\{1\}(2 + \epsilon)} z^{2 + \epsilon} + \phi_{\{1\}(2 + 2 \epsilon)} z^{2 + 2 \epsilon} + O(z^3) \right] + O(\lambda^2).
\end{align}
The source coefficient is $\phi_{\{0\}(1 + \epsilon)} = \phi_0$ while $\phi_{(2 + \epsilon)}$ is the vev term. The fourth coefficient in the expansion, $\phi_{\{1\}(2 + 2 \epsilon)}$ is given locally in terms of the source by equations of motion,
\begin{equation}
\phi_{\{1\}(2 + 2 \epsilon)} = - \frac{\phi_0^2}{\epsilon(1 + \epsilon)}
\end{equation}
and is divergent. From this it follows that the divergence in the vev coefficient is simply
\begin{equation}
\phi_{\{1\}(2 + \epsilon)} = - \phi_{\{1\}(2 + 2 \epsilon)} + O(\epsilon^0) = \frac{\phi_0^2}{\epsilon} + O(\epsilon^0).
\end{equation}
By taking two derivatives with respect to the source $\phi_0$ and remembering about the minus sign in \eqref{e:adscft} we confirm the singularity in \eqref{e:ex2reg3}.

The total divergence of the generating functional can also be extracted from the on-shell action by means of the formula \eqref{e:DivS}. We need to expand the integrand and keep terms of order $z^{-1 + O(\epsilon)}$ only. We find once again
\begin{equation}
S_{\text{div}} = - \frac{\lambda}{3 \epsilon} \int \D^{\dreg} \bs{x} \phi_0^3 + O(\epsilon^0),
\end{equation}
which again agrees with \eqref{e:ex2reg3} as $W = -S$.

\subsubsection{4-point function}

Since no counterterms are available, the 4-point function must be finite on its own. This is a rather important improvement over the method of holographic renormalization, where counterterms are required in almost any case. It does not mean, however, that we may send $\epsilon$ to zero at the very beginning of the calculations. Rather, it signals the fact that the singularity in the 4-point function is removable, so the finite $\epsilon \rightarrow 0$ limit exists after all integrals are evaluated. Nevertheless, in the dimensional regularization method it is physically possible -- and rather simple -- to evaluate the 4-point function at all.

\begin{figure}[ht]
	\includegraphics[width=0.80\textwidth]{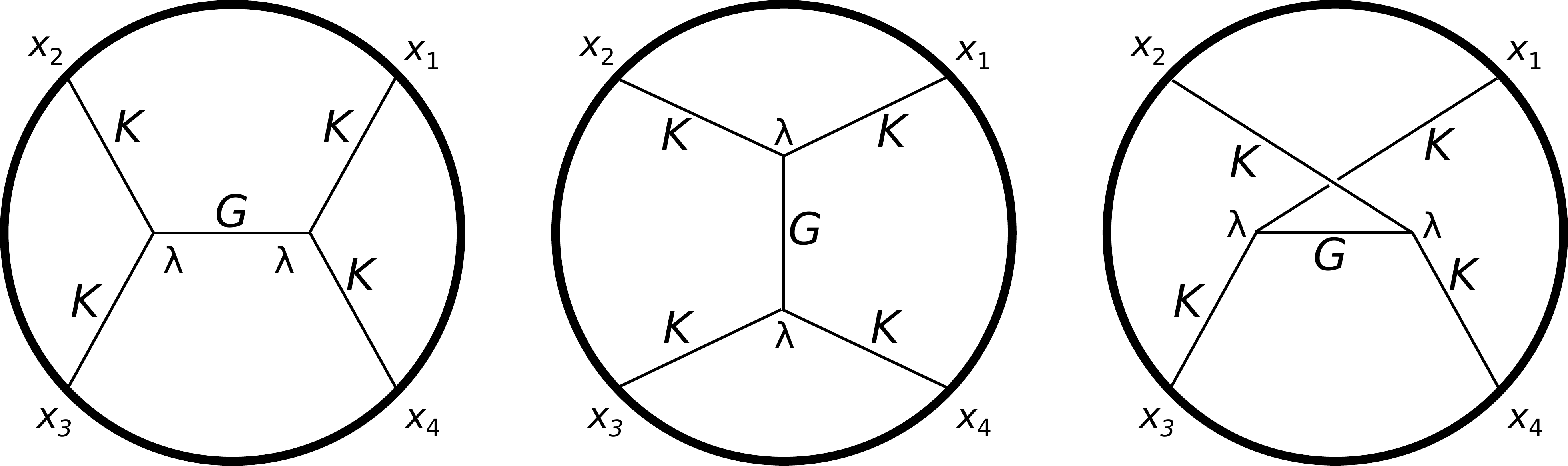}
	\centering
	\caption{Three Witten diagrams contributing to the 4-point function in \eqref{e:reg4pt1}. Permutations of external legs represent three inequivalent permutations of momenta.}
	\label{fig:4}
\end{figure}

The regulated 4-point function reads
\begin{equation} \label{e:reg4pt1}
\lla \O(\bs{k}_1) \O(\bs{k}_2) \O(\bs{k}_3) \O(\bs{k}_4) \rra = 4 \left[ I(\bs{k}_1, \bs{k}_2; \bs{k}_3, \bs{k}_4) + I(\bs{k}_1, \bs{k}_3; \bs{k}_2, \bs{k}_4) + I(\bs{k}_1, \bs{k}_4; \bs{k}_2, \bs{k}_3) \right],
\end{equation}
where the expressions $I(\bs{k}_1, \bs{k}_2; \bs{k}_3, \bs{k}_4)$ correspond to three distinct Witten diagrams in Figure \ref{fig:4}. In terms of propagators one has
\begin{align} \label{e:reg3i1}
I(\bs{k}_1, \bs{k}_2; \bs{k}_3, \bs{k}_4) & = \lim_{\epsilon \rightarrow 0} \int_0^{\infty} \D z z^{-\dreg-1} K_{\dreg,\Dreg}(z, k_1) K_{\dreg,\Dreg}(z, k_2) \times \nn\\
& \qquad\qquad \times \int_0^{\infty} \D \zeta \z^{-\dreg-1} G_{\dreg, \Dreg}(z, |\bs{k}_1 + \bs{k}_2|; \z) K_{\dreg,\Dreg}(\z, k_3) K_{\dreg,\Dreg}(\z, k_4).
\end{align}
By definition the integral is symmetric under two transpositions $\bs{k}_1 \leftrightarrow \bs{k}_2$ or $\bs{k}_3 \leftrightarrow \bs{k}_4$ as well as the joint transposition $(\bs{k}_1, \bs{k}_2) \leftrightarrow (\bs{k}_3, \bs{k}_4)$. Note also that due to the momentum conservation $|\bs{k}_1 + \bs{k}_2| = |\bs{k}_3 + \bs{k}_4|$, and we denote 
\begin{equation}
k_{12} = | \bs{k}_1 + \bs{k}_2 | , \qquad\qquad s_{ij} = k_i + k_j.
\end{equation}
Note that $s_{ij} \geq k_{ij}$ by triangle inequality, $k_{12} = k_{34}$, but $s_{12} \neq s_{34}$ in general.

The bulk-to-bulk propagator in \eqref{e:reg3i1} can be divided into its two parts and each part can be integrated separately. The resulting expression contains the regulator $\epsilon$ but a finite limit $\epsilon \rightarrow 0$ exists. The final expression reads
\begin{align}
I(\bs{k}_1, \bs{k}_2; \bs{k}_3, \bs{k}_4) & = \frac{1}{4 k_{12}} \left[ 2 \Li_2 \left( - \frac{s_{12} - k_{12}}{s_{34} + k_{12}} \right) - 2 \Li_2 \left( - \frac{s_{12} + k_{12}}{s_{34} - k_{12}} \right) \right.\nn\\
& \qquad\qquad\qquad \left. + \log \left( \frac{s_{34} + k_{12}}{s_{34} - k_{12}} \right) \log \left( \frac{s_{34}^2 - k_{12}^2}{(s_{12} + k_{12})^2} \right) \right],
\end{align}
where $\Li_2$ denotes dilogarithm. The full 4-point function is given by \eqref{e:reg4pt1}.

\subsection{Marginal operator} \label{sec:ExampleHard}

Here we consider a marginal operator with $\Delta = d = 3$. For a marginal operator we expect a sequence of singularities in all $n$-point functions for $n \geq 3$. The only way to satisfy the condition \eqref{e:pqr} in our case is $q = 1$ and $r = 0$ with arbitrary $p \geq 2$. This indicates a present of beta functions but excludes any anomalies, which require $q = 0$.

The bulk action is given by \eqref{e:S3}. The propagators in the scheme with $u=2$ and $v=1$ read
\begin{align}
K_{\dreg, \Dreg}(z, k) & = z^{\epsilon} e^{-k z}(1 + k z), \label{e:K3} \\
G_{\dreg, \Dreg}(z, k; \z) & = \frac{(z \zeta)^{\epsilon}}{k^3} \times \left\{ \begin{array}{ll}
e^{-k z} (1 + k z) \left(k \zeta \cosh(k \zeta) - \sinh(k \zeta) \right) & \text{ for } \zeta < z, \\
e^{-k \z} (1 + k \z) \left(k z \cosh(k z) - \sinh(k z) \right) & \text{ for } \zeta > z
\end{array} \right. \label{e:G3}
\end{align}

The 2-point function does not require a regularization. From \eqref{e:K3} the coefficient of $z^{3 + \epsilon}$ equals $-k^3/3$, regardless of the regularization, and hence
\begin{equation}
\lla \O(\bs{k}) \O(-\bs{k}) \rra_{\text{reg}} = \lla \O(\bs{k}) \O(-\bs{k}) \rra = k^3.
\end{equation}

\subsection{3-point function}

By expanding bulk fields in \eqref{e:vevk} one finds the following expression for the 3-point function,
\begin{align} \label{e:toO3}
& \lla \O(\bs{k}_1) \O(\bs{k}_2) \O(\bs{k}_3) \rra_{\text{reg}} = - 2 \lambda \int_0^{\infty} \D z z^{-\dreg-1} K_{\dreg,\Dreg}(z, k_1) K_{\dreg,\Dreg}(z, k_2) K_{\dreg,\Dreg}(z, k_3) \nn\\
& \qquad\qquad = - 2 \lambda \int_0^{\infty} \D z z^{-4 + \epsilon} e^{-(k_1 + k_2 + k_3) z} (1 + k_1 z) (1 + k_2 z) (1 + k_3 z) \nn\\
& \qquad\qquad = \frac{2 \lambda}{3 - \epsilon} \Gamma(-1 + \epsilon) (k_1 + k_2 + k_3)^{-\epsilon} \left[ k_1^3 + k_2^3 + k_3^3 - \epsilon Q_1+ \epsilon^2 k_1 k_2 k_3 \right]
\end{align}
where
\begin{equation} \label{e:Q1}
Q_1 = Q_1(k_1, k_2, k_3) = k_1 k_2 k_3 - ( k_1^2 k_2 + 5 \text{ permutations} ).
\end{equation}
While the integral in the second line is heavily divergent, one can carry it out by assuming $\epsilon$ sufficiently large. The last line remains divergent in $\epsilon \rightarrow 0$ limit and its expansion in the regulator reads
\begin{align} \label{e:O3reg}
& \lla \O(\bs{k}_1) \O(\bs{k}_2) \O(\bs{k}_3) \rra_{\text{reg}} = - \frac{2 \lambda}{3 \epsilon} (k_1^3 + k_2^3 + k_3^3) + \nn\\
& \qquad + \frac{2 \lambda}{3} \left[ Q_1  + (k_1^3 + k_2^3 + k_3^3) \left( \log(k_1 + k_2 + k_3) + \gamma_E - \frac{4}{3} \right) \right] + O(\epsilon).
\end{align}

Let us now discuss in some detail divergences encountered in this 3-point function from the AdS point of view. We will show how the analysis of section \ref{sec:extract_source} is applicable in this case. First we will discuss the extraction of divergences from the integral \eqref{e:toO3} directly, then we will show how this divergence is seen in the radial expansion of the bulk field $\Phi$.

\subsubsection{The integral}

For the singularities in the 3-point function \eqref{e:O3reg} we may use expression in the first line of \eqref{e:toO3} directly, by expanding propagators up to and including terms of order $z^3$. The divergence is then given by
\begin{align}
\lla \O(\bs{k}_1) \O(\bs{k}_2) \O(\bs{k}_3) \rra_{\text{reg}} & = - 2 \lambda \int_0^{\mu^{-1}} \D z \left[ \ldots + \frac{1}{3} z^{-1 + \epsilon} (k_1^3 + k_2^3 + k_3^3) + \ldots \right] \nn\\
& = - \frac{2 \lambda}{3 \epsilon} (k_1^3 + k_2^3 + k_3^3) + \text{finite}.
\end{align}
In the first line, under the integral, we have written explicitly the only relevant term of the form $z^{-1 + O(\epsilon)}$. Despite the fact that the most leading term is $z^{-4 + \epsilon}$, only terms of order close to $-1$ contribute to the divergence, as explained in section \ref{sec:Divs}.

\subsubsection{Radial expansion} \label{sec:exRadial}

The radial expansion of the bulk field $\Phi$ up to order $\lambda$ reads
\begin{align} \label{e:exPhiExp}
\Phi & = z^{\epsilon} (\phi_{0} + \lambda \phi_{\{1\}(\epsilon)}) + z^{2+\epsilon} (\phi_{\{0\}(2+\epsilon)} + \lambda \phi_{\{1\}(2+\epsilon)}) + z^{3 + \epsilon} ( \phi_{\{0\}(3 + \epsilon)} + \lambda \phi_{\{1\}(3 + \epsilon)} ) \nn\\
& \qquad\qquad + \lambda \left[ z^{2 \epsilon} \phi_{\{1\}(2 \epsilon)} + z^{2+2\epsilon} \phi_{\{1\}(2 + 2\epsilon)} + z^{3+2\epsilon} \phi_{\{1\}(3 + 2\epsilon)} \right] + O(\lambda^2, z^4).
\end{align}
The CFT source is $\phi_0 = \phi_{\{0\}(\epsilon)}$ while $\phi_{\{n\}(3+\epsilon)}$ are vev coefficients. We have also included the first subleading in $\lambda$ source $\phi_{\{1\}(\epsilon)}$. All remaining coefficients are determined locally in terms of these by means of the equation of motion,
\begin{align}
\phi_{(2+\epsilon)} & = \frac{\partial^2 \phi_{(\epsilon)}}{2}, \\
\phi_{\{1\}(2 \epsilon)} & = \frac{\phi_0^2}{\epsilon(3-\epsilon)}, \label{e:phi12e} \\
\phi_{\{1\}(2+ 2 \epsilon)} & = \frac{\partial^2 \phi_{\{1\}(2 \epsilon)} + 2 \phi_0 \phi_{\{0\}(2+\epsilon)}}{(1 - \epsilon)(2 + \epsilon)}, \\
\phi_{\{1\}(3 + 2\epsilon)} & = -\frac{2 \phi_0 \phi_{\{0\}(3 + \epsilon)}}{\epsilon(3 + \epsilon) }.
\end{align}

First let us actually see what would happen if one tried to extract the divergence in the vev coefficient $\phi_{\{1\}(3 + \epsilon)}$ without carrying out the renormalization of the source. Equation \eqref{e:DivPi} would imply that
\begin{equation} \label{e:vevDiv}
\phi_{\{1\}(3 + \epsilon)} = - \phi_{\{1\}(3 + 2\epsilon)} + O(\epsilon^0) = \frac{2 \phi_0 \phi_{\{0\}(3 + \epsilon)}}{3 \epsilon} + O(\epsilon^0).
\end{equation}
This would suggest that the divergence in the 3-point function equals
\begin{equation}
- \frac{2 \lambda}{3 \epsilon} (k_2^3 + k_3^3).
\end{equation}
In comparison with \eqref{e:O3reg}, clearly the factor proportional to $k_1^3$ is missing. Such a factor can be obtained by turning on a subleading source $\phi_{\{1\}(\epsilon)}$. To find the value of this source we use equation \eqref{e:RedefSrc}. From the radial expansion \eqref{e:exPhiExp} we see that there is a single term with the power $\epsilon$-close to the source term, namely $\phi_{\{1\}(2 \epsilon)}$. As expected it is locally expressible in terms of the CFT source $\phi_0$ according to \eqref{e:phi12e}. Therefore we need
\begin{equation}
\phi_{\{1\}(\epsilon)} = - \mu^{-\epsilon} \phi_{\{1\}(2\epsilon)} + O(\epsilon^0) = - \frac{\phi_0^2 \mu^{-\epsilon}}{3 \epsilon}  + O(\epsilon^0).
\end{equation}
With the redefined source the full bulk solution $\Phi_{\{1\}}$ is given by the sum \eqref{e:field_decomp} where the inhomogeneous part follows from \eqref{e:Phinexp}, while the homogeneous piece reads,
\begin{equation}
\Phi_{\{0\}} \left[ - \lambda \frac{\mu^{-\epsilon}}{3 \epsilon} \phi_0^2 \right](z, \bs{k}) = - \lambda \frac{\mu^{-\epsilon}}{3 \epsilon} K_{\dreg,\Dreg}(z, \bs{k}) (\phi_0 \ast \phi_0)(\bs{k}).
\end{equation}
This homogeneous solution contains a vev coefficient of order $3 + \epsilon$, which contributes to the 3-point function. Now the argument \eqref{e:vevDiv} can be used, by considering all terms to be sourced by the combined $\phi_{\{0\}(\epsilon)} + \lambda \phi_{\{1\}(\epsilon)}$. One finds
\begin{equation} \label{e:exVevCf}
\phi_{\{1\}(3 + \epsilon)}[\phi_0] - \frac{\mu^{-\epsilon}}{3 \epsilon} \phi_{\{0\}(3 + \epsilon)}[\phi_0^2] = - \phi_{\{1\}(3 + 2\epsilon)}[\phi_0] + O(\epsilon^0),
\end{equation}
which correctly leads to the divergence in $\phi_{\{1\}(3 + \epsilon)}[\phi_0]$ to be equal to that found in \eqref{e:O3reg}.

\subsubsection{Renormalization}

From the point of view of the bulk theory we have turned on the subleading source $\phi_{\{1\}(\epsilon)}$. The total source reads
\begin{equation} \label{e:src3redef}
\phi_{(\epsilon)} = \phi_0 + \lambda \mu^{-\epsilon} \left( - \frac{1}{3 \epsilon} + a_0 \right) \phi_0^2 + O(\lambda^2),
\end{equation}
where $a_0$ is an arbitrary scheme-dependent constant. The bulk solution $\Phi_{\{1\}}$ remains now continuous when $\epsilon \rightarrow 0$. This changes the value of the 1-point function with sources, since according to \eqref{e:1ptRen} we now have
\begin{align}
\frac{\delta \Phi(z, \bs{u})}{\delta \phi_0(\bs{x})} = z^{\dreg - \Dreg} \delta(z) \delta(\bs{u} - \bs{x}) \left[ 1 + 2 \lambda \mu^{-\epsilon} \left( - \frac{1}{3 \epsilon} + a_0 \right) \phi_0(\bs{u}) + O(\lambda^2) \right].
\end{align}
Therefore, the 1-point function $\< \O \>_s$ in the renormalized theory, which is finite in the $\epsilon \rightarrow 0$ limit, is related to the 1-point function $\< \O \>_{s, \text{reg}}$ of the regulated theory,
\begin{equation} \label{e:exO1}
\< \O \>_s = - 3 \left[ 1 + 2 \lambda \mu^{-\epsilon} \left( - \frac{1}{3 \epsilon} + a_0 \right) \phi_0 \right] \phi_{(3 + \epsilon)} [ \phi_{(\epsilon)} ] + O(\lambda^2).
\end{equation}
The vev coefficient $\phi_{(3 + \epsilon)}$ is a functional of the entire source $\phi_{(\epsilon)}$. The renormalized source for the dual operator remains $\phi_0 = \phi_{\{0\}(\epsilon)}$. 

When two functional derivatives of this expression are taken, one finds that all divergences cancel and a finite $\epsilon \rightarrow 0$ limit exists. While the vev coefficient $\phi_{(3 + \epsilon)} [ \phi_{(\epsilon)} ]$ remains divergent according to \eqref{e:exVevCf}, the prefactor emerging from the source renormalization cancels the divergence. In other words, while $W_{\text{div}}$ in \eqref{e:Wdiv} is divergent, the regulated functional after source redefinition \eqref{e:WdivRedef} is finite, due to the local terms in \eqref{e:1ptRenA}. This means that $W_{WZ} = 0$ and no anomalies are present.

Now, however, we would like to discuss how this procedure relates to the CFT renormalization. By looking at the divergence \eqref{e:O3reg} we can write down a counterterm that needs to be added to the regulated generating functional $W_{\text{reg}}$ in order to cancel the divergence in the 3-point function. To be precise,
\begin{equation} \label{e:exW3}
W[\phi_0] = W_{\text{reg}}[\phi_0] - \lambda \mu^{-\epsilon} \left( - \frac{1}{3 \epsilon} + a_0 \right) \int \D^{3 + 2 \epsilon} \bs{x} \phi_0^2 \< \O \>_{\text{reg}, s},
\end{equation}
where $a_0$ is an arbitrary scheme-dependent constant. By taking minus one functional derivative we find \eqref{e:exO1}. When minus three functional derivatives with respect to the source $\phi_0$ are taken, the second term cancels the divergence in the regulated 3-point function following from the first term. 

The above expression is not particularly useful. It shows that the divergence is canceled by a local term, but it is not clear how it is related either to the holographic theory or the dual CFT. From the point of view of the dual field theory we should consider the multiplicative renormalization by an insertion of the $Z$ factor,
\begin{equation}
W = \< \exp \left( - \int \D^{3 + 2 \epsilon} \bs{x} \phi_{0} Z[\lambda \phi_0 \mu^{-\epsilon}] \O \right) \>_{\text{reg}}.
\end{equation}
And indeed, when expanded in $\lambda$ this generating functional agrees with \eqref{e:exW3} provided that
\begin{equation}
Z(g) = 1 + g \left( - \frac{1}{3 \epsilon} + a_0 \right) + O(g^2).
\end{equation}
The counterterm action \eqref{e:exW3} follows then simply from expanding the regularized action $W_{\text{reg}}[\phi_{(\epsilon)}[\phi_0]]$ as expected from \eqref{e:WctZ}.

The final renormalized 3-point function reads
\begin{align}
& \lla \O(\bs{k}_1) \O(\bs{k}_2) \O(\bs{k}_3) \rra = \frac{2 \lambda}{3} \left[ Q_1  + (k_1^3 + k_2^3 + k_3^3) \left( \log \frac{k_1 + k_2 + k_3}{\mu} + \gamma_E - \frac{4}{3} - 3 a_0 \right) \right],
\end{align}
where $Q_1$ is given by \eqref{e:Q1}.

By means of the equation \eqref{e:betaMargin} we find the beta function,
\begin{equation}
\mu \frac{\D}{\D \mu} \phi_0 = \beta_{\phi_0} = - \frac{\lambda}{3} \phi_0^2 \mu^{-\epsilon} + O(\phi_0^3).
\end{equation}
The explicit renormalization scale appears, since $\phi_0$ is a dimensionful coupling. If we define dimensionless quantity $g = \phi_0 \mu^{-\epsilon}$, then the beta function reads
\begin{equation}
\mu \frac{\D}{\D \mu} g = \beta_g = - \epsilon g - \frac{\lambda}{3} g^2 + O(g^3),
\end{equation}
a form more familiar from the perspective of a dimensionally renormalized QFT.

\subsection{4-point function}

Finally we want to apply the method of dimensional renormalization to the 4-point function. The computations are technically more involved but straightforward in execution. The regulated 4-point function reads
\begin{equation} \label{e:reg4pt}
\lla \O(\bs{k}_1) \O(\bs{k}_2) \O(\bs{k}_3) \O(\bs{k}_4) \rra_{\text{reg}} = 4 \left[ I(\bs{k}_1, \bs{k}_2; \bs{k}_3, \bs{k}_4) + I(\bs{k}_1, \bs{k}_3; \bs{k}_2, \bs{k}_4) + I(\bs{k}_1, \bs{k}_4; \bs{k}_2, \bs{k}_3) \right],
\end{equation}
where the expressions $I(\bs{k}_1, \bs{k}_2; \bs{k}_3, \bs{k}_4)$ correspond to three distinct Witten diagrams in Figure \ref{fig:4}. In terms of propagators one has
\begin{align} \label{e:reg3i}
I(\bs{k}_1, \bs{k}_2; \bs{k}_3, \bs{k}_4) & = \int_0^{\infty} \D z z^{-\dreg-1} K_{\dreg,\Dreg}(z, k_1) K_{\dreg,\Dreg}(z, k_2) \times \nn\\
& \qquad\qquad \times \int_0^{\infty} \D \zeta \z^{-\dreg-1} G_{\dreg, \Dreg}(z, |\bs{k}_3 + \bs{k}_4|; \z) K_{\dreg,\Dreg}(\z, k_3) K_{\dreg,\Dreg}(\z, k_4).
\end{align}
By definition the integral is symmetric under two transpositions $\bs{k}_1 \leftrightarrow \bs{k}_2$ or $\bs{k}_3 \leftrightarrow \bs{k}_4$ as well as a joint transposition $(\bs{k}_1, \bs{k}_2) \leftrightarrow (\bs{k}_3, \bs{k}_4)$.

As in the case of the 3-point function one can explicitly evaluate the 4-point function. Its finite part is rather long and unwieldy, so let us concentrate on the divergences.

Divergences in the regulated expression for the 4-point function is best expressed in terms of 2- and 3-point functions. Such a representation is however not unique. This is because one can freely move the contribution proportional to 2-point functions between various terms. The total divergence is obviously fixed and unique, so it is only a problem of the representation that becomes ambiguous. To be specific, the divergence equals
\begin{align} \label{e:O4reg}
& \lla \O(\bs{k}_1) \O(\bs{k}_2) \O(\bs{k}_3) \O(\bs{k}_4) \rra_{\text{reg}} = 2 \lambda a_1 \left[ \lla \O(\bs{k}_1) \O(\bs{k}_2) \O(- \bs{k}_1 - \bs{k}_2) \rra_{\text{reg}} + 5 \text{ permutations} \right] \nn\\
& \qquad\qquad - 4 \lambda^2 a_1^2 \left[ \lla \mathcal{O}(\bs{k}_1 + \bs{k}_2) \mathcal{O}(- \bs{k}_1 - \bs{k}_2) \rra_{\text{reg}} + 2 \text{ permutations} \right] \nn\\
& \qquad\qquad - 6 \lambda^2 a_2 \left[ \lla \mathcal{O}(\bs{k}_1) \mathcal{O}(- \bs{k}_1) \rra_{\text{reg}} + 3 \text{ permutations} \right] + \text{finite},
\end{align}
where
\begin{align}
a_1 & = - \frac{1}{3 \epsilon} + a_0, \label{e:a1} \\
a_2 & = \frac{1}{9 \epsilon^2} - \frac{1 + 18 c}{27 \epsilon}. \label{e:a2}
\end{align}
Here $a_0$ is an arbitrary number that parametrizes the ambiguity of the representation. When the correlation function is expanded in terms of its actual momentum space expressions, $a_0$ cancels against various terms and the actual divergence is fixed, unique and $a_0$-independent. Notice however that we use here the same symbol that we introduced to denote scheme-dependent constant in \eqref{e:src3redef}. Indeed, we will find that $a_1$ is equal to the renormalization constant for the 3-point function that we found in \eqref{e:src3redef}.

Note that this expression contains 3 distinct types of terms: a single terms containing six 3-point functions related by a symmetry and two different terms containing various combinations of 2-point functions. Notice that all terms are in principle of order $\epsilon^{-2}$ as the regulated 3-point function itself has a linear divergence in $\epsilon$.

Using the regulated expression we may immediately write down a counterterm in the genrating function that renormalizes both 3- and 4-point functions. We find
\begin{align} \label{e:exW}
W[\phi_0] & = W_{\text{reg}}[\phi_0] - \lambda a_1 \mu^{-\epsilon} \int \D^{3 + 2 \epsilon} \bs{x} \phi_0^2 \< \O \>_{\text{reg}, s} - \lambda^2 a_2 \mu^{-2\epsilon} \int \D^{3 + 2 \epsilon} \bs{x} \phi_0^3 \< \O \>_{\text{reg}, s} \nn\\
& \qquad\qquad + \: \frac{1}{2} \lambda^2 a_1^2 \mu^{-2 \epsilon} \int \D^{3 + 2 \epsilon} \bs{x} \phi_0^2(\bs{x}) \int \D^{3 + 2 \epsilon} \bs{y} \phi_0^2(\bs{y}) \< \O(\bs{x}) \O(\bs{y}) \>_{\text{reg}, s} + O(\phi_0^5).
\end{align}
When three or four functional derivatives are taken, this expression leads to finite correlation functions. As in the case of 3-point functions, we now want to analyze the bulk and the boundary systems from their respective points of view and recover the correct counterterms.

Formally, counterterm constants in this expression may contain some finite, subleading pieces that can be adjusted by a change of the renormalization scale. Therefore, without loss of generality we may use \eqref{e:a1} and \eqref{e:a2} without any finite pieces. In this way the resulting expressions will be more comprehensive.

\subsubsection{Radial expansion}

For the bulk analysis we have to carry out a radial expansion of the bulk field up to order $\lambda^2$. The zeroth and first order solution was already analyzed in section \ref{sec:exRadial}. Up to the order $\lambda^2$ one finds
\begin{align}
\Phi & = z^{\epsilon} (\phi_{0} + \lambda \phi_{\{1\}(\epsilon)} + \lambda^2 \phi_{\{2\}(\epsilon)}) + z^{2+\epsilon} (\phi_{\{0\}(2+\epsilon)} + \lambda \phi_{\{1\}(2+\epsilon)} + \lambda^2 \phi_{\{2\}(2+\epsilon)} ) \nn\\
& \qquad\qquad\qquad\qquad + z^{3 + \epsilon} ( \phi_{\{0\}(3 + \epsilon)} + \lambda \phi_{\{1\}(3 + \epsilon)} + \lambda^2 \phi_{\{2\}(3 + \epsilon)} ) + O(z^4) \nn\\
& \qquad\qquad + \lambda \left[ z^{2 \epsilon} ( \phi_{\{1\}(2 \epsilon)} + \lambda \phi_{\{2\}(2 \epsilon)} ) + z^{2+2\epsilon} ( \phi_{\{1\}(2 + 2\epsilon)} + \lambda \phi_{\{2\}(2 + 2\epsilon)} ) \right.\nn\\
& \qquad\qquad\qquad\qquad \left. + z^{3+2\epsilon} ( \phi_{\{1\}(3 + 2\epsilon)} + \lambda \phi_{\{2\}(3 + 2\epsilon)} ) + O(z^4) \right] \nn\\
& \qquad\qquad + \lambda^2 \left[ z^{3\epsilon} \phi_{\{2\}(3\epsilon)} + z^{2+3\epsilon} \phi_{\{2\}(2 + 3\epsilon)} + z^{3+3\epsilon} \phi_{\{2\}(3 + 3\epsilon)} + O(z^4) \right] + O(\lambda^3).
\end{align}
The CFT source is $\phi_0 = \phi_{\{0\}(\epsilon)}$ while $\phi_{\{n\}(3+\epsilon)}$ are vev coefficients.

In addition to local coefficients listed in section \ref{sec:exRadial}, we now find
\begin{align}
\phi_{\{2\}(2 \epsilon)} & = \frac{2 \phi_0 \phi_{\{1\}(\epsilon)}}{\epsilon (3 - \epsilon)}, \\
\phi_{\{2\}(3 \epsilon)} & = \frac{\phi_0^3}{\epsilon^2 (3 - \epsilon)(3 - 2 \epsilon)}, \\
\phi_{\{2\}(2+2 \epsilon)} & = \frac{\partial^2 \phi_{\{2\}(2\epsilon)} + 2 \phi_{\{0\}(2+\epsilon)} \phi_{\{1\}(\epsilon)} + 2 \phi_0 \phi_{\{0\}(2+\epsilon)}}{(1 - \epsilon)(2+\epsilon)}, \\
\phi_{\{2\}(2+3 \epsilon)} & = \frac{\partial^2 \phi_{\{2\}(3\epsilon)} + 2 \phi_{\{0\}(2+\epsilon)} \phi_{\{1\}(2\epsilon)} + 2 \phi_0 \phi_{\{1\}(2+\epsilon)}}{2 (1 - 2\epsilon)(1+\epsilon)}, \\
\phi_{\{2\}(3 + 2\epsilon)} & = -\frac{2 (\phi_0 \phi_{\{1\}(3 + \epsilon)} + \phi_{\{1\}(\epsilon)} \phi_{\{0\}(3 + \epsilon)})}{\epsilon(3 + \epsilon)}, \\
\phi_{\{2\}(3 + 3\epsilon)} & = \frac{3(1 - \epsilon) \phi_0^2 \phi_{\{0\}(3 + \epsilon)}}{\epsilon^2 (3 + 2 \epsilon)(3 + \epsilon)(3 - \epsilon)}.
\end{align}

\subsubsection{Source redefinition}

As in the analysis of the 3-point function, the entire divergence can be combined into a source redefinition. First let us observe that the perturbative solution $\Phi_{\{2\}}$ diverges at $\epsilon = 0$. Indeed, by the analysis of the 3-point function we have established that for the continuity of $\Phi_{\{1\}}$ we need $\phi_{\{1\}(\epsilon)} = a_1 \phi_0^2 \mu^{-\epsilon}$, where $a_1$ is given in \eqref{e:a1}. Hence, if we now consider terms of order $z^{\epsilon + O(\epsilon)}$ in $\Phi_{\{2\}}$ we find
\begin{equation}
\phi_{\{2\}(2\epsilon)} z^{2\epsilon} + \phi_{\{2\}(3\epsilon)} z^{3\epsilon} = - \frac{\phi_0^3}{9 \epsilon^2} (1 + \log z ) + \frac{1 + 18 a_0 + 3 \log \mu^2}{27 \epsilon} \phi_0^3 + O(\epsilon^0).
\end{equation}
This divergence will be removed from $\Phi_{\{2\}}$ if we turn on the source of order $\lambda^2$,
\begin{equation}
\phi_{\{2\}(\epsilon)} = a_2 \phi_0^3 \mu^{-2 \epsilon},
\end{equation}
where $a_2$ is given in \eqref{e:a2}. When $\phi_{\{2\}(\epsilon)} z^{\epsilon}$ is added to the expression above, it becomes finite. All in all, the redefined source, up to order $\lambda^2$ reads now
\begin{equation} \label{e:exRedefSrc}
\phi_{(\epsilon)} = \phi_0 + \lambda a_1 \mu^{-\epsilon} \phi_0^2 + \lambda^2 a_2 \mu^{-2 \epsilon} \phi_0^3 + O(\lambda^3),
\end{equation}
where constants $a_1$ and $a_2$ are given by \eqref{e:a1} and \eqref{e:a2}.

For pedagogical reasons we may now check that the renormalized 1-point function with sources is finite up to 4-point functions. We follow the equation \eqref{e:1ptRen}, which now gives
\begin{align}
& - \tfrac{1}{3} \< \O \>_s = \phi_{(3 + \epsilon)}[\phi_0] + \lambda \left[ \phi_{\{1\}(3 + \epsilon)}[\phi_0] + a_1 \mu^{-\epsilon} \left( 2 \phi_0 \phi_{\{0\}(3 + \epsilon)}[\phi_0] + \phi_{\{0\}(3 + \epsilon)}[\phi_0^2] \right) \right] \nn\\
& \qquad\qquad + \lambda^2 \left[ \phi_{\{2\}(3+\epsilon)}[\phi_0] + a_1 \left( 2 \phi_0 \phi_{\{1\}(3 + \epsilon)}[\phi_0, \phi_0] + \phi_{\{1\}(3 + \epsilon)}[\phi_0, \phi_0^2] \right) \right.\nn\\
& \qquad\qquad\qquad\qquad \left. + 2 a_1^2 \phi_0 \phi_{\{0\}(3+\epsilon)}[\phi_0^2] + 3 a_2 \phi_0^2 \phi_{\{0\}(3+\epsilon)}[\phi_0] + a_2 \phi_{\{0\}(3+\epsilon)}[\phi_0^3] \right] + O(\lambda^3).
\end{align}
In this expression we have two similar terms $\phi_{\{1\}(3 + \epsilon)}[\phi_0, \phi_0]$ and $\phi_{\{1\}(3 + \epsilon)}[\phi_0, \phi_0^2]$. What we mean here is as follows. The vev coefficient $\phi_{\{1\}(3 + \epsilon)}$ appears in $\Phi_{\{1\}}$, which is given in terms of two zeroth order solutions $\Phi_{\{0\}}$. Each $\Phi_{\{0\}}$ can be in principle sourced by two different sources, say $\phi_1$ and $\phi_2$, and hence $\phi_{\{1\}(3 + \epsilon)}[\phi_1, \phi_2]$ is a functional of these two sources. The term $\phi_{\{1\}(3 + \epsilon)}[\phi_0, \phi_0^2]$ can be traced back to the redefinition of the source required for the finiteness of $\Phi_{\{1\}}$.

Again, by taking minus three derivatives of the above expression one finds that the divergences cancel, leading to a finite 4-point function.

\subsubsection{Renormalization}

From the point of view of the CFT the redefined source \eqref{e:exRedefSrc} is identified with the bare source. The renormalized generating functional is then given by, 
\begin{equation} \label{e:exW4}
W[\phi_0] = \< \exp \left( - \int \D^{\dreg} \bs{x} \phi_0 Z[\lambda \phi_0 \mu^{-\epsilon}] \O \right) \>_{\text{reg}},
\end{equation}
where
\begin{equation}
Z(g) = 1 + a_1 g + a_2 g^2 + O(g^3),
\end{equation}
where $a_1$ and $a_2$ are given in \eqref{e:a1} and \eqref{e:a2}. At this point one can once again confirm that \eqref{e:exW4} reproduces \eqref{e:exW} up to $\lambda^2$ by expanding the functional in $\lambda$. Notice that a curious double integral term in \eqref{e:exW} appears as a consequence of the expansion of the exponential function that brings down the operator $\O$ twice.

Finally we want to calculate the beta function to this order in $\lambda$. At this point one may be worried by the appearance of a double pole in the counterterm coefficient $a_2$. Intuitively finiteness of a beta function follows from a cancellation between a pole in a renormalization constant and $\epsilon$ coming from a derivative $\mu \partial/(\partial \mu) \mu^{O(\epsilon)} = O(\epsilon)$. On the other hand we do expect the coefficient $a_2$ to have a double pole in $\epsilon$ simply as a composition of two integrals in \eqref{e:reg3i}, both of them resulting in a single pole in $\epsilon$. In fact this effect is responsible for $a_2 = a_1^2 + O(\epsilon^{-1})$. As we will now see it is this non-trivial term of order $\epsilon^{-1}$ that is visible to the beta function.

To calculate the beta function we may simply use equation \eqref{e:betaMargin}. By inverting a derivative of the bare source with respect to $\phi_0$ up to order $\lambda^2$ we arrive at the beta function,
\begin{align}
\beta_{\phi_0} & = \lambda \epsilon a_1 \phi_0^2 \mu^{-\epsilon} + 2 \lambda^2 \epsilon (a_2 - a_1^2) \phi_0^3 \mu^{-2 \epsilon} + O(\phi_0^4) \nn\\
& = - \frac{1}{3} \lambda \phi_0^2 \mu^{-\epsilon} - \frac{2}{27} \lambda^2 \phi_0^3 \mu^{-2 \epsilon} + O(\phi_0^4).
\end{align}
In terms of the dimensionless coupling $g = \phi_0 \mu^{-\epsilon}$,
\begin{equation}
\beta_{g} = - \epsilon g - \frac{1}{3} \lambda g^2 - \frac{2}{27} \lambda^2 g^3 + O(g^4).
\end{equation}
This expression is finite and has the form expected from the dimensionally regulated QFT.

\section*{Acknowledgements}

I would like to thank Fridrik Gautason, Paul McFadden, Marjorie Schillo, and Kostas Skenderis for enlightening discussions and valuable remarks on the manuscript. This work is supported in part by the National Science Foundation of Belgium (FWO) grant G.001.12 Odysseus, by the European Research Council grant no. ERC-2013-CoG 616732 HoloQosmos, and by COST Action MP1210 The String Theory Universe.

\providecommand{\href}[2]{#2}\begingroup\raggedright\endgroup

\end{document}